\pdfoutput=1

\documentclass{vldb}

\usepackage{balance}  
\usepackage{times,amsmath,epsfig}

\usepackage{graphics}
\usepackage{graphicx}
\usepackage[noadjust]{cite}
\usepackage{xspace}
\usepackage{subfigure}
\usepackage{amsfonts}
\usepackage{amssymb}
\usepackage{array}
\usepackage{morefloats}
\usepackage{etoolbox}
\usepackage{float}
\usepackage{caption}
\usepackage{epstopdf}
\usepackage[pdftitle={k-Nearest Neighbors on Road Networks: A Journey in Experimentation and In-Memory Implementation}]{hyperref}
\usepackage{url}
\usepackage{weiwAlgorithm}
\usepackage{multirow}
\usepackage{hhline}

\newcommand{\abvtblskp}{5pt}

\newcommand{\abvfigskp}{-1pt}
\newcommand{\belfigskp}{-1pt}
\newcommand\UB{\mathit{UB}}
\newcommand\LB{\mathit{LB}}

\newcommand{\smallHead}[1]{
	\vspace{0.5mm}\noindent\textbf{#1}}

\newcommand{\smallHeadIndent}[1]{
	\vspace{0.5mm}\textbf{#1}}

\begin{document}


\title{k-Nearest Neighbors on Road Networks: A Journey in Experimentation and In-Memory Implementation\\ \LARGE{\fontfamily{phv}\selectfont{Technical Report}}}

\numberofauthors{1}

\author{
	Tenindra Abeywickrama, Muhammad Aamir Cheema, David Taniar\\
	\fontsize{10}{10}\selectfont\rmfamily\itshape
	Faculty of Information Technology, Monash University, Australia\\
	\fontsize{8}{8}\selectfont\ttfamily\upshape
	\{tenindra.abeywickrama,aamir.cheema,david.taniar\}@monash.edu\\
}

\maketitle

\begin{abstract}
A $k$ nearest neighbor $(k$NN) query on road networks retrieves the $k$ closest points of interest (POIs) by their network distances from a given location. Today, in the era of ubiquitous mobile computing, this is a highly pertinent query. While Euclidean distance has been used as a heuristic to search for the closest POIs by their road network distance, its efficacy has not been thoroughly investigated. The most recent methods have shown significant improvement in query performance. Earlier studies, which proposed disk-based indexes, were compared to the current state-of-the-art in main memory. However, recent studies have shown that main memory comparisons can be challenging and require careful adaptation. This paper presents an extensive experimental investigation in main memory to settle these and several other issues. We use efficient and fair memory-resident implementations of each method to reproduce past experiments and conduct additional comparisons for several overlooked evaluations. Notably we revisit a previously discarded technique (IER) showing that, through a simple improvement, it is often the best performing technique.
\end{abstract}

\section{Introduction}\label{sec:intro}
Cisco reports that more than half a billion mobile devices were activated in $2013$ alone, and $77\%$ of those devices were smartphones. Due to the surge in adoption of smartphones and other GPS-enabled devices, and cheap wireless network bandwidth, map-based services have become ubiquitous. For instance, the GlobalWebIndex reported that Google Maps was the most used smartphone app in 2013 with $54\%$ of smartphone users having used it~\cite{globalwebindex}. Finding nearby facilities (e.g., restaurants, ATMs) are among the most popular queries issued on maps. Due to their popularity and importance, $k$ nearest neighbor ($k$NN) queries, which find the $k$ closest points of interest (objects) to a given query location, have been extensively studied in the past.

While related to the shortest path problem in many ways, the $k$NN problem on road networks introduces new challenges. Since the total number of objects is usually much larger than $k$ it is not efficient to compute the shortest paths (or network distances) to all objects to determine which are $k$NNs. The challenge is to not only ignore the objects that cannot be $k$NNs but also the road network vertices that are not associated with objects. Recently, there has been a large body of work to answer $k$NN queries on road networks. Some of the most notable algorithms include \emph{Incremental Network Expansion} (INE)~\cite{papadias2003ine}, \emph{Incremental Euclidean Restriction} (IER)~\cite{papadias2003ine}, \emph{Distance Browsing}~\cite{samet2008distbrws}, \emph{Route Overlay and Association Directory} (ROAD)~\cite{lee2009road,lee2012road}, and \emph{G-tree}~\cite{zhong2015gtree,zhong2013gtree}. In this paper, we conduct a thorough experimental evaluation of these algorithms. This is the extended technical report of a conference paper \cite{abey2016exp}.

\subsection{Motivation}

\smallHead{1. Neglected Competitor.}
IER \cite{papadias2003ine} was among the first $k$NN algorithms on road networks. It has often been the worst performing method and as a result is no longer included in comparisons. The basic idea of IER is to compute shortest path distances using Dijkstra's algorithm to the closest objects in terms of Euclidean distance. Although many significantly faster shortest path algorithms have been proposed in recent years, surprisingly, IER has never been compared against other $k$NN methods using any algorithm other than Dijkstra. To ascertain the true performance of IER it must be integrated with state-of-the-art shortest path algorithms.

\smallHead{2. Discrepancies in Existing Results.}
We note several discrepancies in the experimental results reported in some of the most notable papers on this topic. ROAD is seen to perform significantly worse than Distance Browsing and INE in~\cite{zhong2015gtree}. But according to~\cite{lee2012road}, ROAD is experimentally superior to both Distance Browsing and INE. The results in both~\cite{lee2012road} and~\cite{zhong2015gtree} show Distance Browsing has worse performance than INE. In contrast, Distance Browsing is shown to be more efficient than INE in~\cite{samet2008distbrws}. These contradictions identify the need for reproducibility.

\smallHead{3. Implementation Does Matter.}
Similar to a recent study~\cite{sidlauskas2014impl}, we observe that simple implementation choices can significantly affect algorithm performance. For example, G-tree utilizes \emph{distance matrices} that can be implemented using either hash-tables or arrays and, on the surface, both seem reasonable choices. However the array implementation in fact performs more than an order of magnitude faster than the hash-table implementation. We show that this is due to data locality in G-tree's index and its impact on cache performance. In short, seemingly innocuous choices can drastically change experimental outcomes. We also believe discrepancies reported above may well be due to different choices made by the implementers. Thus it is critical to provide a fair comparison of existing $k$NN algorithms using careful in-memory implementations.

\smallHead{4. Overlooked Evaluation Measures/Settings.}
All methods studied in this paper decouple the road network index from that of the set of objects, i.e. one index is created for the road network and another to store the set of objects. Although existing studies evaluate the road network indexes, no study evaluates the behaviour of each individual object index. The construction time and storage cost for these \textit{object indexes} may be critical information for developers when choosing methods, especially for object sets that change regularly. Additionally $k$NN queries have not been investigated for travel time graphs (only travel distance), which is also a common scenario in practice. Finally the more recent techniques (G-tree and ROAD) did not include comparisons for real-world POIs.

\subsection{Contributions}
Below we summarize the contributions we make in this paper.

\smallHead{1. Revived IER:}
We investigate IER with several efficient shortest path techniques for the first time (see Section \ref{sec:ier_imprv}). We show that the performance of IER is significantly improved when better shortest path algorithms are used. This occurs to the point that IER is the best performing method in most settings, including travel time road networks where Euclidean distance is a less effective lower bound. 

\smallHead{2. Highly Optimised Algorithms Open-Sourced:}
We present efficient implementations of five of the most notable methods (IER, INE, Distance Browsing, ROAD and G-tree). Firstly we have carefully implemented each method for efficient performance in main memory as described in Section \ref{sec:impl}. Secondly we thoroughly checked each algorithm and made various improvements that are applicable in any setting, as documented in Appendix \ref{sec:app:imprv}. The source code and scripts to run experiments have been released as open-source~\cite{oursource}, making our best effort to ensure it is modular and re-usable.

\smallHead{3. Reproducibility Study:}
With efficient implementations of each algorithm, we repeat many experiments from past studies on many of the same datasets in Section \ref{sec:exp}. Our results provide a deeper understanding of the state-of-the-art with new insights into the weaknesses and strengths of each technique. We also show that there is room to improve $k$NN search heuristics by demonstrating that G-tree can be made more efficient by using Euclidean distances.

\smallHead{4. Extended Experiments and Analysis:}
Our comprehensive experimental study in Section \ref{sec:exp} extends beyond past studies by: 1) comparing object indexes for the first time; 2) revealing new trends by comparing G-tree with another advanced method (ROAD) on larger datasets for the first time; 3) evaluating all methods (including ROAD and G-tree) on real-world POIs; and 4) evaluating applicable methods on travel time road networks.

\smallHead{5. Guidance on Main-Memory Implementations:}
In Section \ref{sec:impl} we also demonstrate how simple choices can severely impact algorithm performance. We share an in-depth case study to give insights into the relationship between algorithms and in-memory performance with respect to data locality and cache efficiency. Additionally we highlight the main choices involved and illustrate them through examples and experimental results, to provide hints to future implementers. Significantly, these insights are potentially applicable to any problem, not just those we study here.

\section{Background}\label{sec:bg}

\subsection{Problem Definition}\label{sec:bg:def}

We represent a road network as a connected undirected graph $G = (V,E)$ where $V$ is the set of vertices and $E$ is the set of edges. For two adjacent vertices $u,v \in V$, we define the edge between them as $e(u,v)$, with weight $w(u,v)$ representing any positive measure such as distance or travel time. We define the shortest path distance, hereafter network distance, between any two vertices $u,v \in V$ as $d(u,v)$, the minimum sum of weights connecting $u$ and $v$. For conceptual simplicity, similar to the existing studies~\cite{zhong2015gtree,samet2008distbrws}, we assume that each object (i.e., POI) and query is located on some vertex in $V$. Given a query vertex $q$ and a set of object vertices $O$, a $k$NN query retrieves the $k$ closest objects in $O$ based on their network distances from $q$.

\subsection{Scope}\label{sec:bg:scope}

We separate existing $k$NN techniques into two broad categories based on the indexing they use: 1) blended indexing; and 2) decoupled indexing. Techniques that use blended indexing~\cite{kolahdouzan2004vknn,hu2006distidx,cho2005unicons} create a single index to store the objects as well the road network. For example, VN\textsuperscript{3} \cite{kolahdouzan2004vknn} is a notable technique that uses a network Voronoi diagram based on the set of objects to partition the network. In contrast, decoupled indexing techniques~\cite{papadias2003ine,lee2012road,zhong2015gtree,samet2008distbrws} use two separate indexes for the object set and road network, which is more practical and has several advantages as explained below.

Firstly, a real-world $k$NN query may be applied to one of many object sets, e.g., return the $k$ closest restaurants or locate the nearest parking space. Blended indexing must repeatedly index the road network for each type of object, entailing huge space and pre-processing time overheads. But decoupled indexing requires only one road network index regardless of the number of object sets, resulting in lower storage and pre-processing cost. Secondly, if there is any change in an object set, blended indexing must update the whole index and reprocess the entire road network, whereas decoupled techniques need only update the object index. For example, the network-based Voronoi diagram must be updated resulting in expensive re-computations \cite{kolahdouzan2004vknn}. Conversely, in decoupled indexing, the object indexes (e.g., R-tree) are typically much cheaper to update. The problem is more serious for object sets that change often, e.g., if the objects are the nearest \emph{available} parking spaces.

Due to these advantages, all recent $k$NN techniques use decoupled indexing. In this paper, we focus on the most notable $k$NN algorithms that employ decoupled indexing. These algorithms either employ an expansion-based method or a heuristic best-first search (BFS).  The expansion-based methods encounter $k$NNs in network distance order. Heuristic BFS methods instead employ heuristics to evaluate the most promising $k$NN candidates, not necessarily in network distance order, potentially terminating sooner. We study the five most notable methods which include two expansion-based methods, INE \cite{papadias2003ine} and ROAD \cite{lee2012road}, and three heuristic BFS methods, IER \cite{papadias2003ine}, Distance Browsing (DisBrw) \cite{samet2008distbrws} and G-tree \cite{zhong2015gtree}.

Given the rapid growth in smartphones and the corresponding widespread use of map-based services, applications must employ fast in-memory query processing to meet the high query workload. In-memory processing has become viable due to the increases in main-memory capacities and its affordability. Thus, we limit our study to in-memory query processing. However, we remark that
disk-based settings are also important but are beyond the scope of this paper mainly due to the space limitation. 
\section{Methods}\label{sec:methods}
We now describe the main ideas behind each method evaluated by our study. Some methods propose a road network index and a $k$NN query algorithm to use it. In some cases, such as G-tree, we refer to both the index and $k$NN algorithm by the same name.

\subsection{Incremental Network Expansion}\label{sec:methods:ine}

Incremental Network Expansion (INE) \cite{papadias2003ine} is a method derived from Dijkstra's algorithm. As in Dijkstra, INE maintains a priority queue of the vertices seen so far (initialised with the query vertex $q$). The search is expanded to the nearest of these vertices $v$. If $v \in O$ then it is added to the result set as one of the $k$NNs and if $v$ is the $k$th object then the search is terminated. Otherwise the edges of $v$ are used to relax the distances to its neighbors and the expansion continues. As in Dijkstra's algorithm, relaxation involves updating the minimum network distances to the neighbors of $v$ using the network distance through $v$. The disadvantage of INE is that it visits all nodes closer to $q$ than the $k$th object, which may be considerable if this object is far from $q$.
\subsection{Incremental Euclidean Restriction}\label{sec:methods:ier}

Incremental Euclidean Restriction (IER) \cite{papadias2003ine} uses Euclidean distance as a heuristic to retrieve candidates from $O$, as it is a lower bound on network distance for road networks with travel distance edges. Firstly, IER retrieves the Euclidean $k$NNs, e.g., using an R-tree \cite{samet2005book}. It then computes the network distance to each of these $k$ objects and sorts them in this order. This set becomes the candidate $k$NNs and the network distance to the furthest candidate (denoted as $D_k$) is an upper bound on the distance to the true $k$th nearest neighbor. Now, IER retrieves the next nearest Euclidean neighbor $p$. If the Euclidean distance to $p$ is $d_E(q,p)$ and $d_E(q,p) \geq D_k$, then $p$ cannot be a better candidate by network distance than any current candidate. Moreover, since it is the \textit{nearest} Euclidean neighbor, the search can be terminated. However, if $d_E(q,p) < D_k$ then $p$ may be a better candidate. In this case, IER computes the network distance $d(q,p)$. If $d(q,p) < D_K$, $p$ is inserted into the candidate set (removing the furthest candidate and updating $D_k$). This continues until the search is terminated or there are no more Euclidean NNs. 
\subsection{Distance Browsing}\label{sec:methods:silc}

Distance Browsing (DisBrw) \cite{samet2008distbrws} uses the Spatially Induced Linkage Cognizance (SILC) index proposed in \cite{sankaranarayanan2005silc} to answer $k$NN queries. \cite{sankaranarayanan2005silc} proposed an incremental $k$NN algorithm, which DisBrw improves upon by making fewer priority queue insertions.

\smallHeadIndent{SILC Index.} 
We first introduce the SILC index used by DisBrw. For a vertex $s \in V$, SILC pre-computes the shortest paths from $s$ to all other vertices. SILC assigns each adjacent vertex of $s$ a unique color. Then, each vertex $u\in V$ is assigned the same color as the adjacent vertex $v$ that is passed through in the shortest path from $s$ to $u$. Figure \ref{method:diagrams:silc} shows the coloring of the vertices for the vertex $s{=}v_6$ where each adjacent vertex of $v_6$ is assigned a unique color and the other vertices are colored accordingly. For example, the vertices $v_9$ to $v_{12}$ have the same color as $v_8$ (blue vertical stripes) because the shortest path from $v_6$ to each of these vertices passes through $v_8$ (for this example assume unit edge weights).

Observe that the vertices close to each other have the same color resulting in several contiguous regions of the same color. These regions are indexed by a region quadtree \cite{samet2005book} to reduce the storage space. The color of a vertex can be determined by locating the region in the quadtree that contains it. SILC applies the coloring scheme and creates a quadtree for each vertex of the road network. This requires $O(|V|^{1.5})$ space in total and, due to the all-pairs shortest path computation, $O(|V|^2\log|V|)$ pre-processing time.

{	
	\setlength{\abovecaptionskip}{\abvfigskp}
	\setlength{\belowcaptionskip}{-10pt}
	\begin{figure}[t]
		\centering
		\begin{minipage}{1\linewidth}
			\centering
			\includegraphics[width=1\linewidth]{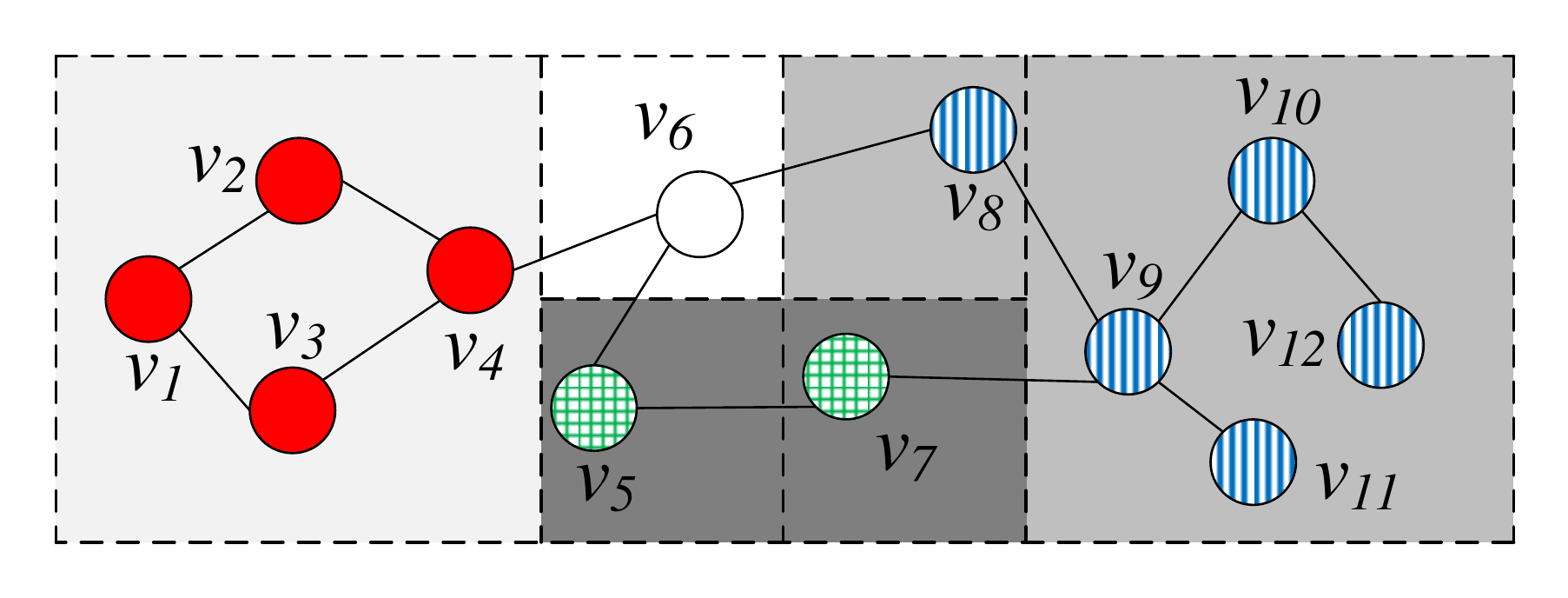}
			\captionof{figure}{SILC: Coloring Scheme and Quadtree for $v_6$}
			\label{method:diagrams:silc}
		\end{minipage}
	\end{figure}
}

To compute the  shortest path from $s$ to $t$, SILC uses the quadtree of $s$ to identify the color of $t$. The color of $t$ determines the first vertex $v$ on the shortest path from $s$ to $t$. To determine the next vertex on the shortest path, this procedure is repeated on the quadtree of $v$. For example, in Figure \ref{method:diagrams:silc}, the first vertex on the shortest path from $v_6$ to $v_{12}$ is $v_8$ because $v_{12}$ has the same color as $v_8$. The color of $v_{12}$ is found by locating the quadtree block containing $v_{12}$. The shortest path can be computed in $O(m\log{|V|})$ where $m$ is the number of edges on the shortest path \cite{samet2008distbrws}.

\smallHeadIndent{$k$NN Algorithm.}
To enable $k$NN search, DisBrw stores additional information in each quadtree. For each vertex $v$ contained in a quadtree block $b$, it computes the ratio of the Euclidean and network distances between the quadtree owner $s$ and $v$. It then stores the minimum and maximum ratios, $\lambda^-$ and $\lambda^+$ respectively, with $b$. Now, given any vertex $t$, DisBrw computes a \textit{distance interval} $[\delta^-,\delta^+]$ by multiplying the Euclidean distance from $s$ to $t$ by the $\lambda^-$ and $\lambda^+$ values of the block containing $t$. This interval defines a lower and upper bound on the network distance from $s$ to $t$ and can be used to prune objects that cannot be $k$NNs. The interval is \textit{refined} by obtaining the next vertex $u$ in the shortest path from $s$ to $t$ (as described earlier), computing an interval for $u$ to $t$, and then adding the known distance from $s$ to $u$ to the new interval. By refining the interval, it eventually converges to the network distance.

DisBrw used an \textit{Object Hierarchy} in \cite{samet2008distbrws} to avoid computing distance intervals for all objects. The basic idea was to compute distance intervals for regions containing objects, then visit the most promising regions (and recursively sub-regions) first. We found this method did not use the SILC index to its full potential. Instead we retrieve Euclidean NNs as candidate objects for which intervals are then computed. Otherwise, the DisBrw $k$NN algorithm proceeds exactly as in \cite{samet2008distbrws}. We refer the readers to Appendix \ref{sec:app:silc:enn} for full details and experimental comparisons with the original method. 

{
	\setlength{\abovecaptionskip}{\abvfigskp}
	\setlength{\belowcaptionskip}{-12pt}
	\begin{figure}[t]
		\centering
		\begin{minipage}{0.8\linewidth}
			\centering
			\includegraphics[width=1\linewidth]{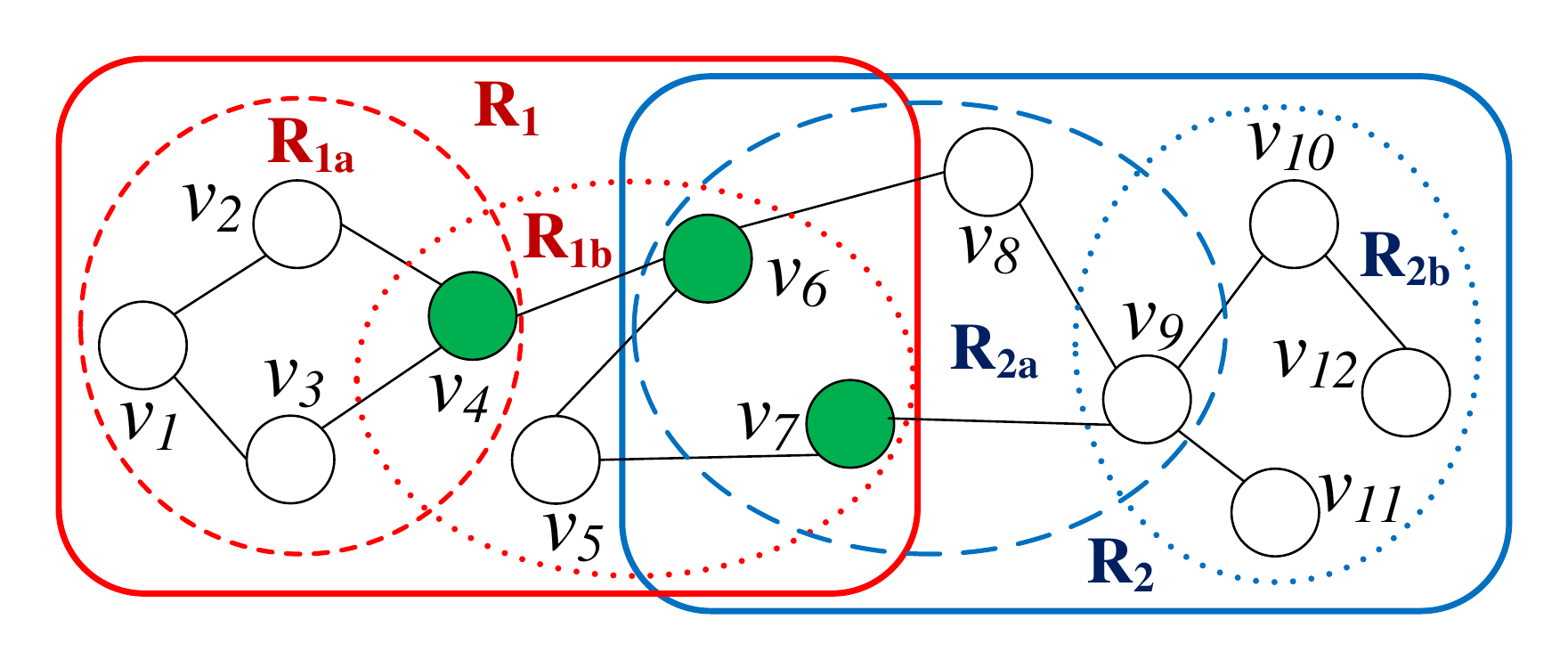}
			\captionof{figure}{ROAD}
			\label{method:diagrams:road}
		\end{minipage}%
	\end{figure}
}

\subsection{Route Overlay \& Association Directory}\label{sec:methods:road}

The search space of INE can be considerably large depending on the distance to the $k$th object. Route Overlay and Association Directory (ROAD) \cite{lee2009road,lee2012road} attempts to remedy this by bypassing regions that do not contain objects by using \textit{search space pruning}.

An \textit{Rnet} is a partition of the road network $G{=}(V,E)$, with every edge in $E$ belonging to at least one Rnet. Thus, an Rnet $R$ represents a set of edges $E_R \subseteq E$. $V_R$ is the set of vertices that are associated with edges in $E_R$. To create Rnets, ROAD partitions the road network $G$ into $f \geq 2$ Rnets, recursively partitioning resulting Rnets until a hierarchy of $l>1$ levels is formed (with $G$ being the root at level 0). Figure \ref{method:diagrams:road} shows Rnets (for $l{=}2$) for the graph in our running example. The enclosing boxes and ovals represent the set $V_R$ of each Rnet. Specifically, $R_1{=}\{v_1,\cdots,v_7\}$ and $R_2{=}\{v_6,\cdots,v_{12}\}$ are the child Rnets of the root $G$. Each of $R_1$ and $R_2$ are further divided into Rnets, e.g., $R_1$ is divided into $R_{1a}{=}\{v_1, v_2, v_3, v_4\}$ and $R_{1b}{=}\{v_4, v_5, v_6, v_7\}$.

For an Rnet $R$, a vertex $b \in V_R$ with an adjacent edge $e(b,v) \notin E_R$ is defined as a \textit{border} of $R$. For instance, $v_4$ is a border of $R_{1b}$ but $v_5$ is not. These borders form the set $B_R \subseteq V_R$, e.g., the border set of $R_{1b}$ consists of $v_4$, $v_6$ and $v_7$. ROAD computes the network distance between every pair of borders $b_i,b_j \in B_R$ in each Rnet and stores each as the \textit{shortcut} $S(b_i,b_j)$. Now any shortest path between two vertices $s,t \notin V_R$ involving a vertex $u \in V_R$ must enter $R$ through a border $b \in B_R$ and leave through a border $b' \in B_R$. So if a search reaches a border $b \in B_R$ the shortcuts associated with $b$, $S(b,b')$ $\forall$ $b' \in B_R$, can be traversed to bypass the Rnet $R$ while preserving network distances. For example, in Figure \ref{method:diagrams:road}, the borders of $R_{1b}$ are $v_4$, $v_6$ and $v_7$ (the colored vertices) and ROAD precomputes the shortcuts between all these borders. Suppose the query vertex is $v_1$ and the search has reached the vertex $v_4$. If it is known that $R_{1b}$ does not contain any object, the algorithm can bypass $R_{1b}$ by quickly expanding the search  to other borders of $R_{1b}$ without the need to access any non-border vertex of $R_{1b}$. E.g., using the shortcut between $v_4$ and $v_7$, the algorithm can compute the distance between $v_1$ to $v_7$ without exploring any vertex in $R_{1b}$.

Since child Rnets are contained by their parent Rnet, a border $b$ of an Rnet must be a border of some child Rnet at each lower level. For example, $v_6$ in Figure \ref{method:diagrams:road} is a border for $R_{1b}$ and its parent $R_1$. This allows the shortcuts to be computed in a bottom-up manner, where shortcuts at level $i$ are computed using those of level $i{+}1$, greatly reducing pre-computation cost. Only leaf Rnets require a Dijkstra's search on the original graph $G$.

ROAD uses a \emph{Route Overlay} index and an \emph{Association Directory} to efficiently compute $k$NNs. Recall that a vertex $v$ may be a border of more than one Rnet. The Route Overlay index stores, for each vertex $v$, the Rnets for which it is a border along with the \emph{shortcut trees} of $v$.  The Association Directory provides a means to check whether a given Rnet or vertex contains an object or not. The $k$NN algorithm proceeds incrementally from the query vertex $q$ in a similar fashion to INE. However, when ROAD expands to a new vertex $v$, instead of inspecting its neighbors, it consults the Route Overlay and Association Directory to find the highest level Rnet associated with it that does not contain any object. ROAD then relaxes all the shortcuts in this Rnet in a similar way to edges in INE, to bypass it. Of course when $v$ is not a border of any Rnet or if all Rnets associated with $v$ contain an object, it relaxes the edges of $v$ exactly as in INE. The search terminates when $k$ objects have been found or there are no further vertices to expand.

{
	\setlength{\abovecaptionskip}{\abvfigskp}
	\setlength{\belowcaptionskip}{-13pt}
	\begin{figure}[t]
		\centering
		\begin{minipage}{.8\linewidth}
			\centering
			\includegraphics[width=1\linewidth]{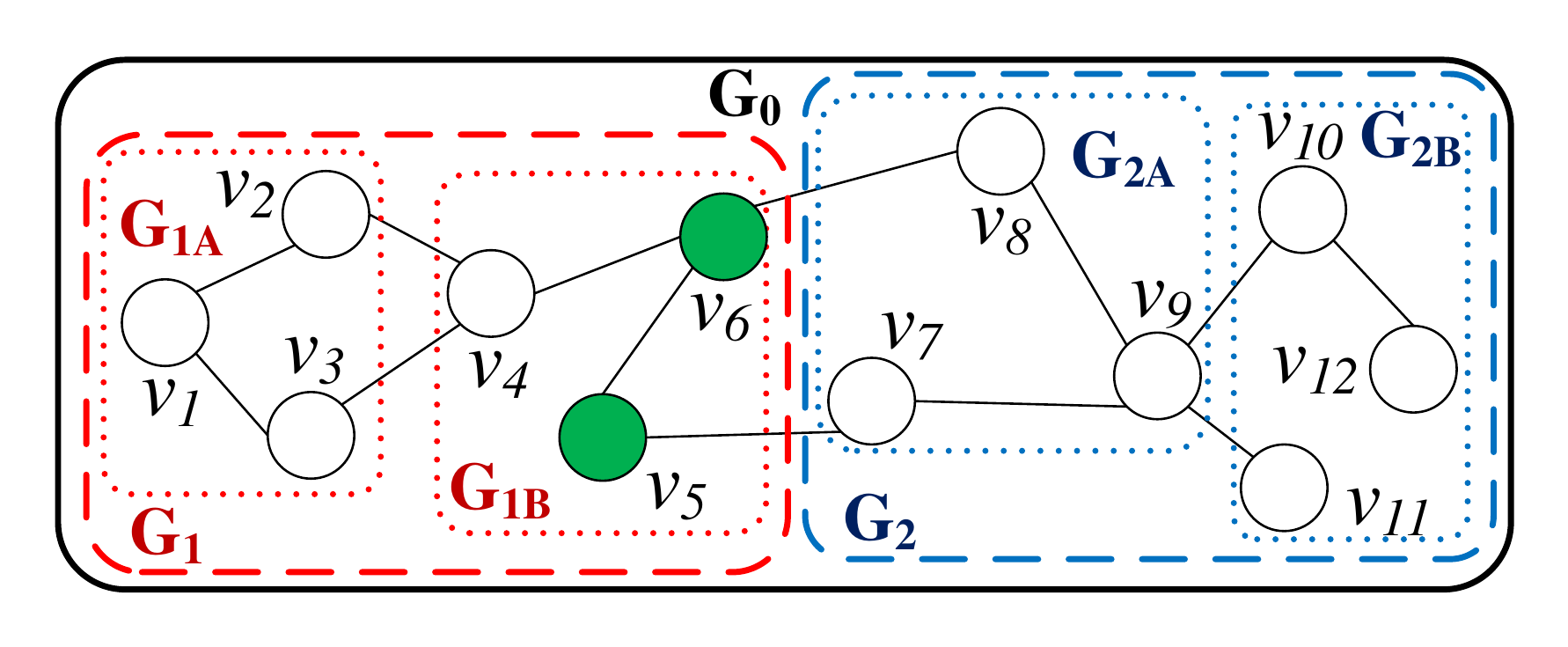}
			\captionof{figure}{G-tree}
			\label{method:diagrams:gtree}
		\end{minipage}
	\end{figure}
}

\subsection{G-tree}\label{sec:methods:gtree}

G-tree \cite{zhong2013gtree,zhong2015gtree} also employs graph partitioning to create a tree index that can be used to efficiently compute network distances through a hierarchy of subgraphs. The partitioning occurs in a similar way to that of ROAD where the input graph $G$ is partitioned into $f \geq 2$ subgraphs. Each subgraph is recursively partitioned until it contains no more than $\tau \geq 1$ vertices. For any subgraph $G_i$, $V_i \subseteq V$ is defined as the set of road network vertices contained within it. Any vertex $b \in V_i$ with an edge $e(b,v)$ where $v \notin V_i$ is defined as a border of $G_i$ and all such vertices form the set of borders $B_i$. Figure \ref{method:diagrams:gtree} shows an example where the colored vertices $v_5$ and $v_6$ are borders for the subgraph $G_1{=}\{v_1,\cdots,v_6\}$.

The partitioned subgraphs naturally form a tree hierarchy with each node in the G-tree associated with one subgraph. Note that we use \textit{node} to refer to the G-tree node while \textit{vertex} refers to road network vertices. Notably a non-leaf node $G_i$ does not need to store subgraph vertices, but only the set of borders $B_i$ and a \textit{distance matrix}. For non-leaf nodes, the distance matrix stores the network distance from each child node border to all other child node borders. For leaf nodes, it stores the network distance between each of its borders and the vertices contained in it.

Similar to the bottom-up computation of shortcuts in ROAD, the distance matrix of nodes at tree level $i$ can be efficiently computed by reducing the graph to only consist of borders at level $i{+}1$ using the distance matrices of that level. Only leaf nodes require a Dijkstra's search on the original graph. Given a planar graph and optimal partitioning method, G-tree is a height-balanced tree with a space complexity of $O(|V| \log |V|)$.
The similarities with ROAD are clear. One major difference is that G-tree uses its border-to-border distance matrices to ``assemble" shortest path distances by the path through the G-tree hierarchy. We refer the reader to the original paper~\cite{zhong2015gtree} for the details of the assembly method. 

Another key difference is the $k$NN algorithm. To support efficient $k$NN queries, G-tree introduces the \textit{Occurrence List}. Given an object set $O$, the Occurrence List of a G-tree node $G_i$ lists its children that contain objects, allowing empty nodes to be pruned. The $k$NN algorithm begins from the leaf node that contains $q$, using an Dijkstra-like search to retrieve leaf objects. However, we found this leaf search could be further optimised and detail our improved leaf search algorithm in Appendix \ref{sec:app:imprv:gtree:leaf_opt}. The algorithm then incrementally traverses the G-tree hierarchy from the source leaf. Elements (nodes or objects) are inserted into a priority queue using their network distances from $q$. The network distance to a G-tree node is computed using the assembly method by finding its nearest border to $q$. Queue elements are dequeued in a loop. If the dequeued element is a node, its Occurrence List is used to insert its children (nodes or object vertices) back into the priority queue. If the dequeued element is a vertex, it is guaranteed to be the next nearest object. The search terminates when $k$ objects are dequeued.

A useful property of assembling distances is that, given a path through the G-tree hierarchy, distances can be \textit{materialized} for already visited G-tree nodes. For example, given a query vertex $q$ and two $k$NN objects in the same leaf node, after locating one of them, the distances to the borders of this leaf need not be recomputed. 

\section{Datasets}\label{sec:datasets}
Here we describe the datasets used to supply the road network $G{=}(V,E)$ and set of object vertices $O \subseteq V$ for $k$NN querying.

{
	\setlength{\abovecaptionskip}{\abvtblskp}
	\setlength{\belowcaptionskip}{-13pt}
	\begin{table}[t]
		\centering
		\begin{tabular}{|c|c|c|c|} \hline
			\textbf{Name} & \textbf{Region} & \textbf{\# Vertices} & \textbf{\# Edges} \\ \hline
			DE & Delaware & 48,812 & 119,004 \\ \hline
			VT & Vermont & 95,672 & 209,288 \\ \hline
			ME & Maine & 187,315 & 412,352 \\ \hline
			CO & Colorado & 435,666 & 1,042,400 \\ \hline
			\textbf{NW} & North-West US & 1,089,933 & 2,545,844 \\ \hline
			CA & California \& Nevada & 1,890,815 & 4,630,444 \\ \hline
			E & Eastern US & 3,598,623 & 8,708,058 \\ \hline
			W & Western US & 6,262,104 & 15,119,284 \\ \hline
			C & Central US & 14,081,816 & 33,866,826\\ \hline
			\textbf{US} & United States & 23,947,347 & 57,708,624 \\ \hline
		\end{tabular}
		\caption{Road Network Datasets}
		\label{tab:datasets}
	\end{table}
}

\subsection{Real Road Networks}
We study $k$NN queries on 10 real-world road network graphs as listed in Table \ref{tab:datasets}. These were created for the 9th DIMACS Challenge \cite{dimacs9} from data publicly released by the US Census Bureau. Each network covers all types of roads, including local roads, and contains real edge weights for travel distances and travel times (both are used in our experiments). We also conduct in-depth studies for the United States (US) and North-West US (NW) road networks. The US dataset, covering the entire continental United States, is the largest with 24 million vertices. The NW road network (with 1 million vertices), covering Oregon and Washington, represents queries limited to a smaller region or  country. Notably this is the first time DisBrw has been evaluated on a network with more than $500,000$ vertices, previously not possible due to its high pre-processing cost (in terms of both space and time).

\subsection{Real and Synthetic Object Sets}\label{sec:datasets:obj}
We create object sets based on both real-world points of interest (POIs) and synthetic methods as described below.

\smallHeadIndent{Real-World POI Sets.} We created 8 real-world object sets (listed in Table \ref{tab:pois}) using data extracted from OpenStreetMap (OSM)~\cite{osm} for locations of real-world POIs in the United States. Each object set is associated with a particular type of POI, e.g., all fast food outlets. POIs were mapped to road network vertices on both the US and NW road networks using their coordinates. While real POIs can be obtained freely from OSM, it is not a propriety system. As a result the data quality can vary, e.g., the largest object sets in OSM may not be representative of the true largest object sets and the completeness of POI data may vary between regions. So, in addition to real-world object sets, we generate synthetic sets to make generalizable and repeatable observations for all road networks.

{
	\setlength{\abovecaptionskip}{\abvtblskp}
	\setlength{\belowcaptionskip}{-13pt}
	\begin{table}[t]
		\centering
		\begin{tabular}{|c|c|c|c|c|c|} \hline
			
			\multirow{2}{*}{\textbf{Object Set}} & \multicolumn{2}{c|}{\textbf{United States}} & \multicolumn{2}{c|}{\textbf{North-West US}} \\
			\hhline{~----}
			& Size & Density & Size & Density \\ \hline
			Schools & 160,525 & 0.007 & 4,441 & 0.004 \\ \hline
			Parks & 69,338 & 0.003 & 5,098 & 0.005 \\ \hline
			Fast Food & 25,069 & 0.001 & 1,328 & 0.001 \\ \hline
			Post Offices & 21,319 & 0.0009 & 1,403 & 0.001 \\ \hline
			Hospitals & 11,417 & 0.0005 & 258 & 0.0002 \\ \hline
			Hotels & 8,742 & 0.0004 & 460 & 0.0004 \\ \hline
			Universities & 3,954 & 0.0002 & 95 & 0.00009 \\ \hline
			Courthouses & 2,161 & 0.00009 & 49 & 0.00005 \\ \hline
		\end{tabular}
		\caption{Real-World Object Sets}
		\label{tab:pois}
	\end{table}
}

\smallHeadIndent{Uniform Object Sets.} A uniform object set is generated by selecting uniformly random vertices from the road network. As these objects are randomly selected road network vertices, they are likely to simulate real POIs, e.g., areas with more vertices have more POIs (e.g., cities) while those with fewer roads have less (e.g., rural areas). The density of objects sets $d$ is varied from $0.0001$ to $1$, where $d$ is the ratio of the number of objects $|O|$ to the number of vertices $|V|$ in the road network. High densities can simulate larger object sets which are common occurrences, e.g., ATM machines, parking spaces. Low densities correspond to the sparsely located POIs, e.g., post offices or restaurants in a particular chain. By decreasing the density we can simulate more difficult queries, as fewer objects imply longer distances and therefore larger search spaces. Uniform objects were used to evaluate G-tree in~\cite{zhong2013gtree,zhong2015gtree}.

\smallHeadIndent{Clustered Object Sets.} While some POIs may be uniformly distributed other types, such as fast food outlets, occur in clusters. To create such clustered object sets, given a number of clusters $|C|$, we select $|C|$ central vertices uniformly at random (as above). For each central vertex, we select several vertices (up to a maximum cluster size $C_{max}$) in its vicinity, by expanding outwards from it. This distribution was used to evaluate ROAD in~\cite{lee2012road}.

\smallHeadIndent{Minimum Object Distance Sets.} The worst-case $k$NN query occurs when the query location is remote. To simulate this we create minimum distance object sets as follows. We choose an approximate centre vertex $v_c$ by using the nearest vertex to the Euclidean centre of the road network. We find the furthest vertex $v_f$ from $v_c$ and set $D_{max}$ as the network distance from $v_c$ to $v_f$. For an object set $R_i$, $i \in [1,m]$, we choose $|O|$ objects such that the network distance from $v_c$ to each object in $R_i$ is at least $\frac{D_{max}}{2^{m-i{+}1}}$. For example for $m{=}5$, the set $R_1$ contains objects within the range $(\frac{D_{max}}{32},\allowbreak D_{max}]$. Thus we investigate the effect of increasing minimum object distance by comparing query time on $R_i$ with increasing $i$.

\section{IER Revisited}\label{sec:ier_imprv}

Network distance computation is a critical part of IER. However, to the best of our knowledge, all existing studies~\cite{papadias2003ine,samet2008distbrws,lee2009road,lee2012road} employ Dijkstra's algorithm to compute network distances. Dijkstra's algorithm is not only slow but it must also revisit the same vertices for subsequent network distance computations. Even if Dijkstra's algorithm is suspended and resumed for subsequent Euclidean NNs, this is necessarily no better than INE, which uses Dijkstra-like expansion until $k$NNs are found.

To understand the true potential of IER, we combined it with several fast techniques. \textit{Pruned Highway Labelling} \cite{kawata2014phl} is amongst the fastest techniques. It boasts fast construction times despite being a labelling method, but has similarly large index sizes. The G-tree assembly-based method mentioned earlier can also compute network distances. Notably, in a similar manner to G-tree's $k$NN search, the ``materialization" property can be used to optimise repeated network distance queries from the same source (as in IER). The Dijkstra-like leaf-search can also be suspended and resumed. This is doubly advantageous for IER, as it becomes more robust to ``false hits" (Euclidean NNs that are not real $k$NNs), especially if they are in the vicinity of a real $k$NN. We refer to this version of G-tree as MGtree. Finally we combined IER with \textit{Contraction Hierarchies} (CH) \cite{geisberger2008ch} and \textit{Transit Node Routing} (TNR) \cite{bast2007tnr} using implementations made available by a recent experimental paper \cite{wu2012shortest}. We use a grid size of 128 for TNR as in \cite {wu2012shortest}.

{
	\setlength{\abovecaptionskip}{\abvfigskp}
	\setlength{\belowcaptionskip}{-12pt}
	\begin{figure}[t]
		\vspace{-2mm}
		\subfigure[Varying $k$]{
			\includegraphics[width=0.49\linewidth]{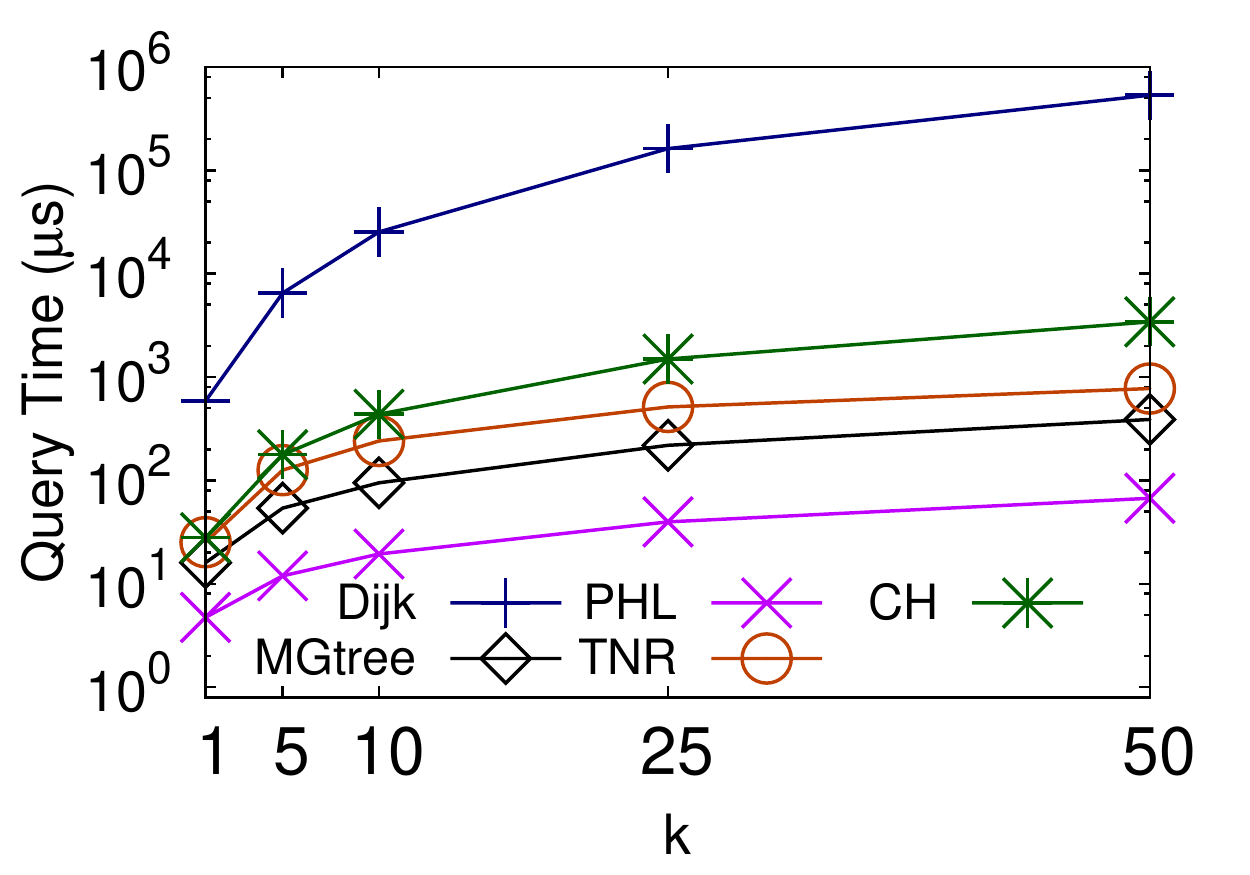}
			\label{exp:imprv:ier_comparison::varyk}
		}
		\hspace*{-5mm}
		\subfigure[Varying Density]{
			\includegraphics[width=0.49\linewidth]{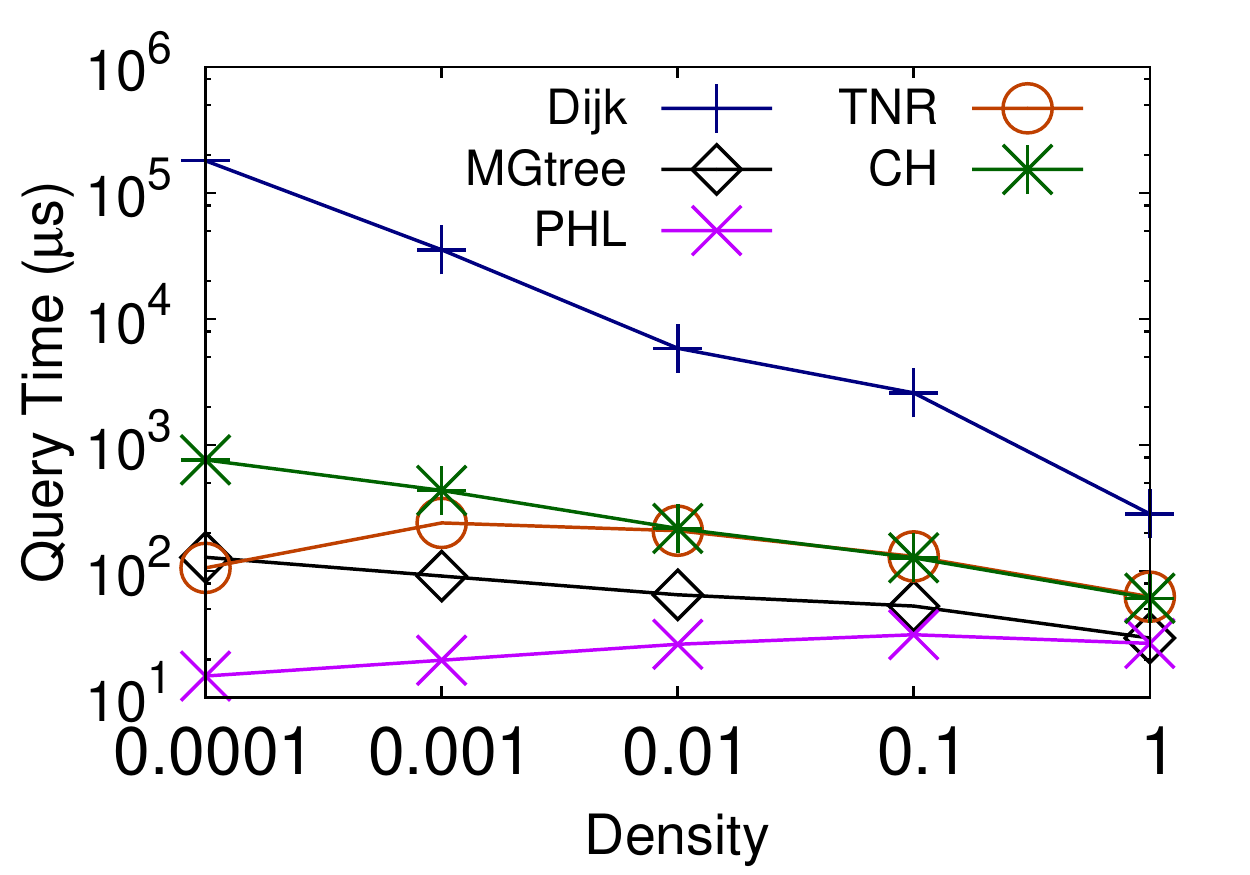}
			\label{exp:imprv:ier_comparison:varyd}
		}
		\caption{IER Variants \textmd{(NW, $d{=}0.001$, $k{=}10$, uniform objects)}}\label{exp:imprv:ier_comparison}
	\end{figure}
}

We compare the performance of IER using each method in Figure \ref{exp:imprv:ier_comparison}. PHL is the consistent winner, being 4 orders of magnitude faster than Dijkstra and an order of magnitude better than the next fastest method at its peak. G-tree, assisted by materialization, is the next best method. All methods converge with increasing density, as the search space becomes smaller. Note that CH is the technique used to answer local queries in TNR, which explains why TNR and CH are so similar for high densities as the distances are too small to use transit nodes. At lower densities, transit nodes are used more often, leading to a larger speed up. Given these results, in our main experiments, we include the two fastest versions of IER, i.e., PHL and MGtree. Note that the superiority of PHL and MGtree is also observed for other road networks and object sets.

\section{Implementation in Main Memory}\label{sec:impl}

Given the affordability of memory, the capacities available and the demand for high performance map-based services, memory-resident query processing is a realistic and often necessary requirement. However, we have seen in-memory implementation efficiency can affect performance to the point that algorithmic efficiency becomes irrelevant \cite{sidlauskas2014impl}. Firstly, this identifies the need to understand how this can happen so that guidelines for efficient implementation may be developed. Secondly, it implies that some algorithms may possess intrinsic qualities that make them superior in-memory. The utility of the latter cannot be ignored. We first illustrate both aforementioned points using a case study and then outline typical choices and our approach to settle them.

\subsection{Case Study: G-tree Distance Matrices}\label{sec:imprv:gtree_matrices}
G-tree's distance matrices store certain pre-computed graph distances (between borders of sub-graphs), allowing ``assembly" of longer distances in a piece-wise manner. We firstly describe the G-tree assembly method below, then show how the implementation of distance matrices can significantly impact its performance.

Every G-tree node has a set of borders. From our running example in Figure \ref{method:diagrams:gtree}, $v_5$ and $v_6$ are borders of $G_1$. Each non-leaf node also has a set of children, for example $G_{1A}$ and $G_{1B}$ are the children of $G_{1}$. These in turn have their own borders, which we refer to as ``child borders" of $G_1$. A distance matrix stores the distances from every child border to every other child border. For example for $G_1$, its child borders are $v_2,v_3,v_4,v_5,v_6$, and its distance matrix is shown in Figure \ref{impl:diagrams:gtree_dist_mtrx1}. But recall that a border of a G-tree node must necessarily be a border of a child node, e.g., the borders of $G_1$, $v_5$ and $v_6$, are also borders of $G_{1B}$. This means the distance matrix of $G_{1}$ repeatedly stores some border-to-border distances already in the distance matrix of $G_{1B}$, a redundancy that can become quite large for bigger graphs. To avoid this repetition and utilise, in general, $O(1)$ random retrievals, a practitioner may choose to implement the distance matrix as a hash-table. This has the added benefit of being able to retrieve distances for any two arbitrary borders.\looseness=-1 

{
	\setlength{\abovecaptionskip}{\abvfigskp}
	\setlength{\belowcaptionskip}{-13pt}
	\begin{figure}[t]%
		\vspace{-2mm}
		\subfigure[$G_1$]{
			\includegraphics[width=0.5\linewidth]{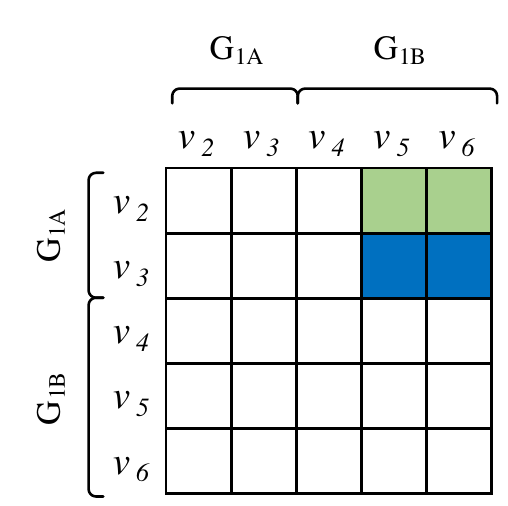}
			\label{impl:diagrams:gtree_dist_mtrx1}
		}
		\hspace*{-5mm}
		\subfigure[$G_0$]{
			\includegraphics[width=0.5\linewidth]{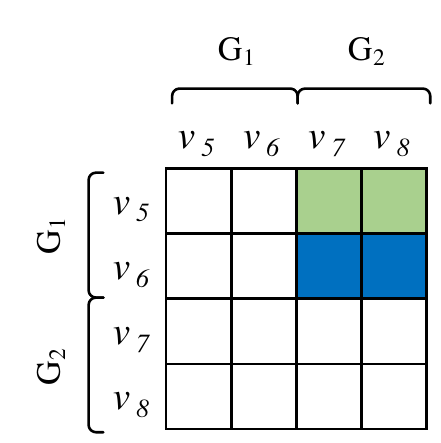}
			\label{impl:diagrams:gtree_dist_mtrx2}
		}
		\caption{Distance Matrices}\label{impl:diagrams:gtree_dist_mtrx}
	\end{figure}
}

Given a source vertex $s$ and target $t$, G-tree's assembly method firstly determines the \textit{tree path} through the G-tree hierarchy. This is a sequence of G-tree nodes starting from the leaf node containing $s$ through its immediate parent and each successive parent node up to the least-common ancestor (LCA) node. From the LCA, the path traces through successive child nodes until reaching the leaf node containing $t$. The assembly method then computes the distances from all borders from the $i$th node in the path, $G_i$, to all borders in $i{+}1$th node, $G_{i+1}$. These two nodes are necessarily either both children of the LCA or have a parent-child relationship. In either case the parent node's distance matrix contains values for all border-to-border distances. Assuming we have computed all distances from $s$ to the borders of $G_i$, we compute the distances to the borders of $G_{i+1}$ by iterating over each border of $G_{i}$ and computing the minimum distance through them to each border of $G_{i+1}$.

From our running example in Figure \ref{method:diagrams:gtree}, let $v_1$ be the source and $v_{12}$ be the target. In this case the beginning of the tree path will contain the child node $G_{1A}$ and then its parent node $G_1$. Assume we have computed the distances to the borders of $G_{1A}$ (easily done by using the distance matrix of leaf node $G_{1A}$, which stores leaf vertex to leaf border distances). Now we compute the distance to each border of $G_{1}$ from $v_1$, by finding the minimum distance through one of $G_{1A}$'s borders. To do this, for each of $G_{1A}$'s borders, we iterate over $G_{1}$'s borders, retrieving distance matrix values for each pair (updating the minimum when a smaller distance is found). This is shown by the shaded cells in Figure \ref{impl:diagrams:gtree_dist_mtrx1}. Similarly $G_{1}$ and its sibling $G_2$ are the next nodes in the tree path, and we again retrieve distance matrix values by iterating over two lists of borders. These values are retrieved from the matrix of the LCA node, $G_0$, and the values accessed are shaded in Figure \ref{impl:diagrams:gtree_dist_mtrx2}.

As we are iterating over lists (i.e., arrays) of borders, the distance matrix does not need to be accessed in an arbitrary order, as we observed in the G-tree authors' implementation. This is made possible by grouping the borders of child nodes as shown in Figure \ref{impl:diagrams:gtree_dist_mtrx} and storing the starting index for each child's borders. Additionally we create an offset array indicating the position of the nodes' own borders in its distance matrix. For example, the offset array for $G_1$ indicates its borders ($v_5$ and $v_6$) are at the 3rd and 4th index of each row in its distance matrix shown in Figure \ref{impl:diagrams:gtree_dist_mtrx1}. While Figure \ref{impl:diagrams:gtree_dist_mtrx} shows the distance matrix as a 2D array, it is best implemented as a 1D array. This and the previously described accessed method, allow all shaded values to be accessed from sequential memory locations, thus displaying excellent spatial locality. This is shown in Figure \ref{impl:diagrams:gtree_dist_mtrx} as the shaded cells are either contiguous or very close to being so. Spatial locality makes the code cache-friendly, allowing the CPU to easily predict and pre-fetch data into cache that will be read next. Otherwise the data would need to be retrieved from memory, which is $20{-}200\times$ slower than CPU cache (depending on the level). This effect is amplified in real road networks as they contain significantly larger numbers of borders per node.

{
	\setlength{\abovecaptionskip}{\abvfigskp}
	\setlength{\belowcaptionskip}{-12pt}
	\begin{figure}[t]
		\vspace{-2mm}
		\subfigure[Varying $k$]{
			\includegraphics[width=0.49\linewidth]{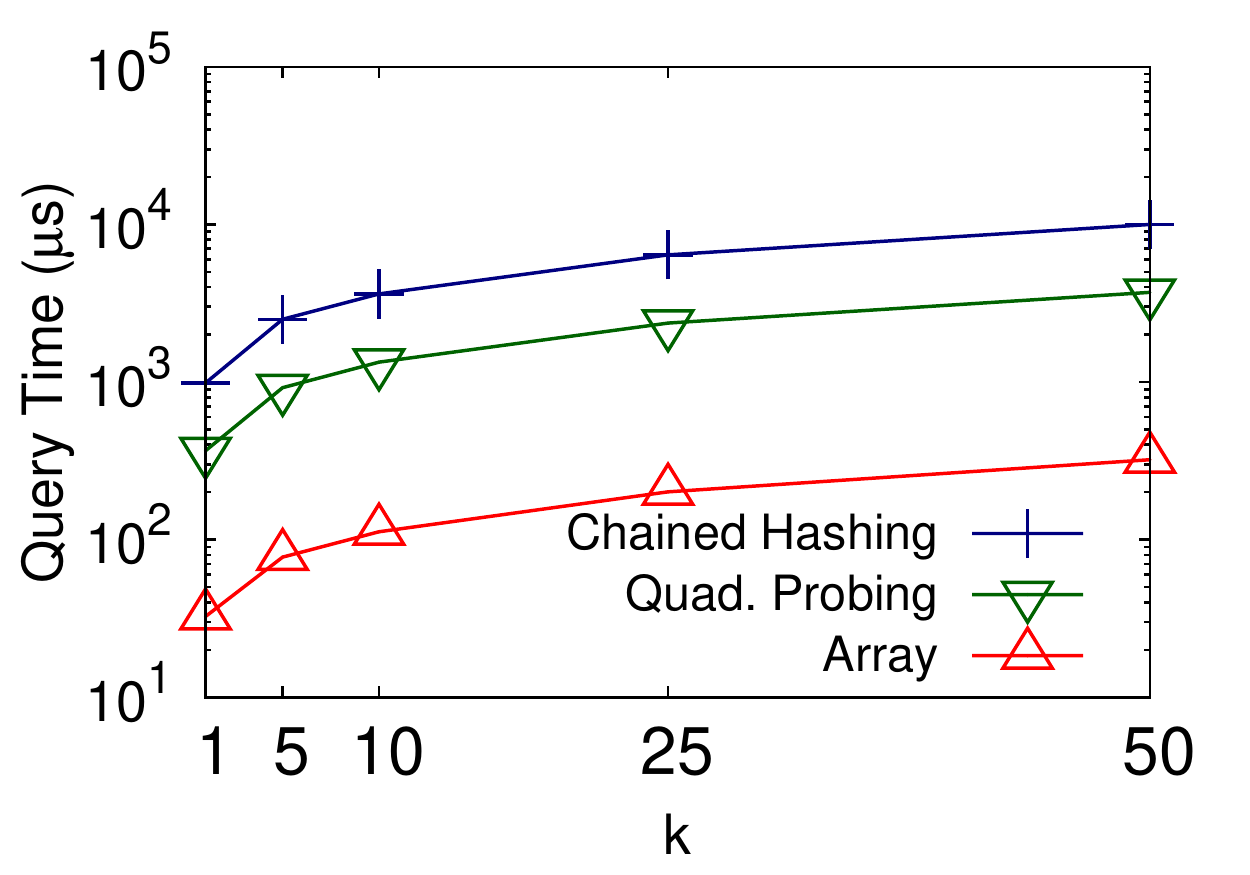}
			\label{exp:impl:gtree_distmtrx:varyk}
		}
		\hspace*{-5mm}
		\subfigure[Varying Density]{
			\includegraphics[width=0.49\linewidth]{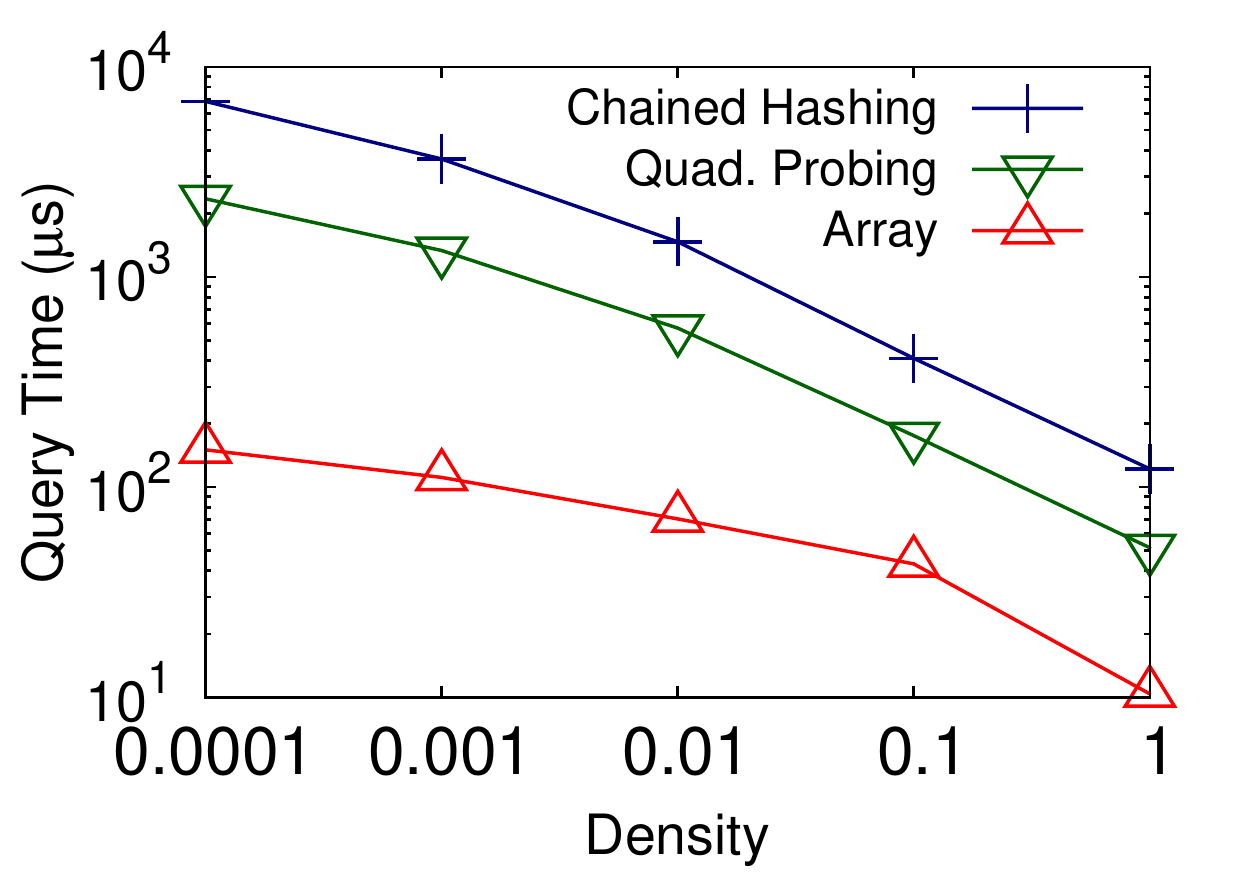}
			\label{exp:impl:gtree_distmtrx:varyd}
		}
		\caption{Distance Matrix Variants \textmd{(NW, $d{=}0.001$, $k{=}10$)}}\label{exp:impl:gtree_distmtrx}
	\end{figure}
}

We compare three implementations of distance matrices, including the 1D array described above and two types of hash-tables: chained hashing~\cite{cormen2011book} (STL \texttt{unordered\_map}); and quadratic probing~\cite{cormen2011book} (Google \texttt{dense\_hash\_map}). In Figure~\ref{exp:impl:gtree_distmtrx}, chained hashing is a staggering 30 times slower than the array. While quadratic probing is an improvement, it is still an order of magnitude slower. Had we used either of the hash-table types, we would have unfairly concluded that G-tree was the worst performing algorithm.

{
	\setlength{\abovecaptionskip}{\abvtblskp}
	\setlength{\belowcaptionskip}{-2pt}
	\begin{table}[!htpb]
		\centering
		\begin{tabular}{|c|c|c|c|c|c|} \hline
			
			\multirow{2}{*}{\textbf{Distance Matrix}} & \multirow{2}{*}{\textbf{INS}} & \multicolumn{3}{c|}{\textbf{Cache Misses (Data)}} \\
			\hhline{~~---}
			& & L1 & L2 & L3 \\	\hline
			Chained Hashing & 953 B & 28.8 B & 20.5 B & 13 B \\ \hline
			Quadratic Probing & 1482 B & 11.2 B & 7.5 B & 5.3 B \\ \hline
			Array & 151 B & 1.5 B & 0.4 B & 0.3 B \\ \hline
		\end{tabular}
		\caption{Hardware Profiling: 250,000 Queries on NW Dataset}
		\label{tab:gtree_cache}
	\end{table}
}

We investigate the cache efficiency of each implementation in CPU cache misses at each level in billions in Table \ref{tab:gtree_cache} (also showing INS, no. of instructions in billions) using \texttt{perf} hardware profiling of 250,000 varied queries on NW. Chained hashing uses indirection to access data, resulting in poor locality and the highest number of cache misses. Quadratic probing improves locality at the expense of more costly collision resolution, hence it uses more instructions than chained hashing. However, it cannot achieve better locality than storing data in an array sorted in the order it will be accessed. This ordering means the next value we retrieve from the array is far more likely to be in some level of cache. Unsurprisingly, it suffers from the fewest cache misses. This is a unique strength of G-tree's distance matrices and shows, while in-memory implementation is challenging, it is still possible to design algorithms that work well.

\subsection{Guidelines for Implementation Choices}\label{sec:imprv:choices}

In-memory implementation requires careful consideration, or experimental outcomes can be drastically affected as seen with G-tree's distance matrices and in \cite{sidlauskas2014impl}. Many choices are actually quite simple, but their simplicity can lead to them being overlooked. Here we outline several choices and options to deal with them to assist future implementers. To illustrate the impact of these choices we progressively improve a first-cut in-memory implementation of INE. Each plot line in Figure \ref{exp:impl:ine_choices} shows the effect of one improved choice. Each roughly halves the query time, with the final implementation of INE being $6{-}7\times$ faster.

{
	\setlength{\abovecaptionskip}{\abvfigskp}
	\setlength{\belowcaptionskip}{-12pt}
	\begin{figure}[t]
		\vspace{-2mm}
		\subfigure[Varying $k$]{
			\includegraphics[width=0.49\linewidth]{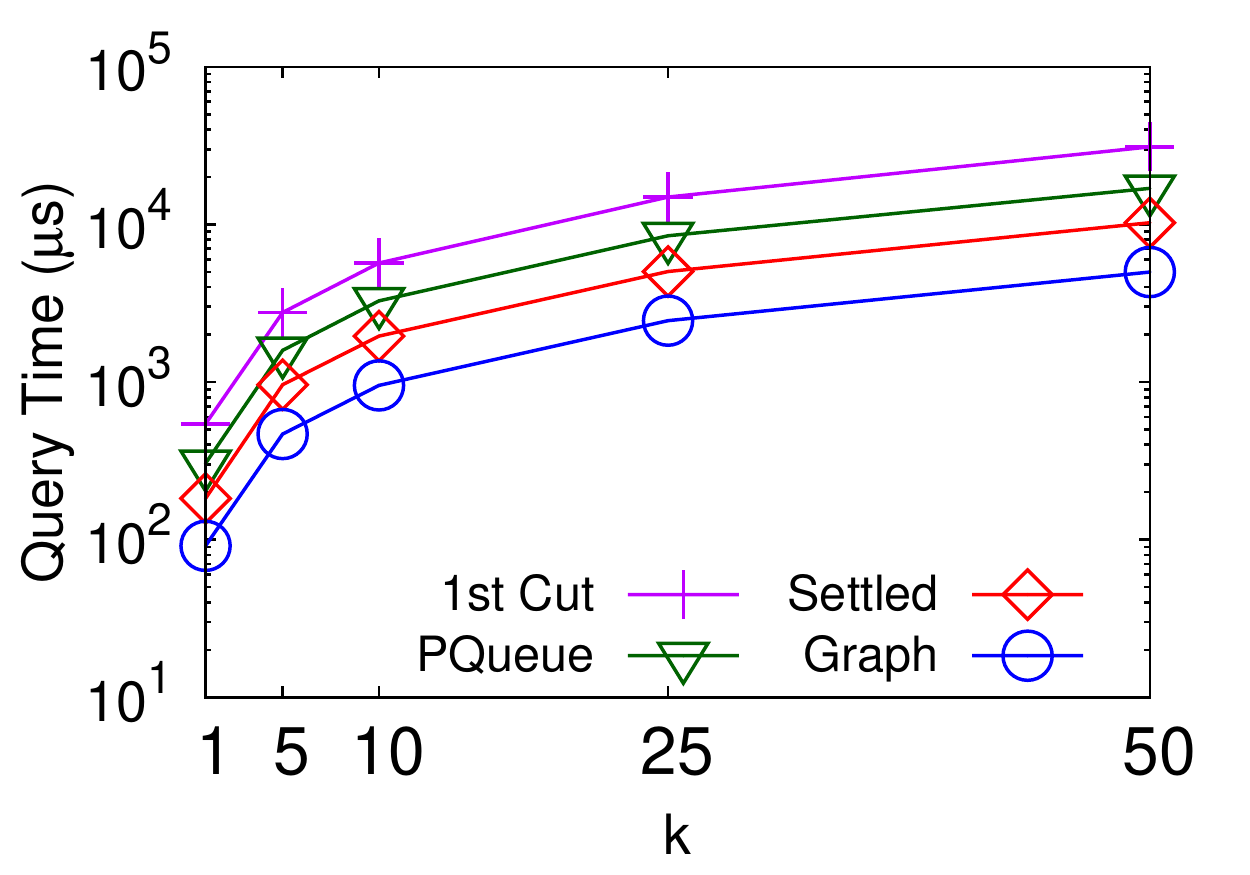}
			\label{exp:impl:ine_choices:varyk}
		}
		\hspace*{-5mm}
		\subfigure[Varying Density]{
			\includegraphics[width=0.49\linewidth]{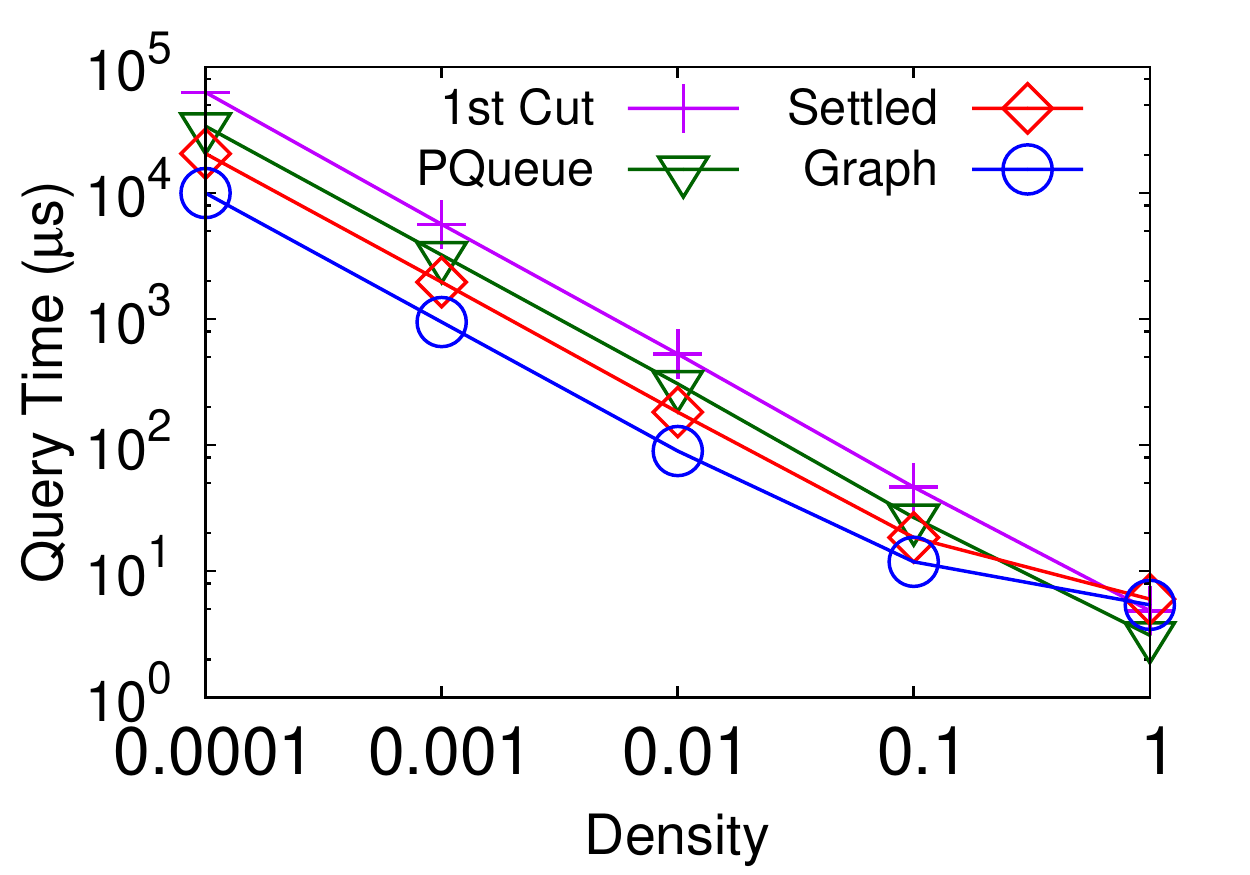}
			\label{exp:impl:ine_choices:varyd}
		}
		\caption{INE Improvement \textmd{(NW, $d{=}0.001$, $k{=}10$)}}\label{exp:impl:ine_choices}
	\end{figure}
}

\smallHead{1. Priority Queues}. All methods in our study employ priority queues. In particular, INE and ROAD involve many queue operations and thus rely on their efficient implementation. Binary heaps are most commonly used, but we must choose whether to allow duplicate vertices in the queue or not. Without duplicates, the queue is smaller and queue operations involve less work. But this means the heap index of each vertex must be looked up to update keys e.g., through a hash-table. On degree-bounded graphs, such as road networks, the number of duplicates is small, and removing them is simply not worth the lost locality and increased processing time incurred with hash-tables. As a result, we see a 2$\times$ improvement when INE is implemented without decreasing keys (see \textit{PQueue} in Figure \ref{exp:impl:ine_choices}). Note that we use this binary heap for all methods.

\smallHead{2. Settled Vertex Container}. Recall INE and ROAD must track vertices that have been dequeued from their priority queues (i.e., settled). The scalable choice is to store vertices in a hash-table as they are settled. However we observe an almost 2$\times$ improvement, as shown by \textit{Settled} in Figure \ref{exp:impl:ine_choices} by using a \textit{bit-array} instead. This is despite the need to allocate memory for $|V|$ vertices for each query. The bit-array has the added benefit of occupying 32$\times$ less space than an integer array, thus fitting more data in cache lines. This does add a constant pre-allocation overhead for each query, which is proportionally higher for small search spaces (i.e, for high density). But the trade-off is worth it due to the significant benefit on larger search spaces (i.e., low density).

\smallHead{3. Graph Representation}. A disk-optimised graph data structure was proposed for INE in \cite{papadias2003ine}. In main memory, we may choose to replace it with an array of node objects, with each object containing an adjacency list array. However by combining all adjacency lists into a single array we are able to obtain another 2$\times$ speed-up (refer to \textit{Graph} in Figure \ref{exp:impl:ine_choices}). Firstly, we assign numbers to vertices from $0$ to $|V|{-}1$. An \textit{edges} array stores the adjacency list of each vertex consecutively in this order. The \textit{vertices} array stores the starting index of each vertex's adjacency list in \textit{edges}, also in order. Now for any vertex $u$ we can find the beginning of its adjacency list in \textit{edges} using \textit{vertices}$[u]$ and its end using \textit{vertices}$[u{+}1]$. This contiguity increases the likelihood of a cache hit during expansion. We similarly store ROAD's shortcuts in a global shortcut array, with each shortcut tree node storing an offset to this array. The principle demonstrated here is that recommended data structures in past studies cannot be used verbatim. It is necessary to replace IO-oriented data structures e.g., we replaced the B\textsuperscript{+}-trees, recommended in the originally disk-based DisBrw and ROAD, with sorted arrays.

\smallHead{4. Language}. C++ presently allows more low-level tuning, such as specifying the layout of data in memory for cache benefits, making it preferable in high performance applications. Implementers may consider other languages such as Java for its portability and design features. But when we implemented INE with all aforementioned improvements in Java (Oracle JDK 7), we found it was at least 2$\times$ slower than the equivalent C++ implementation. One possible reason is that Java does not guarantee contiguity in memory for collections of objects. Also, the same objects take up more space in Java. Both factors lead to lower cache utilisation, which may penalise methods that are better able to exploit it.

\section{Experiments}\label{sec:exp}

\subsection{Experimental Setting}\label{sec:exp:settings}\label{sec:exp:settings:params}

\smallHeadIndent{Environment.}
We conducted experiments on a 3.2GHz Intel Core i5-4570 CPU and 32GB RAM running 64-bit Linux (kernel 4.2). Our program was compiled with g++ 5.2 using the O3 flag, and all query algorithms use a single thread. To ensure fairness, we used the same subroutines for common tasks between the algorithms whenever possible. We implemented INE, IER, G-tree and ROAD from scratch.  We obtained the authors code for G-tree, which we used to further improve our implementation, e.g., by selecting the better option when our choices disagreed with the authors' choice of data structures. For Distance Browsing, we partly based our SILC index on open-source code from \cite{wu2012shortest}, but being a shortest path study this implementation did not support $k$NN queries. As a result, we implemented the $k$NN algorithms ourselves from scratch, modifying the index to support them, taking the opportunity to make significant improvements (as discussed in Section \ref{sec:impl} and Appendix \ref{sec:app:imprv}). We used a highly efficient open-source implementation of PHL made available by its authors \cite{kawata2014phl}. All source code and scripts to generate datasets, run experiments, and draw figures have been released as open-source~\cite{oursource} for readers to reproduce our results or re-use in future studies.

{
	\setlength{\abovecaptionskip}{\abvtblskp}
	\setlength{\belowcaptionskip}{-12pt}
	\begin{table}[t]
		\centering
		\begin{tabular}{|c|c|} \hline
			\textbf{Parameter} & \textbf{Values} \\ \hline 
	Road Networks & DE, VT, ME, CO, \textbf{NW}, CA, E, W, C, \textbf{US} \\ \hline
			$k$ & 1, 5, \textbf{10}, 25, 50 \\ \hline
			Density ($d$) & 1, 0.1, 0.01, \textbf{0.001}, 0.0001 \\ \hline
			Synthetic POIs & \textbf{uniform}, clustered, min. obj. distance \\ \hline
			Real POIs & Refer to Table \ref{tab:pois} \\ \hline
		\end{tabular}
		\caption{Parameters (Defaults in Bold)}
		\label{tab:parameters}
	\end{table}
}

\smallHeadIndent{Index Parameters.}
The performance of the G-tree and ROAD indexes are highly dependent on the choice of leaf capacity $\tau$ (G-tree), hierarchy levels $l$ (ROAD) and fanout $f$ (both) \cite{zhong2015gtree,lee2012road,lee2009road}. We experimentally confirmed trends observed in those studies and computed parameters for new datasets. As such, we use fanout $f{=}4$ for both methods. For G-tree we set $\tau$ to 64 (DE), 128 (VT, ME, CO), 256 (NW, CA, E), and 512 (W, C, US). For ROAD, we set $l$ to 7 (DE), 8 (VT, ME), 9 (CO, NW), 10 (CA, E) and 11 (W, C, US). We chose values of $l$ for ROAD in accordance with the results reported in~\cite{lee2009road} that show query performance of ROAD improves for larger $l$. Specifically, for each dataset, we increased $l$ until either the query performance did not improve or further partitioning was not possible due to too few vertices in the leaf levels.

\smallHeadIndent{Query Variables.}
Table~\ref{tab:parameters} shows the range of each variable used in our experiments (defaults in bold). Similar to past studies \cite{zhong2015gtree}, we vary $k$ from $1$ to $50$ with a default of $10$. We used 8 real-world object sets as discussed Section \ref{sec:datasets}. We vary uniform object set density $d$ from $0.0001$ to $1$ where $d{=}|O|/|V|$ with a default value of $0.001$. We choose this default density as it closely matches the typical density for real-world object sets as shown in Table \ref{tab:pois}. Furthermore this density creates a large enough search space to reveal interesting performance trends for methods. We vary over 10 real road networks (listed in Table \ref{tab:datasets}) with median-sized NW and largest US road networks as defaults. We use distance edge weights in Sections \ref{sec:exp:indexes} and \ref{sec:exp:query} for comparison with past studies, and because IER and DisBrw were developed for such graphs. But we repeat experiments on travel times later in Section \ref{sec:exp:tt}.

\smallHeadIndent{Query and Object Sets.}
All query times are averaged over $10{,}000$ queries. For real-world object sets, we tested each set with $10{,}000$ random query vertices. For uniform and clustered object sets, we generate $50$ different sets for each density and number of clusters (resp.) combined with $200$ random query vertices. For minimum distance object sets (described in Section \ref{sec:datasets:obj}), we generated $50$ sets for each distance set $R_i$ with $i \in [1,m]$. We also chose 200 random query vertices with distances from the centre vertex in range $[0, \frac{D_{max}}{2^m})$ (i.e., vertices closer than $R_1$) for use with all sets. We use $m{=}6$ for NW and $m{=}8$ for US to ensure there were enough objects in each set to satisfy the default density $0.001$.

\subsection{Road Network Index Pre-Processing Cost}\label{sec:exp:indexes}

Here we measure the construction time and size of the index used by each technique for all road networks in Table \ref{tab:datasets}.

\smallHeadIndent{Index Size.} Figure \ref{exp:index:size} shows the index size for each algorithm. INE only uses the original graph data structure, so its size can be seen as the lower bound on space. DisBrw could only be built for the first 5 road networks before exceeding our memory capacity. This is not surprising given the $O(|V|^{1.5})$ storage complexity. However, in our implementation, we were able to build DisBrw for an index with 1 million vertices (NW) consuming 17GB. PHL also exhibits large indexes, however it can still be built for all but the 2 largest datasets. We note that PHL experiences larger indexes on travel distance graphs because they do not exhibit prominent hierarchies needed for effective pruning (on travel time graphs we were able to build PHL for all indexes). G-tree consumed less space than ROAD. E.g., for the US dataset G-tree used 2.9GB compared to ROAD's 4.4GB. As explained in past studies \cite{zhong2015gtree}, ROAD's Route Overlay contains significant redundancy as multiple shortcut trees repeatedly store a subset of the Rnet hierarchy.

{
	\setlength{\abovecaptionskip}{\abvfigskp}
	\setlength{\belowcaptionskip}{-12pt}
	\begin{figure}[t]
		\vspace{-2mm}
		\subfigure[Index Size]{
			\includegraphics[width=0.48\linewidth]{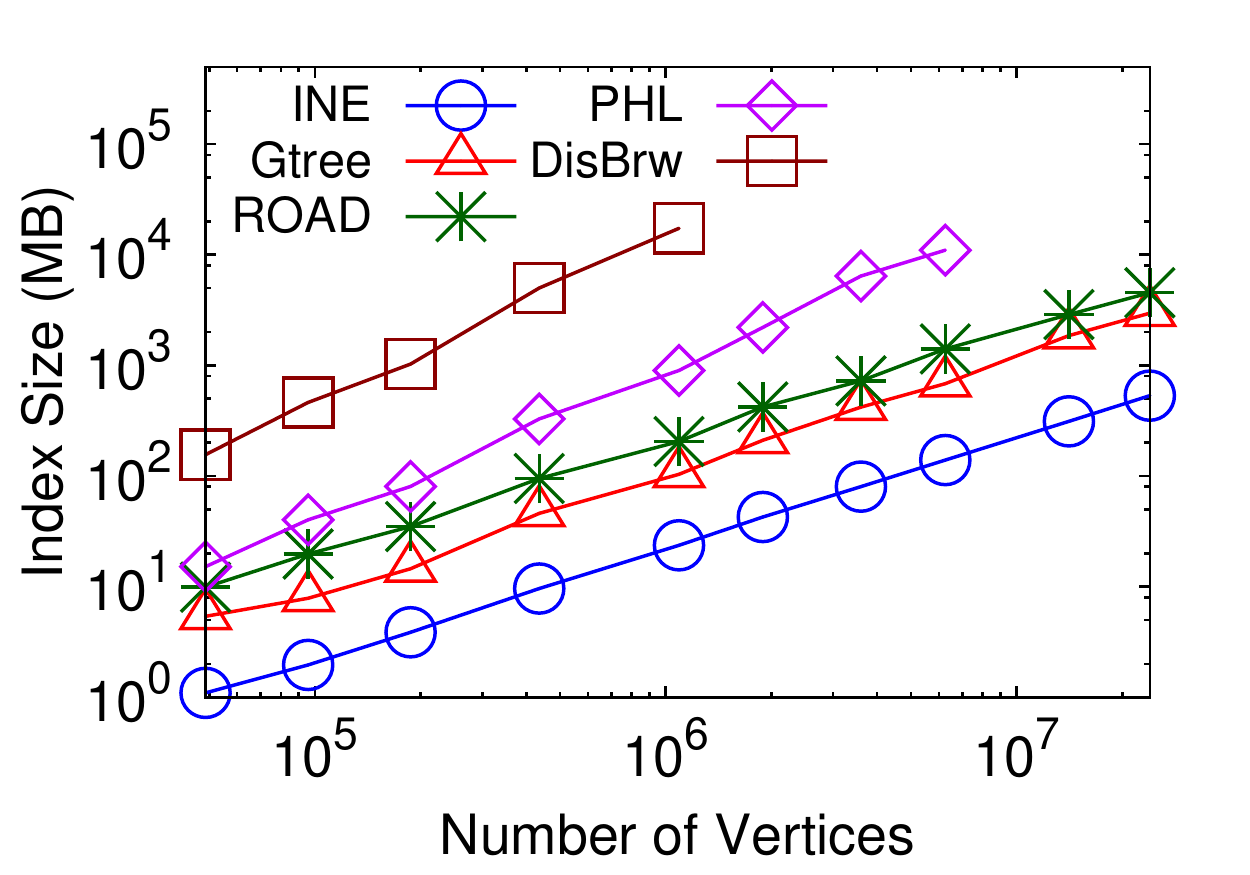}
			\label{exp:index:size}
		}
		\hspace*{-5mm}
		\subfigure[Construction Time]{
			\includegraphics[width=0.48\linewidth]{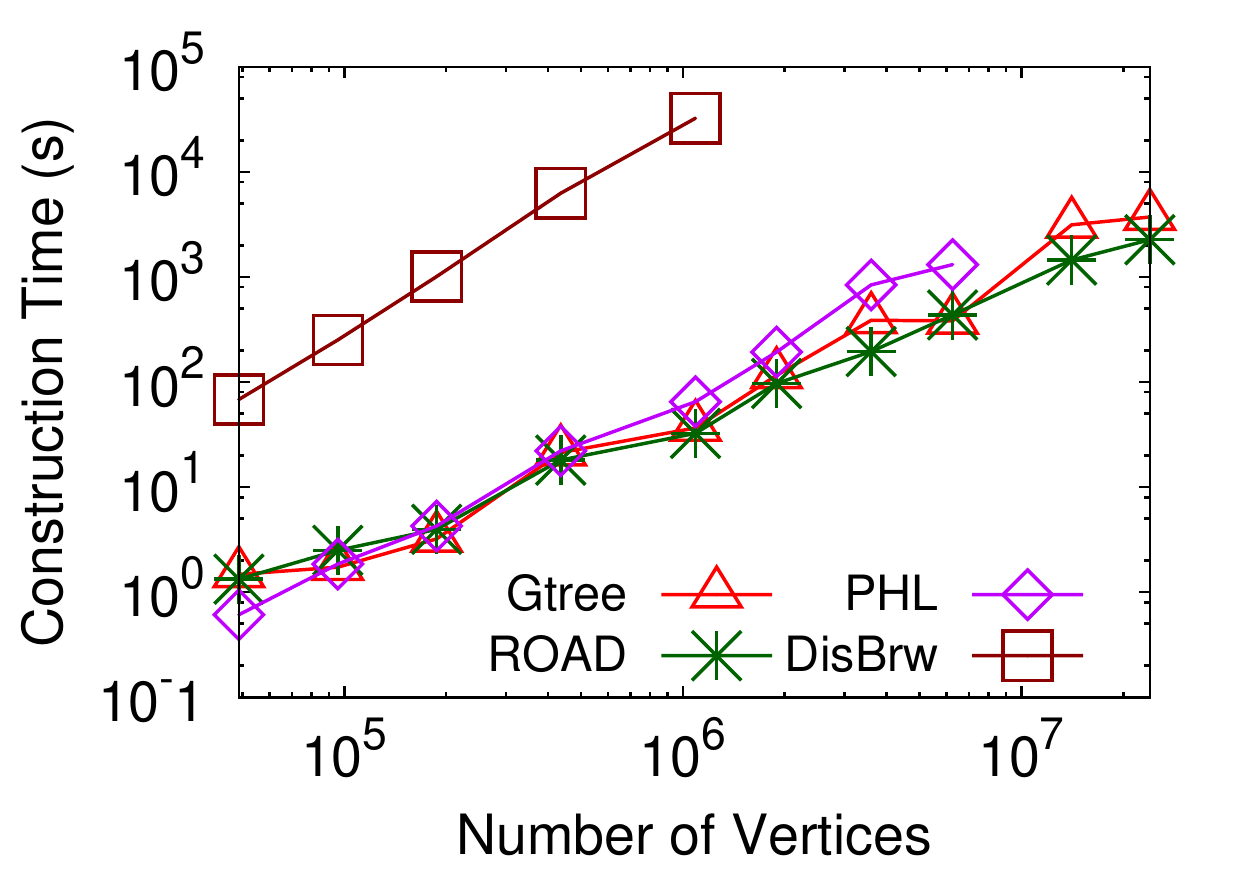}
			\label{exp:index:time}
		}
		\caption{Effect of Road Network Size $|V|$}\label{exp:index}
	\end{figure}
}

\smallHeadIndent{Construction Time.} Figure \ref{exp:index:time} compares the construction time of each index for increasing network sizes. DisBrw again stands out as its index (SILC \cite{sankaranarayanan2005silc}) requires an all-pairs shortest path computation. However, the computation of each SILC quadtree is independent and can be easily parallelized. We observed a speed-up factor of very close to 4$\times$ with our quad-core CPU using OpenMP. Note that other methods cannot be so easily parallelized. Despite this DisBrw still required $9$ hours on NW, while parallelization is useful it does not change the asymptotic behaviour. PHL takes longer than G-tree and ROAD but surprisingly not significantly so, thanks to pruned labelling \cite{kawata2014phl}. IER's index performance depends on the network distance method it employs (i.e., G-tree or PHL).

Recall that both ROAD and G-tree must partition the road network. Since the network partitioning problem is known to be NP-complete, ROAD and G-tree both employ heuristic algorithms. As both methods require the same type of partitioning we use the same algorithm, the multilevel graph partitioning algorithm~\cite{karypis1998fast} used in G-tree. This method uses a much faster variant of the Kernighan-Lin algorithm recommended in ROAD~\cite{lee2012road}. Consequently, we are able to evaluate ROAD for much larger datasets for the first time, with ROAD being constructed in less than one hour for even the largest dataset (US) containing $24$ million vertices. The construction time of ROAD is comparable to G-tree, because both use the same partitioning method, and employ bottom-up methods to compute shortcuts and distance matrices, respectively. 

We remark that, while most existing studies have focused on improving query processing time, there is a need to develop algorithms and indexes providing comparable efficiency with a focus on reducing memory usage and construction time.

\subsection{Query Performance}\label{sec:exp:query}

We investigated $k$NN query performance over several variables: road network size, $k$, density, object distance, clusters, and real-world POIs. Implementations have been optimized according to Section~\ref{sec:impl}. We have applied numerous improvements to each algorithm, as detailed in Appendix~\ref{sec:app:imprv}. IER network distances are computed using both PHL \cite{kawata2014phl} (when its index fits in memory) and G-tree with materialization (shown as IER-PHL and IER-Gt, resp.).

\subsubsection{Varying Network Size}\label{exp:knn:network_size}

Figure~\ref{exp:knn:nw_network_density001_k10} shows query times with increasing numbers of road network vertices $|V|$ for all 10 road networks in Table \ref{tab:datasets} on uniform objects. We observe the consistent superiority of IER-based methods. Figure \ref{exp:knn:nw_network_density001_k10} clearly shows the reduced applicability of DisBrw. Even though its performance is close to ROAD, its large index size makes it applicable on only the first 5 datasets.

{
	\setlength{\abovecaptionskip}{\abvfigskp}
	\setlength{\belowcaptionskip}{-12pt}
	\begin{figure}[t]
		\vspace{-2mm}
		\subfigure[Query Time]{
			\includegraphics[width=0.5\linewidth]{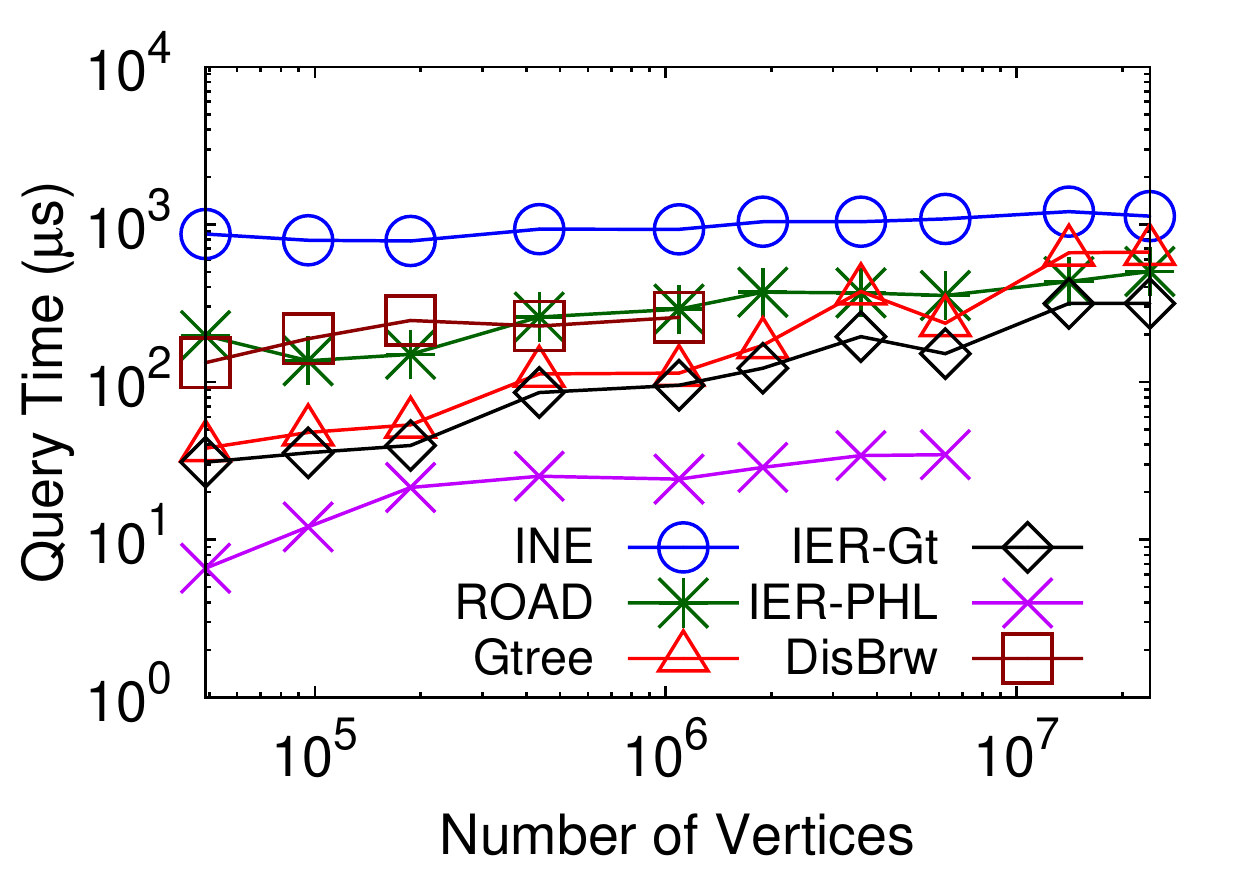}
			\label{exp:knn:nw_network_density001_k10}
		}
		\hspace*{-5mm}
		\subfigure[G-tree \& ROAD Stats]{
			\includegraphics[width=0.5\linewidth]{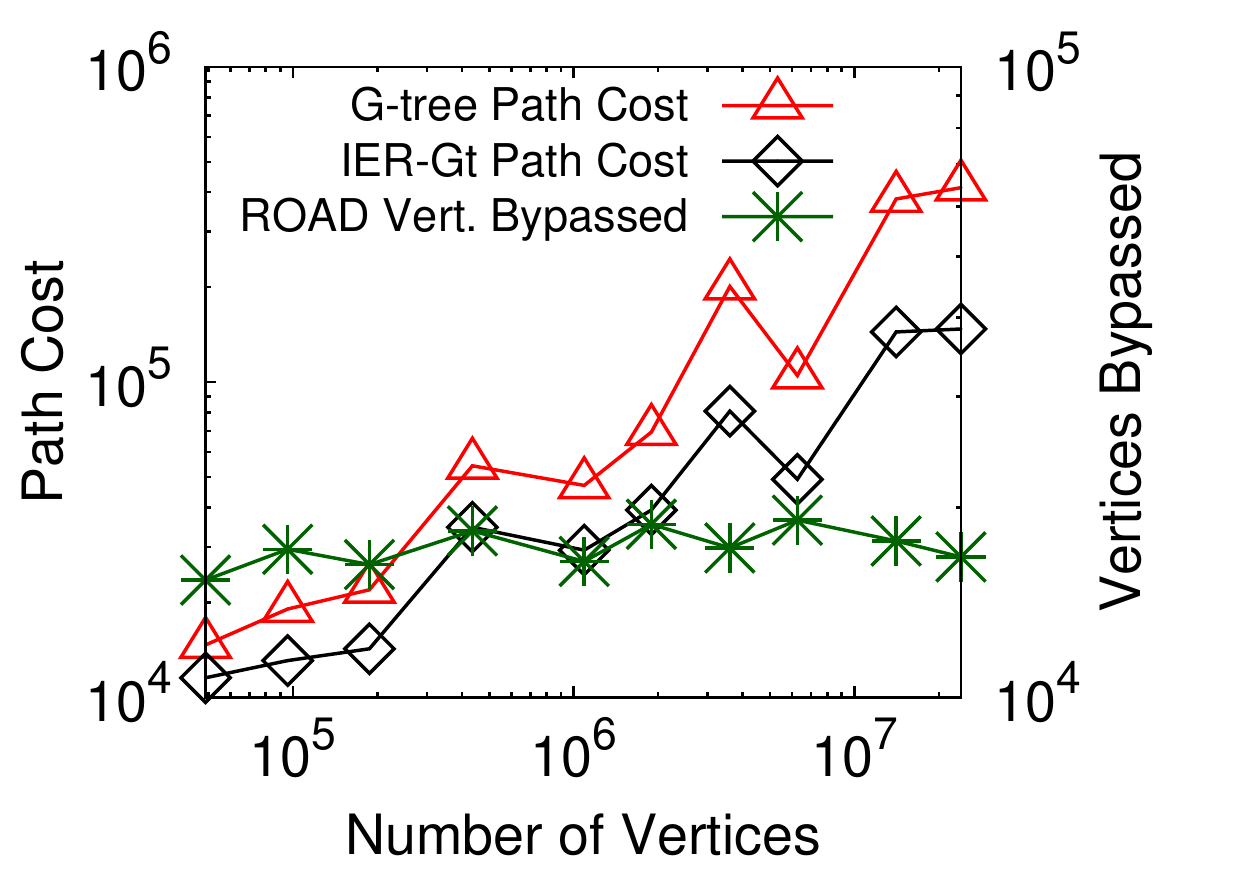}
			\label{exp:knn:gtree_vs_road_stats}
		}		
		\caption{Effect of Road Network Size $\boldsymbol{|V|}$ \textmd{($d{=}0.001$, $k{=}10$)}}\label{exp:knn_size}
	\end{figure}
}

Surprisingly G-tree's advantage over ROAD decreases with increasing network size $|V|$. Recall that ROAD can be seen as an optimisation on INE, where the expansion can bypass object-less regions (i.e., Rnets). Thus ROAD's relative improvement over INE depends on the time saved bypassing Rnets versus additional time spent descending shortcut trees. In general, given the same density, we can expect a similar sized region to contain the same number of objects irrespective of the network size $|V|$. This explains why INE remains relatively unaffected by $|V|$. It also means that regions without objects are similarly sized. Although Rnets may grow, the size of the Rnets we do bypass also grow, so ROAD bypasses similar numbers of vertices. So the time saved bypassing regions does not increase greatly. Thus ROAD's query time with increasing $|V|$ mainly depends on the depth of shortcut trees. But the depth is bounded by $l$, which we know does not increase greatly, and as a result ROAD scales extremely well with increasing $|V|$.

G-tree's non-materialized distance computation cost is a function of the number of borders of G-tree nodes (i.e., subgraphs) involved in the tree path to another node or object. With increasing network size, a G-tree node at the same depth has more borders and the path cost is consequently higher. Thus, we see G-tree ``catch-up" to ROAD on the US dataset. These trends are demonstrated in \ref{exp:knn:gtree_vs_road_stats}. G-tree's path cost (in  border-to-border computations) increases while the number of vertices ROAD bypasses remains stable with increasing $|V|$ (note these are not directly comparable).

\subsubsection{Varying k}

Figures \ref{exp:knn:nw_k_density001} and \ref{exp:knn:usa_k_density001} show the results for varying $k$ for the NW and US datasets, respectively, on uniform objects. Significantly, IER-PHL is 5$\times$ faster than any other method on NW. While PHL could not be constructed for the US dataset for travel distances, IER-Gt takes its place as the fastest method, being twice as fast as G-tree. Interestingly, this is despite both using the same index, also materializing intermediate results, and IER-Gt having the additional overhead of retrieving Euclidean NNs. So this is really an examination of heuristics used by G-tree. Essentially G-tree visits the closest subgraph (i.e., by one of its borders) while IER-Gt visits the subgraph with the next Euclidean NN. IER-Gt can perform better because its heuristic incorporates an estimate on distances to objects within subgraphs while G-tree does not. Each time G-tree visits a subgraph not containing a $k$NN it pays a penalty in the cost of non-materialized distance computations. We have seen this cost increases with network size, which explains why the improvement of IER-Gt is greater on the US than on NW. This is verified in Figure \ref{exp:knn:gtree_vs_road_stats}, which shows IER-Gt involves fewer computations than G-tree and the gap increases with network size.

{
	\setlength{\abovecaptionskip}{\abvfigskp}
	\setlength{\belowcaptionskip}{-15pt}
	\begin{figure}[t]%
		\vspace{-2mm}
		\subfigure[NW Dataset]{
			\includegraphics[width=0.5\linewidth]{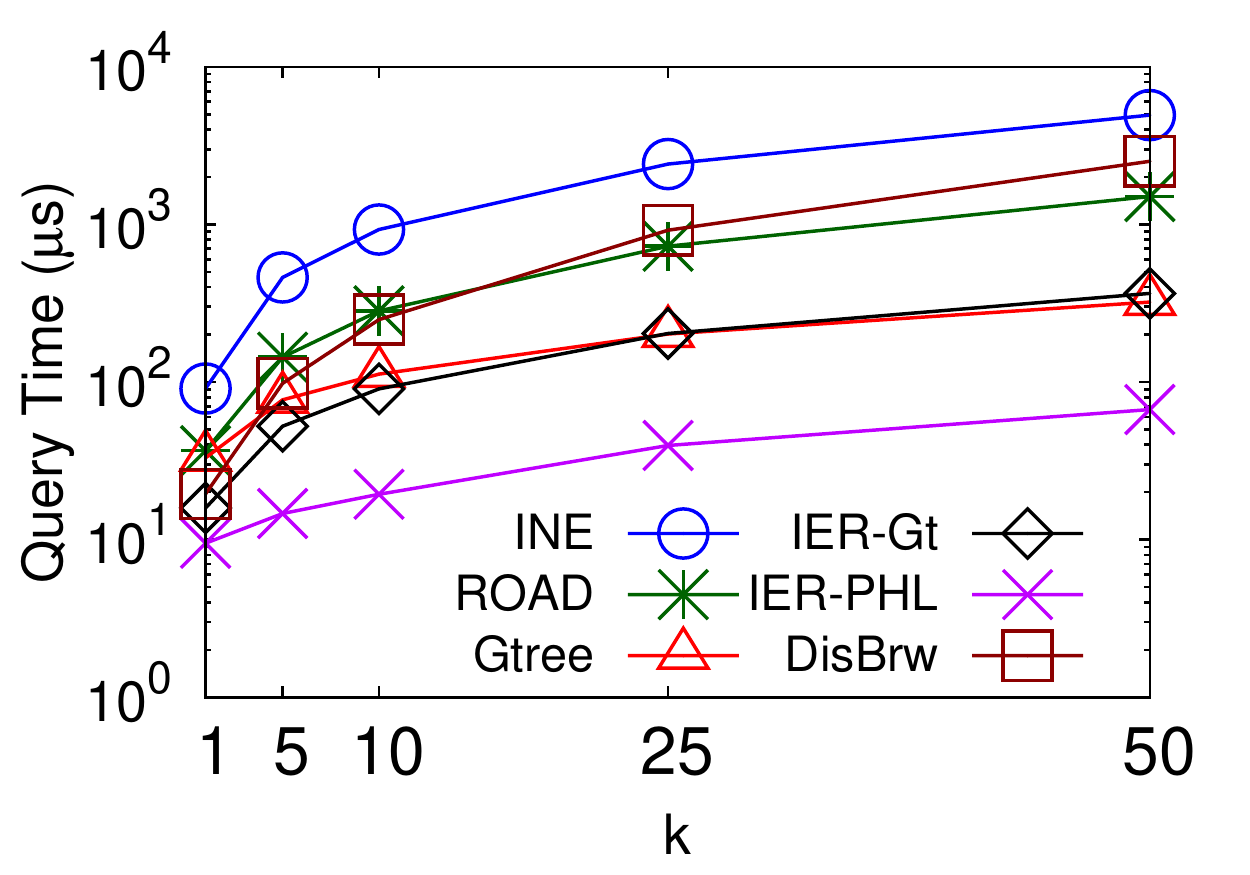}
			\label{exp:knn:nw_k_density001}
		}
		\hspace*{-5mm}
		\subfigure[US Dataset]{
			\includegraphics[width=0.5\linewidth]{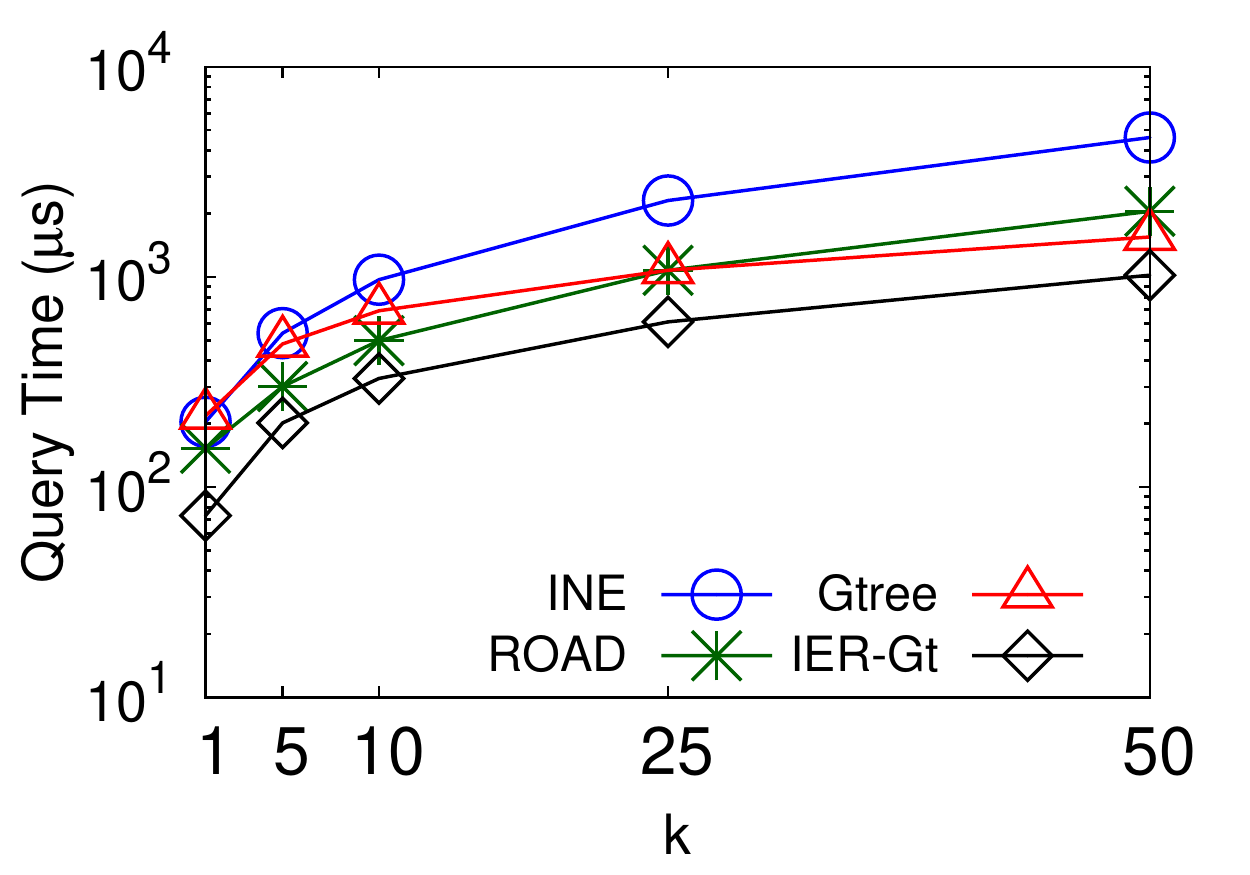}
			\label{exp:knn:usa_k_density001}
		}
		\caption{Effect of $\boldsymbol{k}$ \textmd{($d{=}0.001$)}}\label{exp:knn_k}
	\end{figure}
}

We observe that G-tree outperforms ROAD, DisBrw and INE on NW, with a trend similar to previous studies~\cite{zhong2015gtree}. INE is the slowest as it visits many vertices. For $k=1$ the ROAD, DisBrw and G-tree methods are indistinguishable as a small area is likely to contain the NN. ROAD and DisBrw scale very similarly with $k$. G-tree scales better than both, at its peak nearly an order of magnitude better than ROAD and DisBrw. As more objects are located, more paths in the G-tree hierarchy are traversed, allowing greater numbers of subsequent traversals to be materialized. As explained in Section \ref{exp:knn:network_size}, we again see G-tree's relative improvement over ROAD decrease in Figure \ref{exp:knn:usa_k_density001} for the larger US dataset.

\subsubsection{Varying Density}

We evaluate performance for varying uniform object densities in Figure \ref{exp:knn_d}. With increasing density the average distance between objects decreases and in general query times are lower. The rate of improvement for heuristic-based methods (DisBrw, G-tree, IER) is slower because they are less able to distinguish better candidates. For IER this means more false hits, explaining why IER-PHL's query times increase (slightly) as it has no means to re-use previous computations like IER-Gt does. The rate of improvement is higher for expansion-based methods as their search spaces become smaller. ROAD falls behind INE beyond density 0.01 indicating the tipping point at which the time spent traversing shortcut trees exceeds the time saved bypassing Rnets (if any). The query times plateau at high densities on the US dataset for ROAD and INE because it is dominated by the bit-array initialization cost (refer to Section \ref{sec:imprv:choices}). G-tree performs well at high densities as more $k$NNs are found in the source leaf node. In this case it reverts to a Dijkstra-like search (which we improved as in Appendix \ref{sec:app:imprv:gtree:leaf_opt}) providing comparable performance to INE and ROAD on NW. G-tree exceeds them on the US as a bit-array is not required due to G-tree's leaf search being limited to at most $\tau$ vertices.

{
	\setlength{\abovecaptionskip}{\abvfigskp}
	\setlength{\belowcaptionskip}{-13pt}
	\begin{figure}[t]
		\vspace{-2mm}
		\subfigure[NW Dataset]{
			\includegraphics[width=0.5\linewidth]{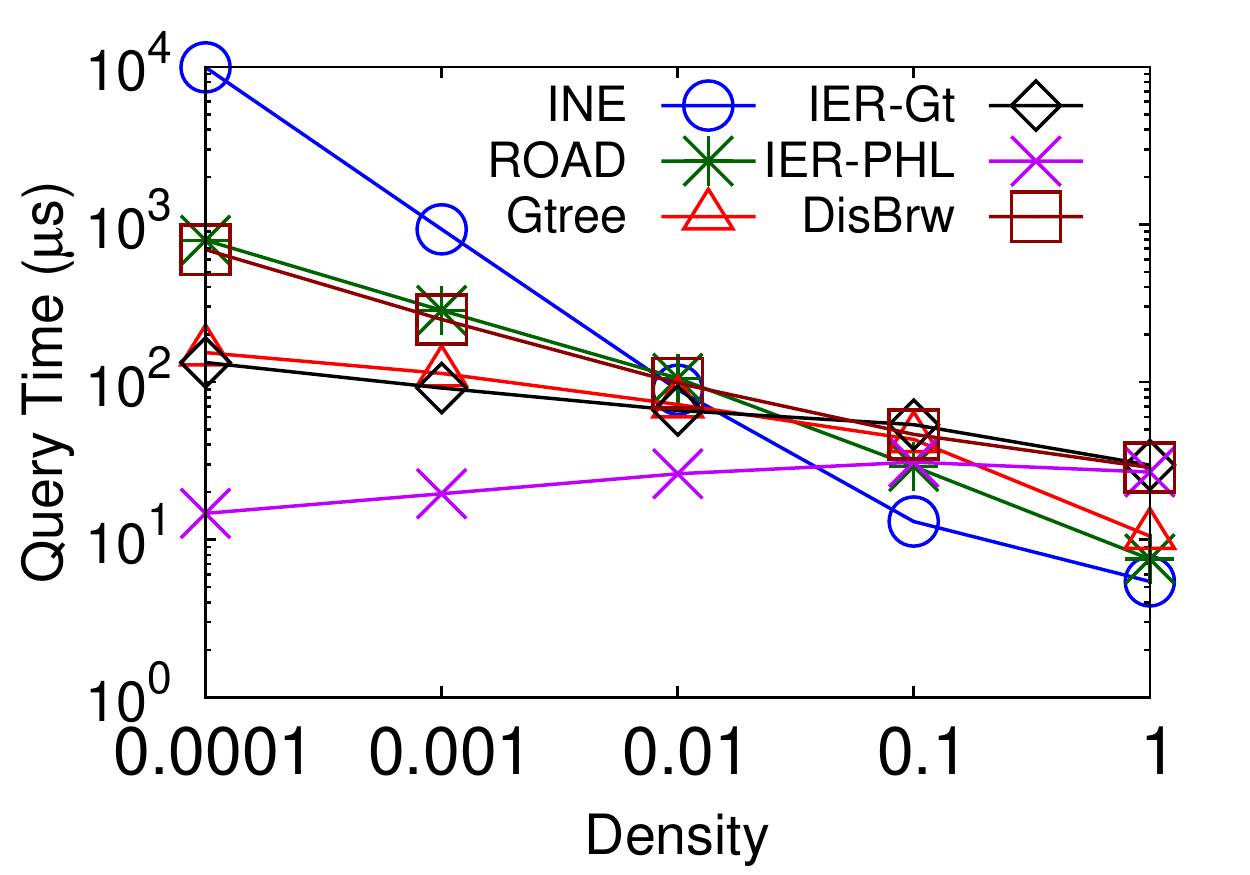}
			\label{exp:knn:nw_density_k10}
		}		
	    \hspace*{-5mm}
		\subfigure[US Dataset]{
			\includegraphics[width=0.5\linewidth]{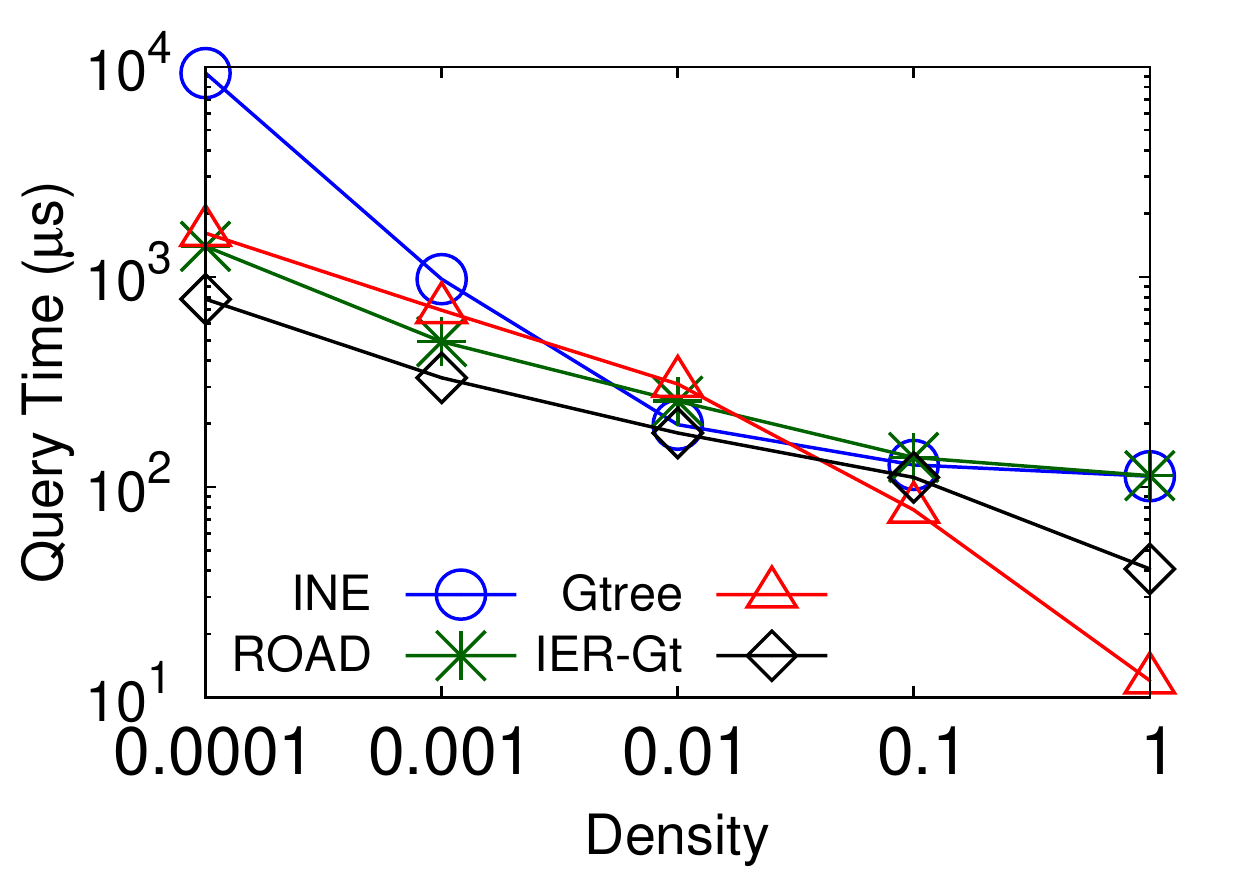}
			\label{exp:knn:usa_density_k10}
		}
	    \caption{Effect of Density \textmd{($k{=}10$)}}\label{exp:knn_d}
	\end{figure}
}

\subsubsection{Varying Clusters}\label{sec:exp:obj-clusters}

In this section we evaluate performance on clustered object sets proposed in Section \ref{sec:datasets:obj}. Figure \ref{exp:knn:clusters} shows the query time with increasing numbers of clusters and varying $k$. In both cases cluster size is at most $5$. Figure \ref{exp:knn:clusters:varyk} uses an object density of $0.001$. As the number of clusters increases the average distance between objects decreases leading to faster queries. This is analogous to increasing density, thus showing the same trend as for uniform objects.
IER-PHL's superiority is again apparent. One difference to uniform objects is IER-based methods find it more difficult to differentiate between candidates as the number of clusters increases, and query times increase (but not significantly). Similarly in Figure \ref{exp:knn:clusters:varyk}, as $k$ increases, IER-PHL visits more clusters, causing its performance lead to be slightly smaller than for uniform objects. IER-Gt on the other hand is more robust to this, as it is able to materialize most results. G-tree again performs better than DisBrw and ROAD. Due to clustering, objects in the same cluster will likely be located in the same G-tree leaf node. After finding the first object, G-tree can quickly retrieve other objects without recomputing distances to the leaf node, thus remaining relatively constant.

{
	\setlength{\abovecaptionskip}{\abvfigskp}
	\setlength{\belowcaptionskip}{-10pt}
	\begin{figure}[!htbp]
		\subfigure[Varying No. of Clusters]{
			\includegraphics[width=0.49\linewidth]{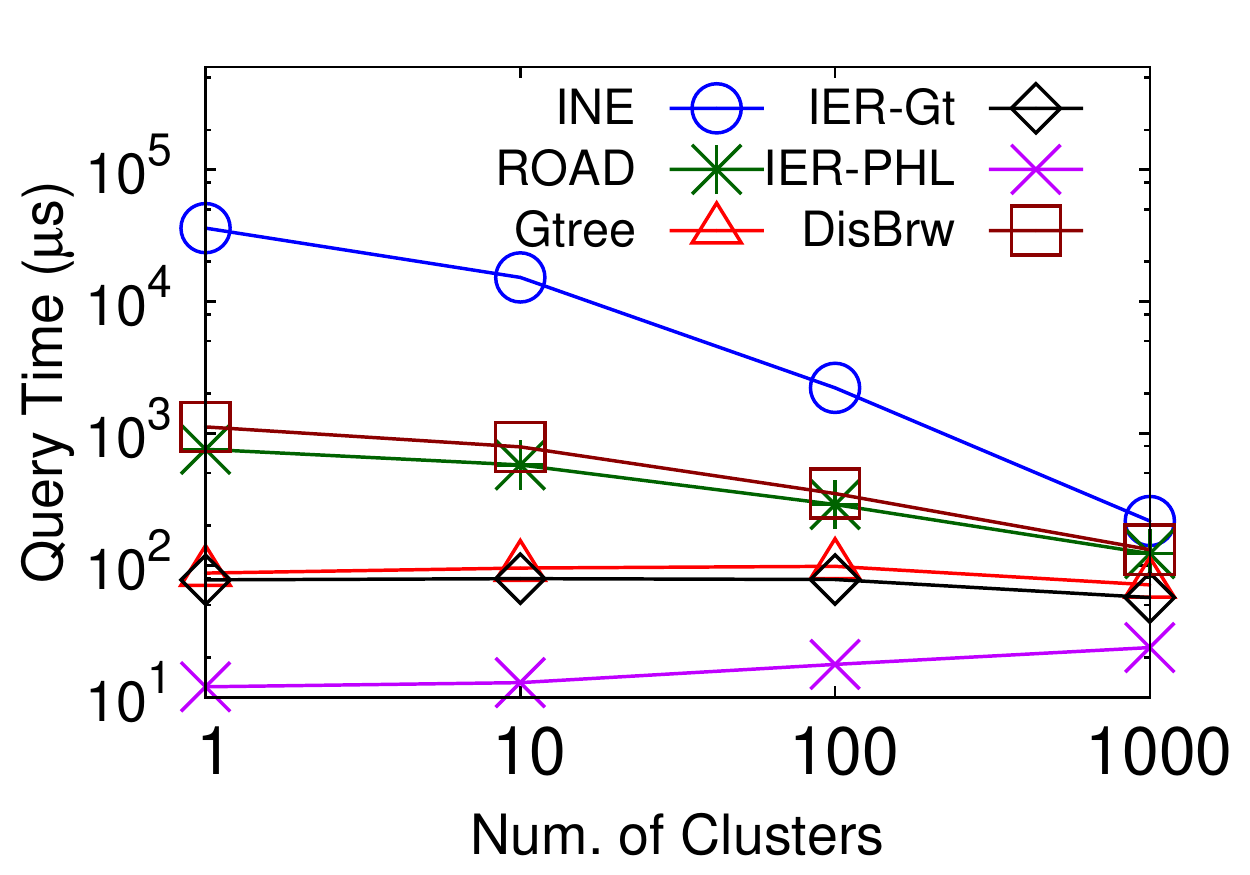}
			\label{exp:knn:clusters:varyc}
		}
		\hspace*{-5mm}
		\subfigure[Varying $k$]{
			\includegraphics[width=0.49\linewidth]{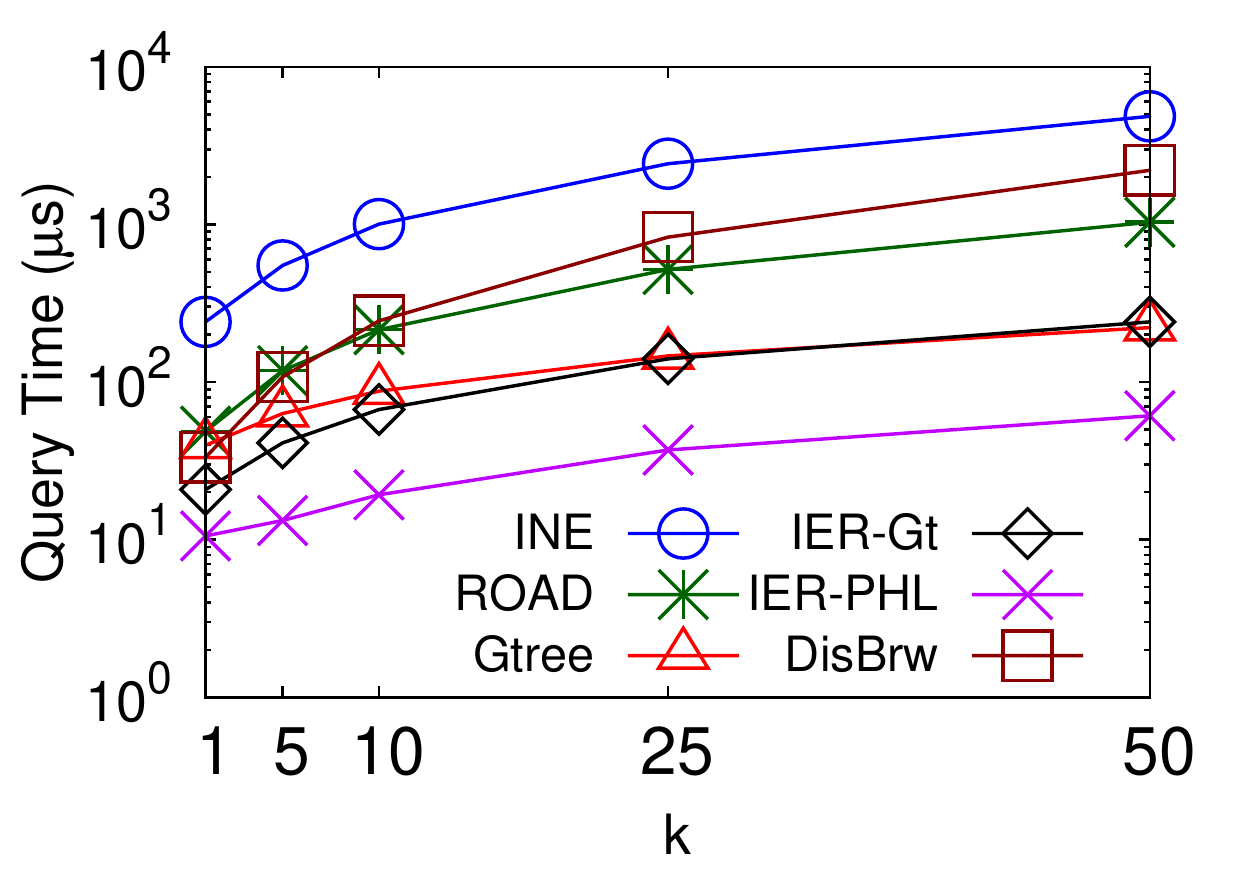}
			\label{exp:knn:clusters:varyk}
		}
		\caption{Effect of Clustered Objects \textmd{(NW, $|C|{=}0.001$, $k{=}10$)}}\label{exp:knn:clusters}
	\end{figure}
}

{
	\setlength{\abovecaptionskip}{\abvfigskp}
	\setlength{\belowcaptionskip}{-13pt}
	\begin{figure*}[!htbp]
		\subfigure[NW Road Network]{
			\includegraphics[width=0.55\linewidth]{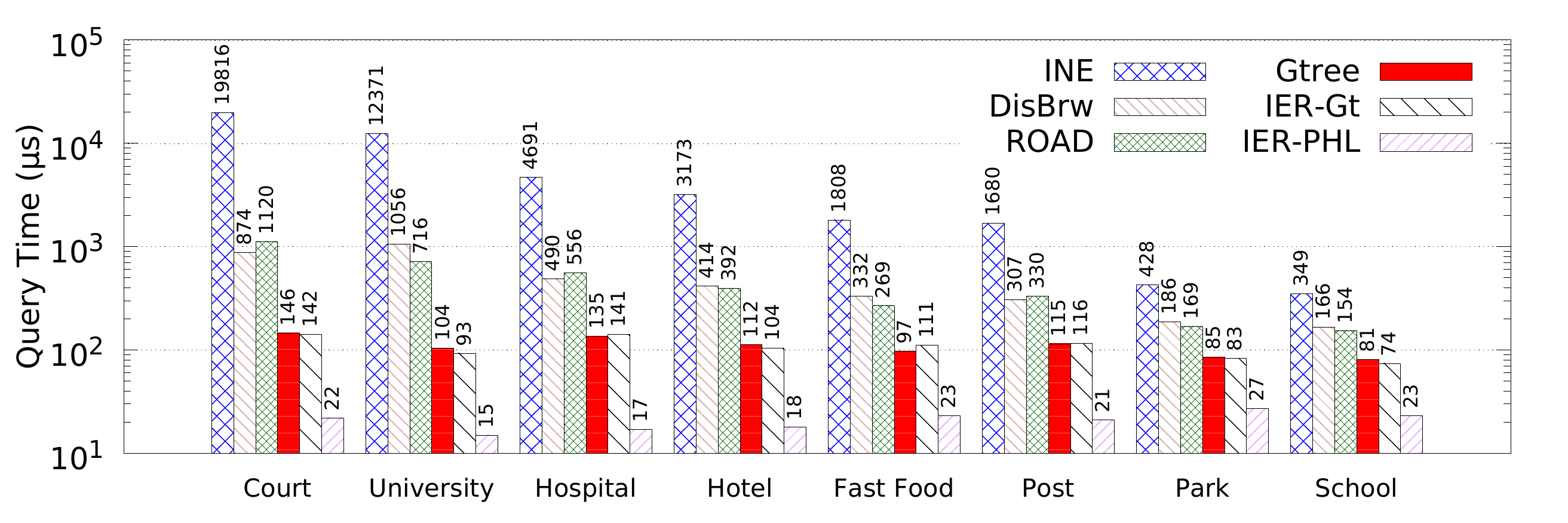}
			\label{exp:knn:rw:NW}
		}
		\hspace*{-6mm}
		\subfigure[US Road Network]{
			\includegraphics[width=0.48\linewidth]{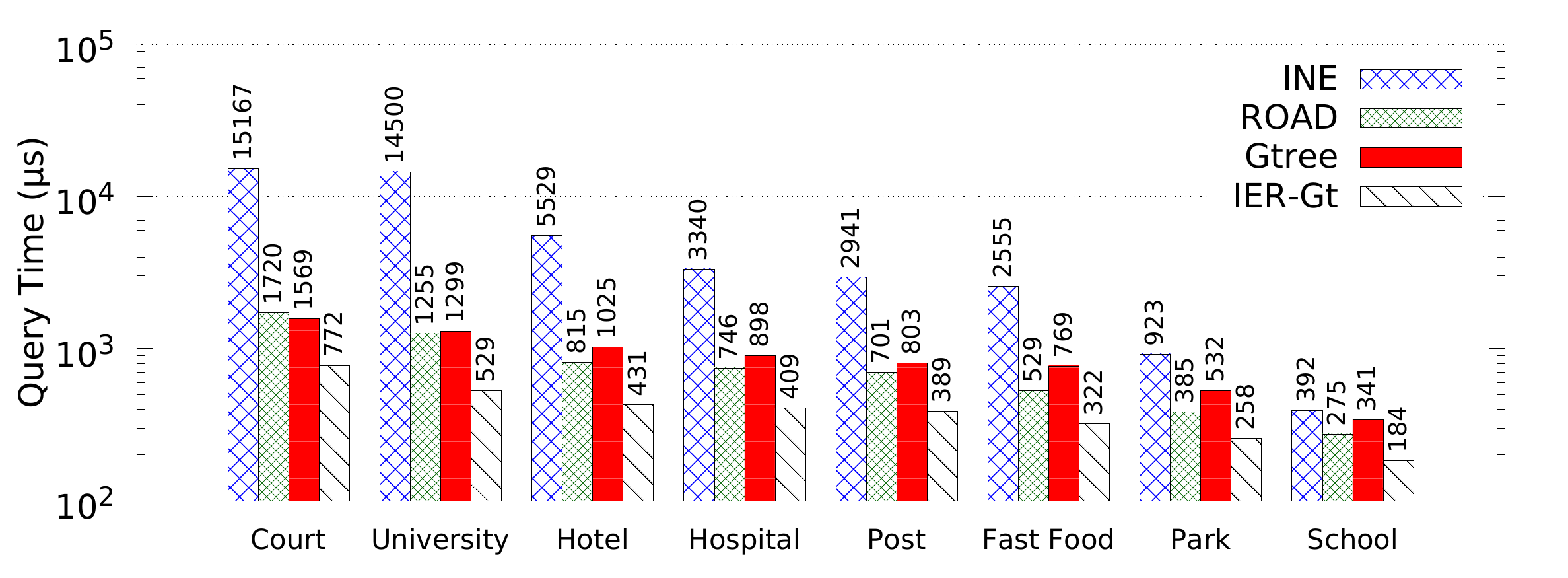}
			\label{exp:knn:rw:USA}
		}		
		\caption{Varying Real-World Object Sets \textmd{(Defaults: $k{=}10$)}}\label{exp:knn:rw}
	\end{figure*}
}

\subsubsection{Varying Minimum Object Distance}\label{sec:exp:obj-dist}

Each set $R_i$ in Figure \ref{exp:knn:obj_minmaxdist} represents an exponentially increasing network distance to the closest object with increasing $i$, as described in Section \ref{sec:datasets:obj}. For the smallest sets, objects still tend to be found further away, as there are fewer closer vertices.
However as distance increases further, we see the effect of ``remoteness". INE scales badly due to the increasing search space. IER-based methods scale poorly as the Euclidean lower bounds becomes less accurate with increasing network distance. This is particularly noticeable in Figure \ref{exp:knn:usa_obj_minmaxdist} as G-tree eventually overtakes IER-Gt on the US. But IER-PHL still outperforms all methods on NW. DisBrw performs poorly for a similar reason, making many interval refinements. G-tree scales extremely well in both cases, as more paths are visited through the G-tree hierarchy, more computations can be materialized for subsequent traversals.

{
	\setlength{\abovecaptionskip}{\abvfigskp}
	\setlength{\belowcaptionskip}{-10pt}
	\begin{figure}[t]
		\vspace{-2mm}
		\subfigure[NW Dataset]{
		\includegraphics[width=0.49\linewidth]{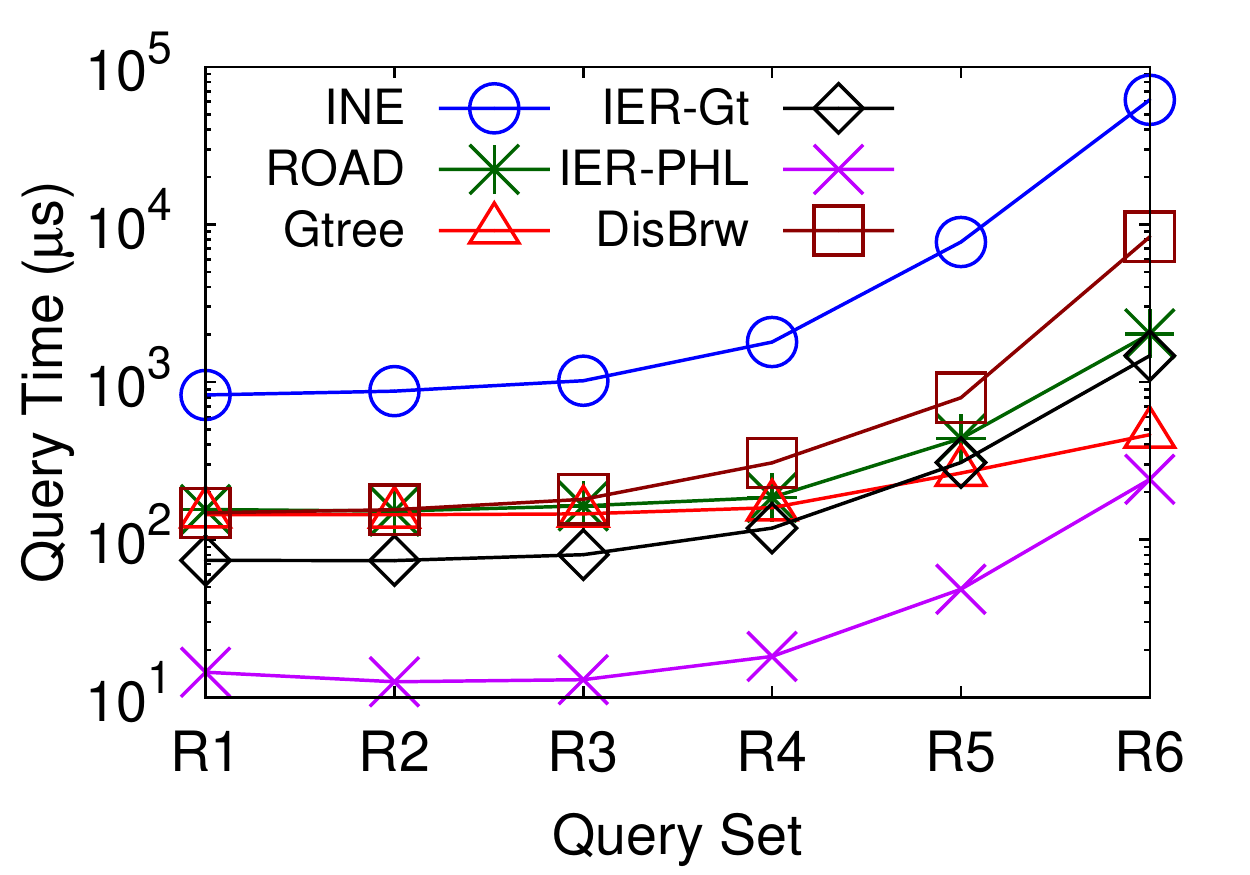}
			\label{exp:knn:nw_obj_minmaxdist}
		}
		\hspace*{-5mm}
		\subfigure[US Dataset]{
		\includegraphics[width=0.49\linewidth]{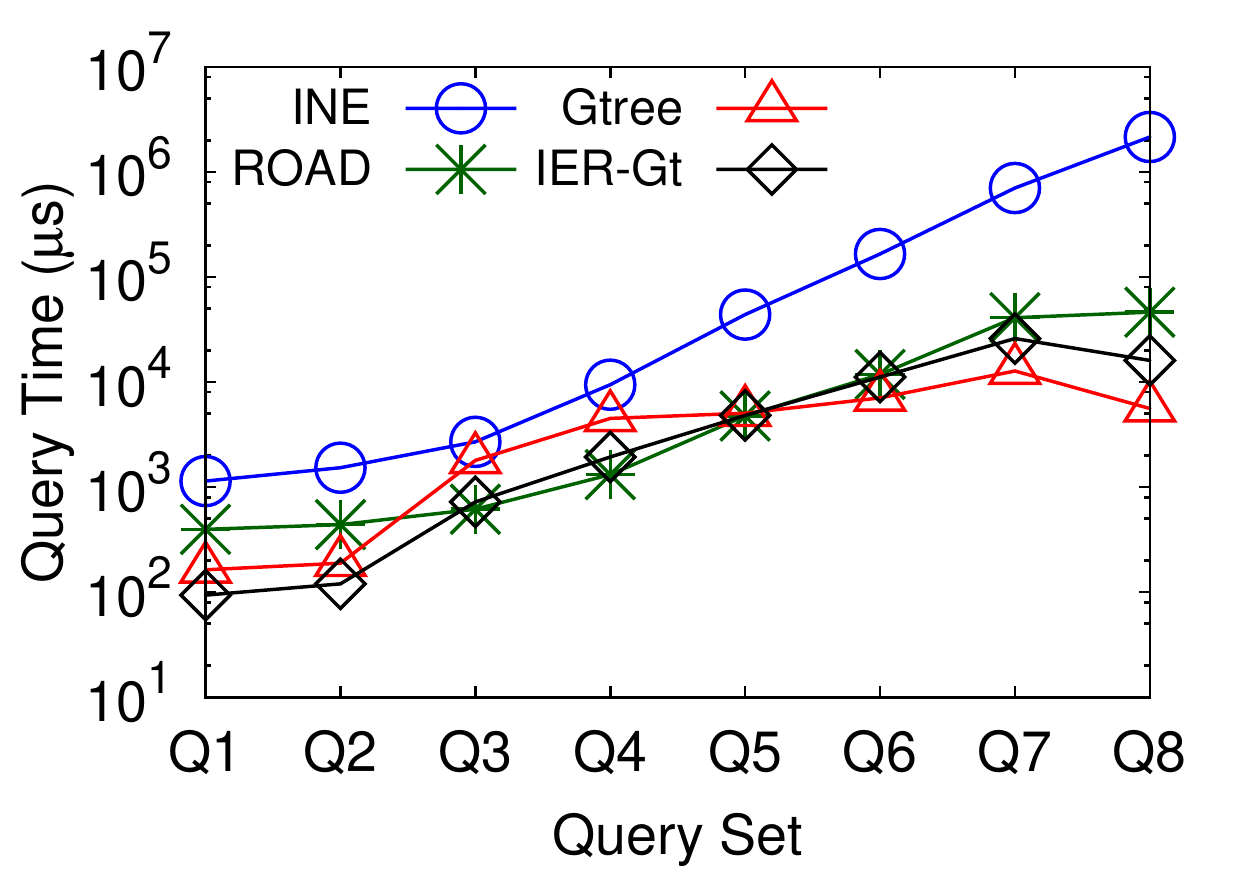}
			\label{exp:knn:usa_obj_minmaxdist}
		}
		\caption{Varying Min. Obj. Distance (\textmd{$d{=}0.001$, $k{=}10$)}}\label{exp:knn:obj_minmaxdist}
	\end{figure}

\subsubsection{Real-World Object Sets}\label{sec:exp:rw_pois}

\smallHeadIndent{Varying Object Sets.} In Figure \ref{exp:knn:rw}, we show query times of each technique on typical real-world object sets from Table \ref{tab:pois}. These are ordered by decreasing size, which is analogous to decreasing density, showing the same trend as in Figure \ref{exp:knn_d}. Schools represent the largest object set and all methods are extremely fast as seen for high density. A more typical POI, like hospitals, are less numerous and show the differences between methods more clearly. Regardless, IER-PHL on NW and IER-Gt on US consistently and significantly outperform other methods on most real-world object sets. Also note query times for G-tree are higher on US than NW for the same sets, confirming our observations in Section \ref{exp:knn:network_size}.

{
	\setlength{\abovecaptionskip}{\abvfigskp}
	\setlength{\belowcaptionskip}{-10pt}
	\begin{figure}[!htbp]
		\subfigure[Hospitals]{
			\includegraphics[width=0.49\linewidth]{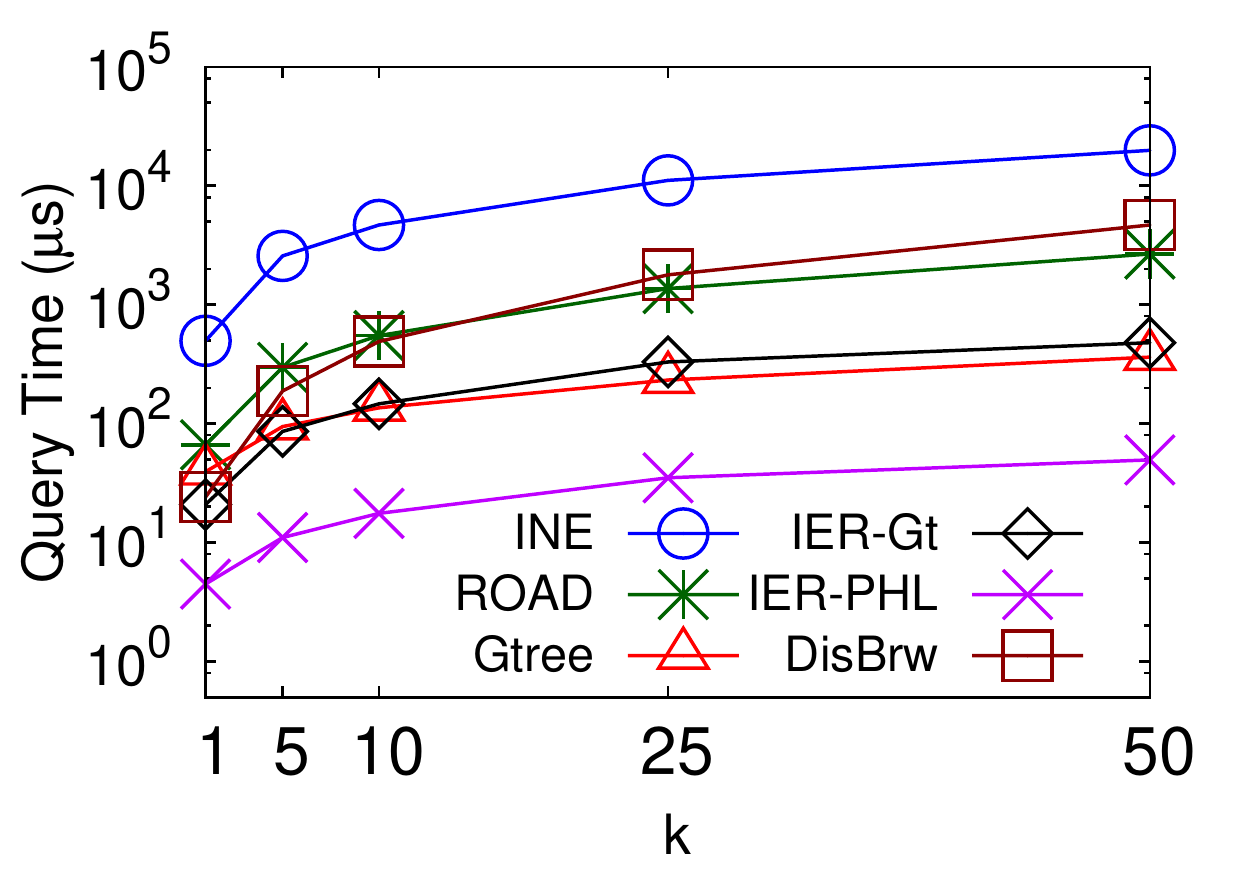}
			\label{exp:knn:rw_k:NW_rw_k_hospital}
		}
		\hspace*{-5mm}
		\subfigure[Fast Food]{
			\includegraphics[width=0.49\linewidth]{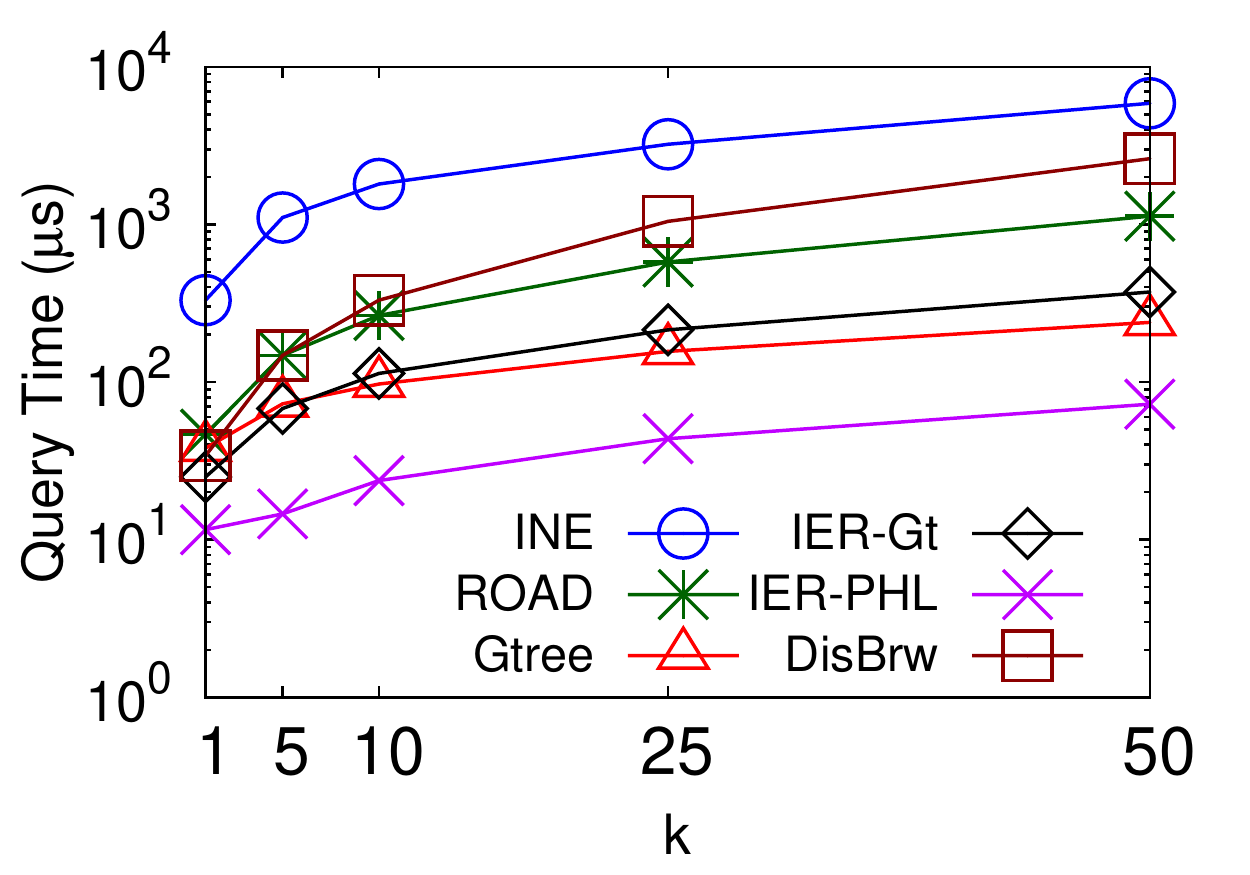}
			\label{exp:knn:rw_k:NW_rw_k_fastfood}
		}
		\caption{Varying $k$ for Real-World Objects (\textmd{NW, $k{=}10$)}}\label{exp:knn:rw_k}
	\end{figure}
}

\smallHeadIndent{Varying $k$.} Figure \ref{exp:knn:rw_k} shows the behaviour of two typically searched POIs, fast food outlets and hospitals, on the NW dataset. Hospitals display a trend similar to that of uniform objects for increasing $k$, as they tend to be sparse. IER-PHL is again significantly faster than G-tree. Although still fastest, IER-PHL has slightly lower performance for fast food outlets as these tend to appear in clusters where Euclidean distance is less able to distinguish better candidates, similar to synthetic clusters in Figure \ref{exp:knn:clusters:varyk}. Thus trends observed for equivalent synthetic object sets in previous experiments are also observed for real-world POIs.

\subsubsection{Original Settings}

A recent experimental comparison~\cite{zhong2015gtree} used a higher default density of $d{=}0.01$. While we choose a more typical default density, we reproduce results using $d{=}0.01$ in Figure~\ref{exp:knn_defaults} for varying $k$ and network size. Note that we use the smaller Colorado dataset in Figure~\ref{exp:knn_defaults:k_density01} for direct comparison with~\cite{zhong2015gtree}. Firstly, all methods compared in~\cite{zhong2015gtree} now answer queries in less than 1ms. While our CPU is faster, it cannot account for such a large difference. This suggests our implementations are indeed efficient.
Secondly, most methods are difficult to differentiate, as such a high density implies a very small search space (i.e., queries are ``easy" for all methods).

{
	\setlength{\abovecaptionskip}{\abvfigskp}
	\setlength{\belowcaptionskip}{-5pt}
	\begin{figure}[!htbp]
		\subfigure[Varying $k$]{
			\includegraphics[width=0.49\linewidth]{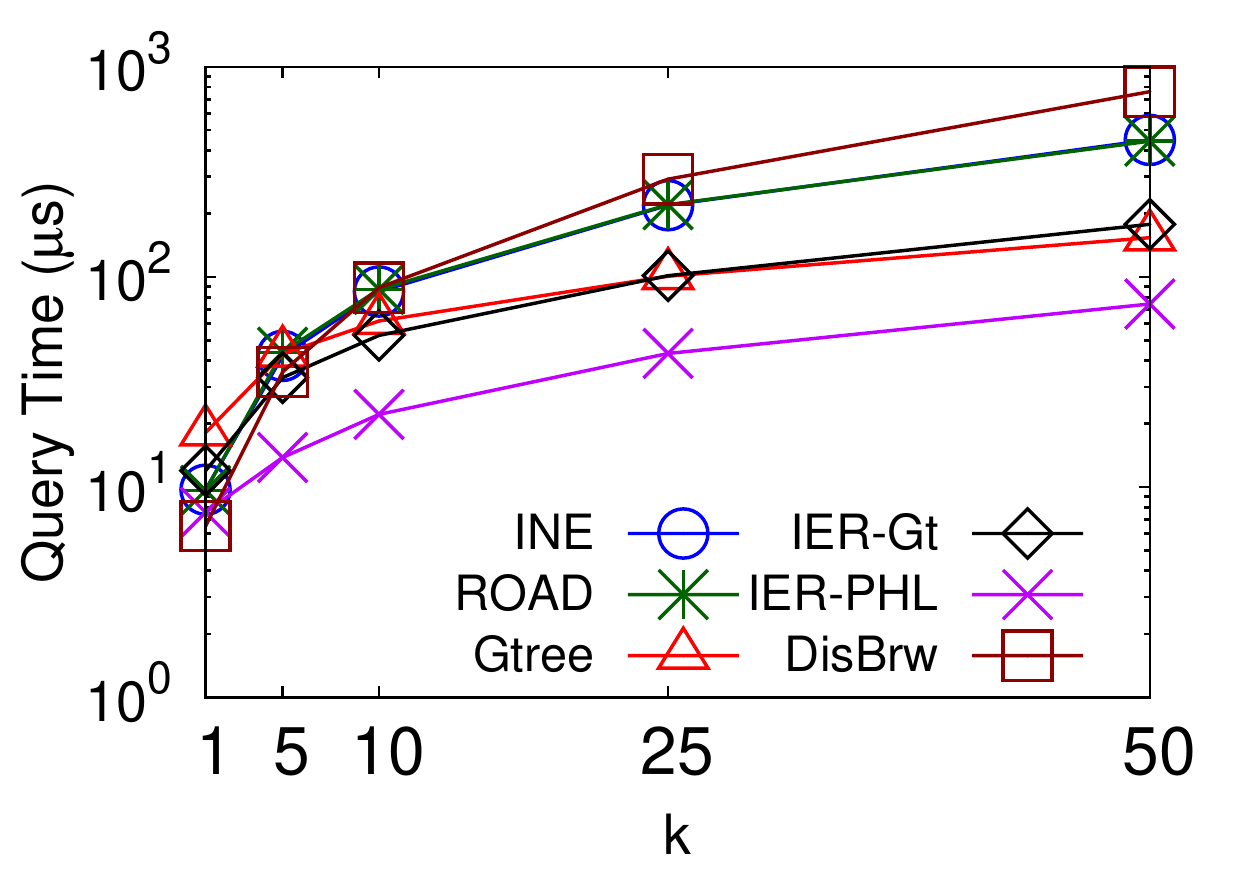}
			\label{exp:knn_defaults:k_density01}
		}
		\hspace*{-5mm}
		\subfigure[Varying $|V|$]{
			\includegraphics[width=0.49\linewidth]{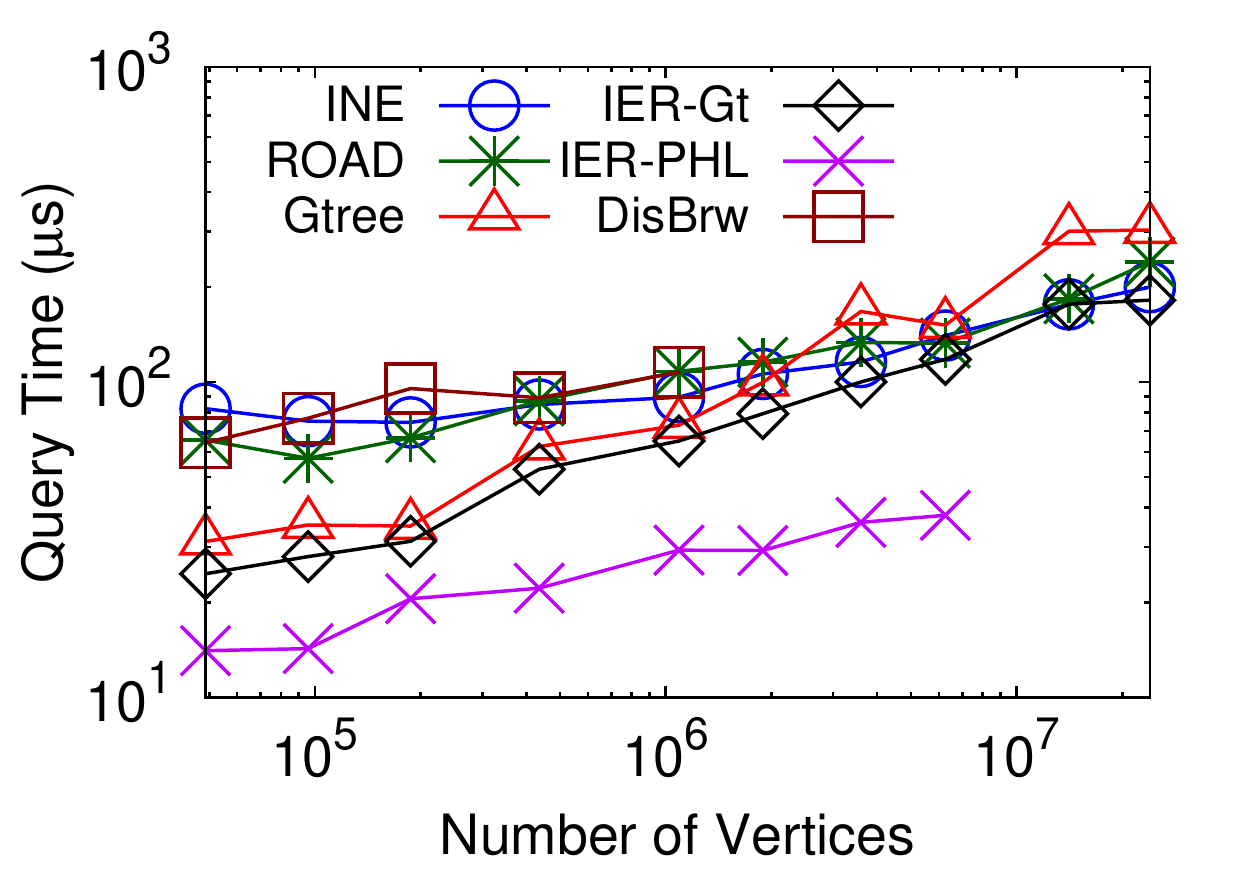}
			\label{exp:knn_defaults:network_density01_k10}
		}
		\caption{$k$NN Queries \textmd{(CO, $d{=}0.01$, $k{=}10$)}}\label{exp:knn_defaults}
	\end{figure}
}

{
	\setlength{\abovecaptionskip}{\abvfigskp}
	\setlength{\belowcaptionskip}{-13pt}
	\begin{figure*}[!htbp]%
		\subfigure[Varying $k$]{
			\includegraphics[width=0.25\linewidth]{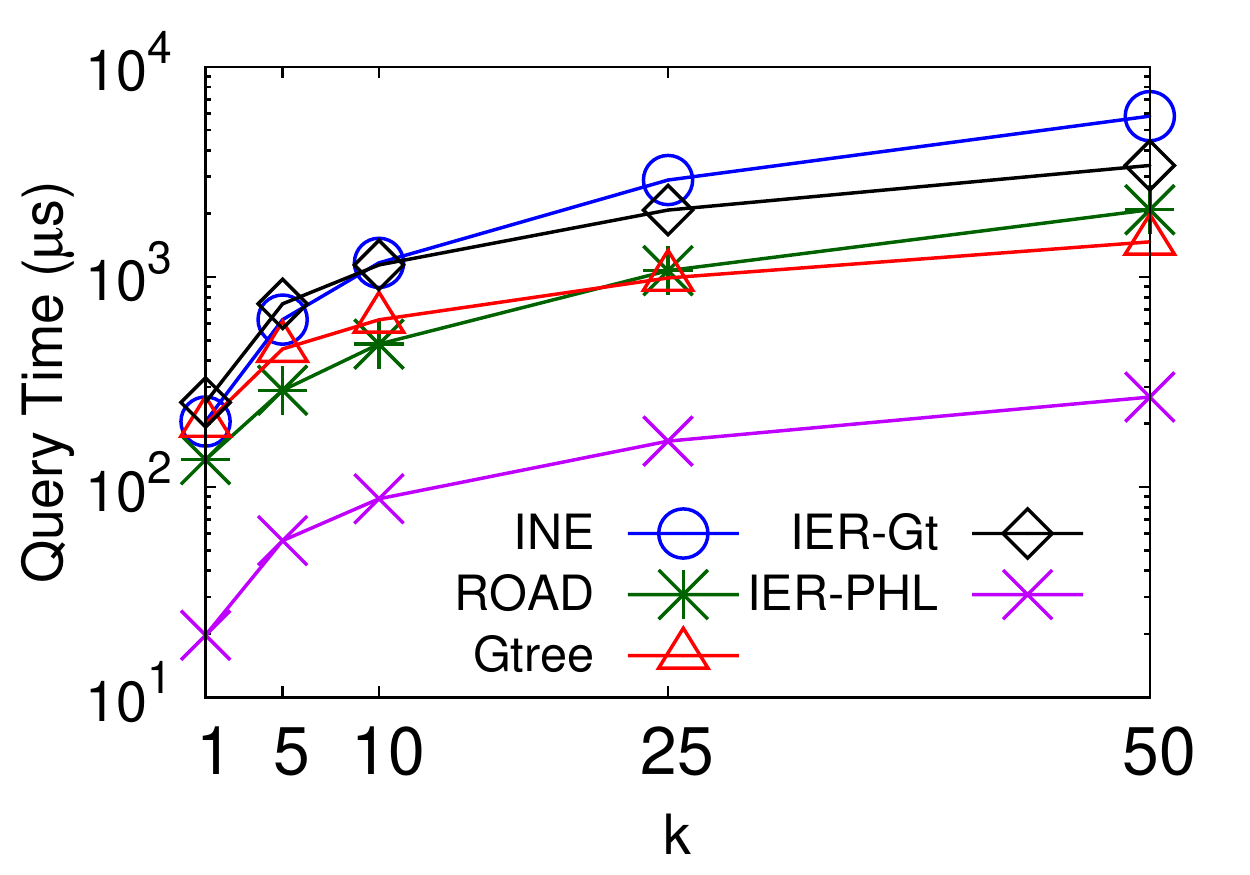}
			\label{exp:knn:tt_usa_k_density001}
		}
		\hspace*{-5mm}
		\subfigure[Varying Density $d$]{
			\includegraphics[width=0.25\linewidth]{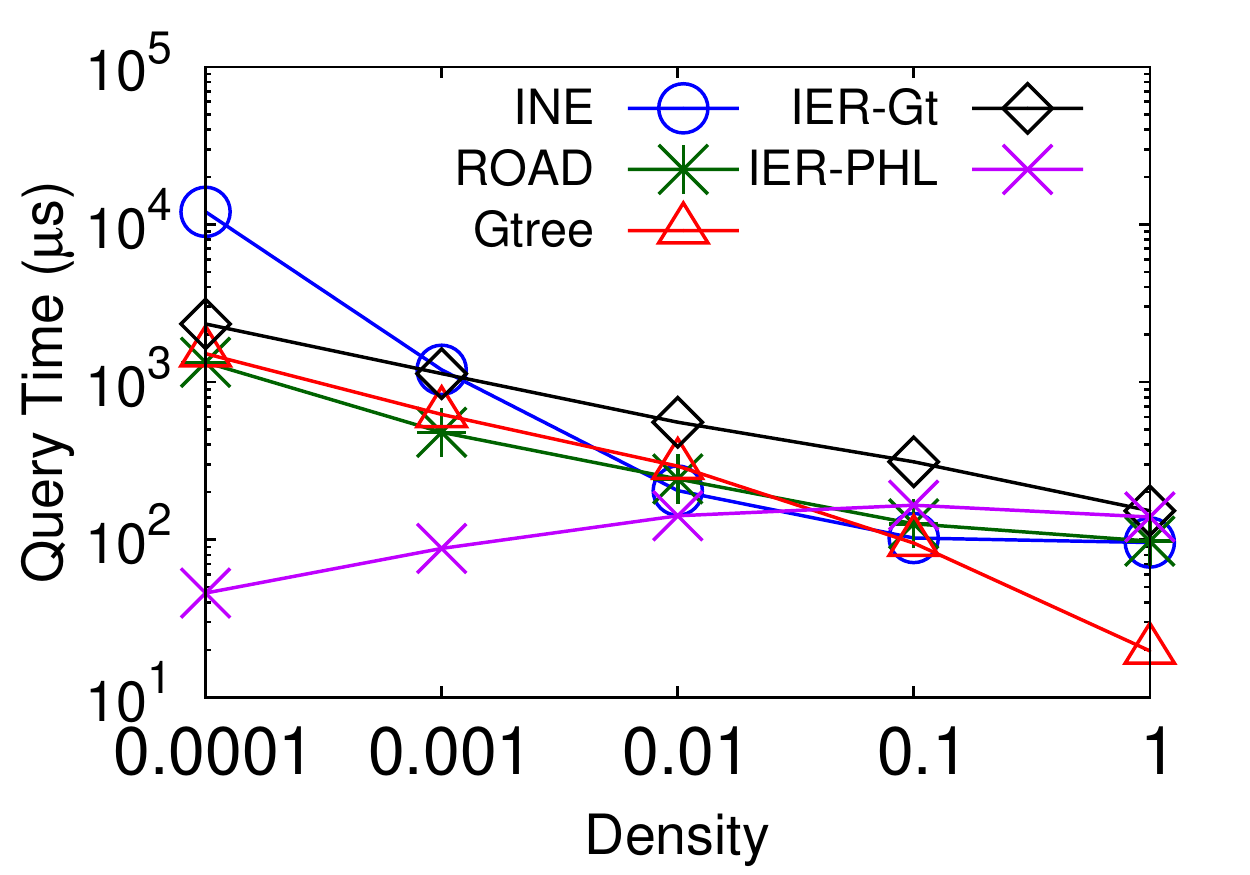}
			\label{exp:knn:tt_usa_density_k10}
		}		
		\hspace*{-5mm}
		\subfigure[Varying $|V|$]{
			\includegraphics[width=0.25\linewidth]{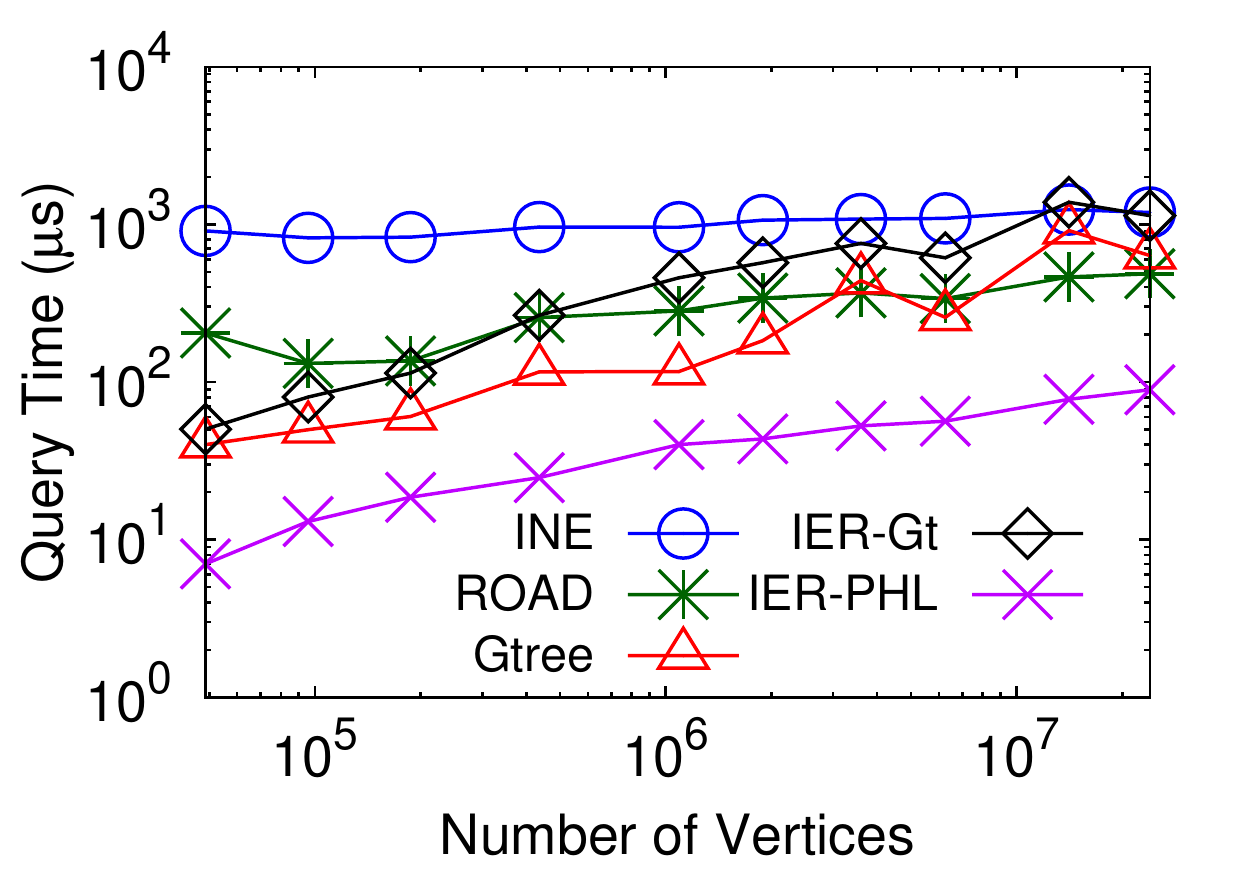}
			\label{exp:knn:tt_usa_network_density001_k10}
		}		
		\hspace*{-5mm}
		\subfigure[Varying Min. Obj. Distance]{
			\includegraphics[width=0.25\linewidth]{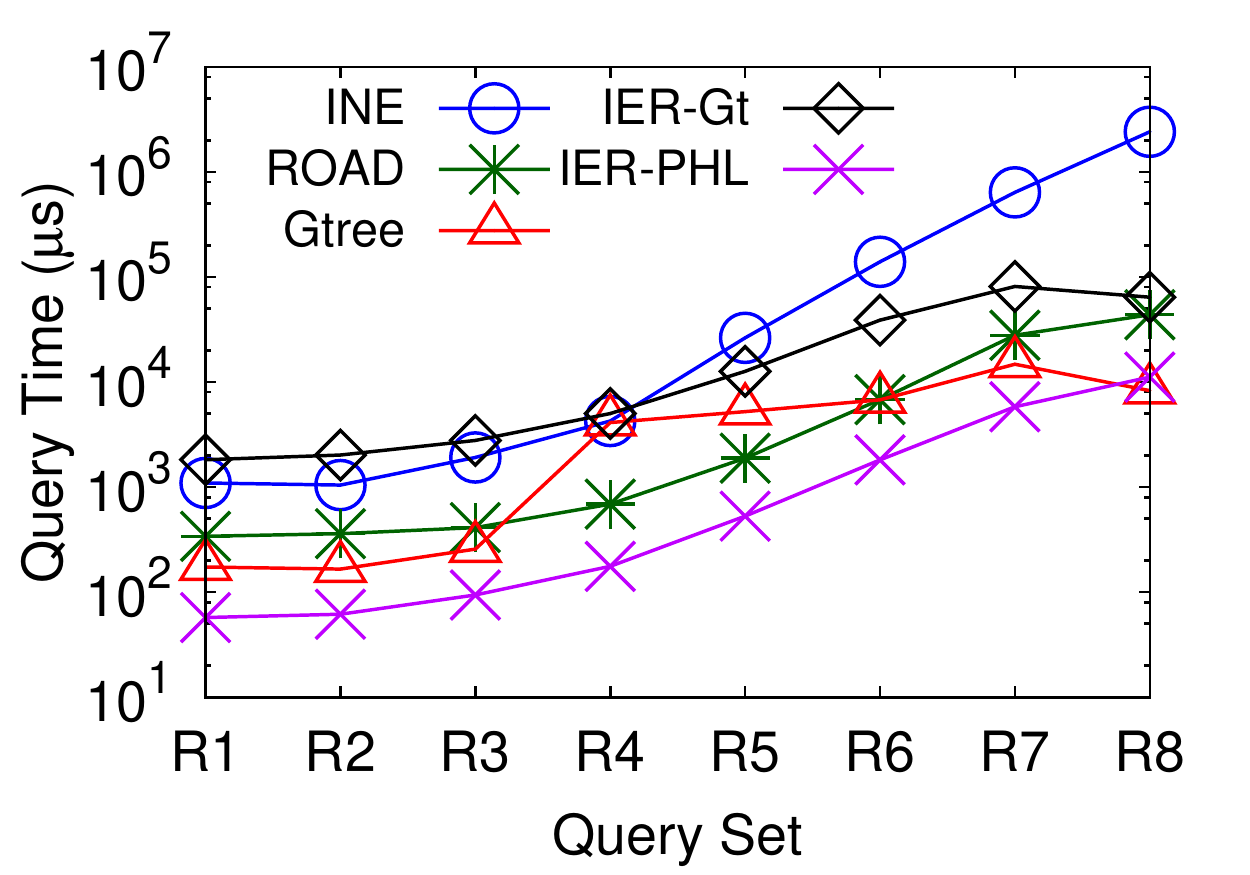}
			\label{exp:knn:tt_usa_obj_mindist}
		}
		\caption{Query Performance on Travel Time Graphs \textmd{(US, $k{=}10$, $d{=}0.001$, uniform objects)}}\label{exp:knn:travel_times}
	\end{figure*}
}

\subsection{Object Set Index Pre-Processing Cost}\label{sec:exp:obj_indexes}

The original ROAD paper \cite{lee2012road} included pre-processing of a fixed object set in its road network index statistics. But there may be many object sets (e.g., one for each type of restaurant) or objects may need frequent updating (e.g., hotels with vacancies). So we are interested in the performance of individual object indexes over varying size (i.e., density). We evaluate 3 object indexes on the US dataset, namely: \textit{R-trees} used by IER, \textit{Association Directories} used by ROAD and \textit{Occurrence Lists} used by G-tree. Note that in our study DisBrw also uses R-trees (see Appendix \ref{sec:app:silc:enn}).

{
	\setlength{\abovecaptionskip}{\abvfigskp}
	\setlength{\belowcaptionskip}{-12pt}
	\begin{figure}[!htbp]
		\subfigure[Index Size vs. $d$]{
			\includegraphics[width=0.48\linewidth]{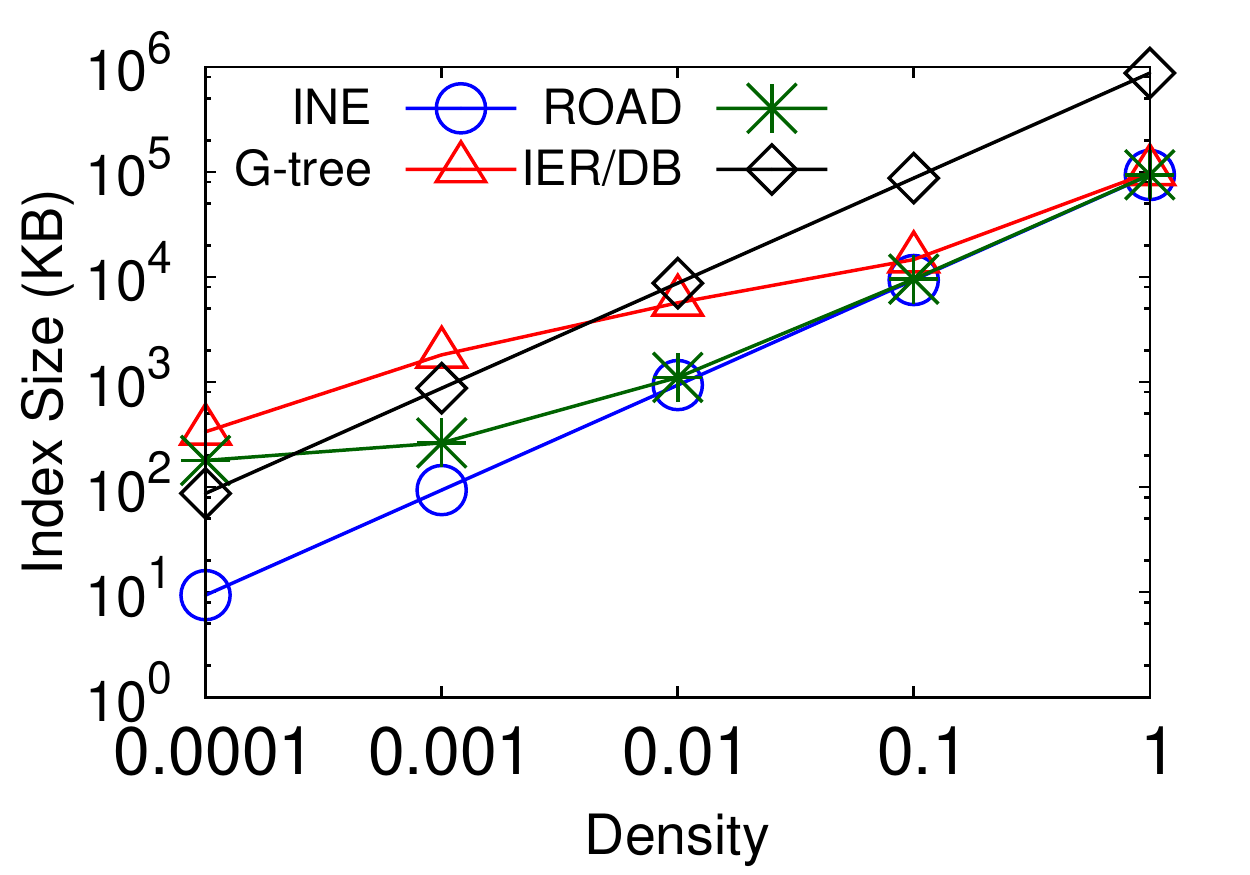}
			\label{exp:obj_idx:varyd_vs_size}
		}
		\hspace*{-5mm}
		\subfigure[Build Time vs. Density $d$]{
			\includegraphics[width=0.48\linewidth]{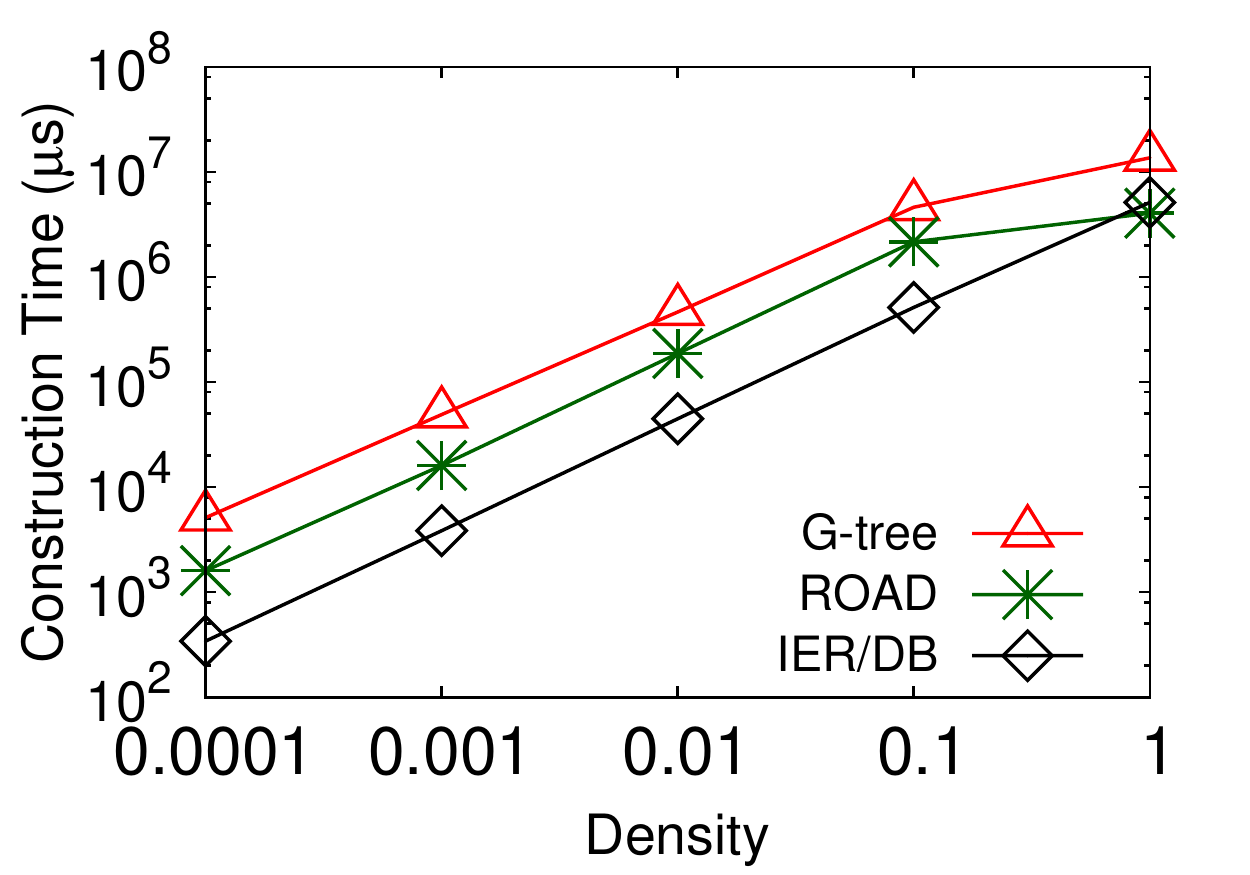}
			\label{exp:obj_idx:varyd_vs_time}
		}
		\caption{Object Indexes for US \textmd{(uniform objects)}}\label{exp:obj_idx}
	\end{figure}
}

\smallHeadIndent{Index Size.} In practice object indexes for all object sets would be constructed offline, loaded into memory and the appropriate one injected at query time. We investigate the index sizes (in KB) in Figure \ref{exp:obj_idx:varyd_vs_size} to gauge what effect each density has on the total size. The size of the input object set used by INE is the lower bound storage cost. ROAD's object index is smaller than G-tree's because it need only store whether an Rnet contains an object or not, which is easily done in a low memory bit-array. G-tree's object index must additionally store the child nodes containing objects. Both indexes must however store the actual objects, which gradually dominates the index size with increasing density. Note that we chose R-tree parameters (e.g., node capacity) for best performance. As a result R-trees fall behind after density 0.01, but this can be remedied by increasing the node capacity at the expense of Euclidean $k$NN performance. We note that object indexes are much smaller than road network indexes, as they are simpler data structures, and real-world object sets with high densities are less frequent.

\smallHeadIndent{Construction Time.} Figure \ref{exp:obj_idx:varyd_vs_time} similarly shows the object index construction times. Again, they are all constructed much faster than road network indexes, due to being simpler data structures. The ROAD and G-tree object indexes incur the largest build time due to bottom-up propagation of the presence of objects through their respective hierarchies. However, the R-trees used by IER are significantly faster to build. As R-trees support updates, this suggests the possibility of use in real-time settings.

\subsection{Travel Time Road Networks}\label{sec:exp:tt}

$k$NNs may just as commonly be required in terms of travel time. In this section we reproduce query results for the US road network with travel time edge weights. Results for other experiments are found in Appendix \ref{sec:app:tt}, but we note any significant differences here.

\smallHeadIndent{Extending IER}. IER uses the Euclidean distance as a lower bound on the network distance between two points for travel distance edge weights. This can easily be extended for other edge weights. Let $w_i$ (resp. $d_i$) represent the edge weight (resp. Euclidean length) of an edge $e_i$. We compute $S=max_{\forall e_i\in E} \allowbreak(d_i/w_i)$. E.g., if $w_i$ represents travel time, $S$ corresponds to the maximum speed on any edge in the network. Let $d_E(p,q)$ be the Euclidean distance between two points $p$ and $q$. It is easy to see that $d_E(p,q)/S$ is a lower bound on the network distance between $p$ and $q$, e.g., the time it takes to travel the Euclidean distance at the maximum possible speed. Thus, we compute $S$ for the network and use the new lower bound in IER. Landmarks are known to provide better lower bounds on travel time graphs \cite{goldberg2005alt}. However there is no equivalent data structure, such as an R-tree, to incrementally retrieve candidates by their lower bound, making them undesirable for use here.

Unlike travel distances, we were able to construct the PHL index for \textit{all} datasets, with the largest requiring 16GB. This is due to ``highway" properties exhibited in travel time graphs (e.g., an edge with a higher speed is more likely to be on a shortest path) leading to smaller label sizes. Figure \ref{exp:knn:travel_times} shows the query times for travel times with varying $k$, density, network size $|V|$ and object distance on the US dataset. In general, IER experiences more false hits due to the looser lower bound on travel times, explaining why IER-Gt is now significantly outperformed by G-tree. But, surprisingly, IER-PHL still remains the fastest method in most situations. The penalty in false hits is partly offset by the reduced label sizes for PHL. The looser lower bound also aggravates cases where Euclidean distance was less effective on travel distances. For example, IER was already less able to distinguish better candidates with increasing density, and as result IER-PHL degrades faster on travel times in Figure \ref{exp:knn:tt_usa_density_k10}. This is similarly observed for increasing network distance in Figure \ref{exp:knn:tt_usa_obj_mindist} for the same reason. Despite this, IER-PHL remains the fastest method in most cases. Other trends observed for travel distances are similarly observed for travel times (e.g., G-tree degrades with increasing $|V|$ in Figure \ref{exp:knn:tt_usa_network_density001_k10}).

\vspace{1.5mm}

{
	\setlength{\abovecaptionskip}{\abvtblskp}
	\setlength{\belowcaptionskip}{-10pt}
	\begin{table}[!htbp]
		\centering
		\small
		\begin{tabular}{|l|c|c|c|c|c|}
			\hline
			\textbf{Criteria} & \textbf{INE} & \textbf{G-tree} & \textbf{ROAD} & \textbf{IER} & \textbf{DisBrw} \\ \hline 
			\multicolumn{6}{|c|}{\textbf{Query Performance}} \\ \hline
			Default Settings & 5th & 2nd  & =3rd & 1st & =3rd \\ \hline
			Small $k$ & 5th & =3rd & =3rd & 1st & 2nd \\ \hline
			Large $k$ & 5th  & 2nd & 3rd & 1st & 4th \\ \hline
			Low Density & 5th & 2nd & =3rd & 1st & =3rd \\ \hline
			High Density & 1st & 3rd & 2nd & 4th & 5th \\ \hline
			Small Networks & 5th & 2nd & =3rd & 1st  & =3rd \\ \hline
			Large Networks & 4th & =3rd & 2nd & 1st  & N/A \\ \hline
			\multicolumn{6}{|c|}{\textbf{Network and Object Index Pre-Processing}} \\ \hline
			Time (Network) & 1st & 3rd & 2nd & 4th & 5th \\ \hline
			Time (Objects) & 1st & 5th & 4th & =2nd & =2nd \\ \hline		
			Space (Network) & 1st & 2nd & 3rd & 4th & 5th \\ \hline
			Space (Objects) & 1st & 5th & 2nd & =3rd & =3rd \\ \hline		
		\end{tabular}
		\caption{Ranking of Algorithms Under Different Criteria}
		\label{tab:conclusion}
	\end{table}
}

\section{Conclusions}\label{sec:conclusions}

We have presented an extensive experimental study for the $k$NN problem on road networks, settling unanswered questions by evaluating object indexes, travel time graphs and real-world POIs. We verify that G-tree generally outperforms INE, DisBrw and ROAD, but the relative improvement is much smaller and at times reversed, demonstrating the impact of implementation efficiency.
Table~\ref{tab:conclusion} provides the ranking of the algorithms under different criteria.

Our most significant conclusions are regarding IER, which we investigated with fast network distance techniques for the first time. IER-PHL significantly outperformed every competitor in all but a few cases, even on travel time graphs where Euclidean distance is less effective. IER provides a flexible framework that can be combined with the fastest shortest path technique allowed by the users' memory capacity and must be included in future comparisons. Additionally, on travel distances, we saw that IER-Gt often outperformed the original G-tree $k$NN algorithm despite using the same index. As this suggests Euclidean NN can be a better heuristic, it identifies room for improvement in $k$NN search heuristics. Perhaps more information can be incorporated into object indexes.

Finally, we investigated the effect of implementation choices using G-trees distance matrices and data structures in INE. By investigating simple choices, we show that even small improvements in cache-friendliness can significantly improve algorithm performance. As such there is a need to pay careful attention when implementing and designing algorithms for main memory, and our insights are applicable to any technique not just those we study.

\vspace{1.5mm}
\noindent
\textbf{Acknowledgements}. We sincerely thank the authors of G-tree~\cite{zhong2015gtree}, PHL \cite{kawata2014phl} and the shortest path experimental paper \cite{wu2012shortest} for providing source code of their algorithms, in particular Y. Kawata for valuable assistance in improving the code of PHL. We also thank the anonymous reviewers for their feedback through which we significantly improved our work. The research of Muhammad Aamir Cheema is supported by ARC DE130101002 and DP130103405. 


\bibliographystyle{abbrv}
\bibliography{nd_knn}  

\begin{thebibliography}{10}

\bibitem{globalwebindex}
\url{https://www.globalwebindex.net/blog/top-global-smartphone-apps}.

\bibitem{oursource}
\url{https://github.com/tenindra/RN-kNN-Exp}.

\bibitem{dimacs9}
\url{http://www.dis.uniroma1.it/\%7Echallenge9/}.

\bibitem{osm}
\url{http://www.openstreetmap.org}.

\bibitem{lifeifei}
\url{http://www.cs.utah.edu/\%7Elifeifei/SpatialDataset.htm}.

\bibitem{abey2016exp}
T.~Abeywickrama, M.~A. Cheema, and D.~Taniar.
\newblock K-nearest neighbors on road networks: A journey in experimentation
  and in-memory implementation.
\newblock {\em PVLDB}, 9(6):492--503, Jan. 2016.

\bibitem{kawata2014phl}
T.~Akiba, Y.~Iwata, K.-i. Kawarabayashi, and Y.~Kawata.
\newblock Fast shortest-path distance queries on road networks by pruned
  highway labeling.
\newblock In {\em ALENEX}, pages 147--154, 2014.

\bibitem{bast2007tnr}
H.~Bast, S.~Funke, D.~Matijevic, P.~Sanders, and D.~Schultes.
\newblock In transit to constant time shortest-path queries in road networks.
\newblock In {\em WEA}, pages 46--59, 2007.

\bibitem{bauer2010chg}
R.~Bauer, D.~Delling, P.~Sanders, D.~Schieferdecker, D.~Schultes, and
  D.~Wagner.
\newblock Combining hierarchical and goal-directed speed-up techniques for
  dijkstra's algorithm.
\newblock {\em JEA}, 15:2.3:2.1--2.3:2.31, Mar. 2010.

\bibitem{chen2009pathknn}
Z.~Chen, H.~T. Shen, X.~Zhou, and J.~X. Yu.
\newblock Monitoring path nearest neighbor in road networks.
\newblock In {\em SIGMOD}, pages 591--602, 2009.

\bibitem{cho2005unicons}
H.-J. Cho and C.-W. Chung.
\newblock An efficient and scalable approach to cnn queries in a road network.
\newblock In {\em VLDB}, pages 865--876, 2005.

\bibitem{cormen2011book}
T.~H. Cormen, C.~Stein, R.~L. Rivest, and C.~E. Leiserson.
\newblock {\em Introduction to Algorithms}.
\newblock McGraw-Hill, 2nd edition, 2001.

\bibitem{geisberger2008ch}
R.~Geisberger, P.~Sanders, D.~Schultes, and D.~Delling.
\newblock Contraction hierarchies: Faster and simpler hierarchical routing in
  road networks.
\newblock In {\em WEA}, pages 319--333, 2008.

\bibitem{goldberg2005alt}
A.~V. Goldberg and C.~Harrelson.
\newblock Computing the shortest path: A search meets graph theory.
\newblock In {\em SODA}, 2005.

\bibitem{hu2006distidx}
H.~Hu, D.~L. Lee, and V.~C.~S. Lee.
\newblock Distance indexing on road networks.
\newblock In {\em VLDB}, pages 894--905, 2006.

\bibitem{hu2006nd}
H.~Hu, D.~L. Lee, and J.~Xu.
\newblock Fast nearest neighbor search on road networks.
\newblock In {\em EDBT}, pages 186--203, 2006.

\bibitem{huang2005islands}
X.~Huang, C.~S. Jensen, and S.~\v{S}altenis.
\newblock The islands approach to nearest neighbor querying in spatial
  networks.
\newblock In {\em SSTD}, pages 73--90, 2005.

\bibitem{karypis1998fast}
G.~Karypis and V.~Kumar.
\newblock A fast and high quality multilevel scheme for partitioning irregular
  graphs.
\newblock {\em SIAM J. Sci. Comput.}, 20(1):359--392, 1998.

\bibitem{kolahdouzan2004vknn}
M.~Kolahdouzan and C.~Shahabi.
\newblock Voronoi-based k nearest neighbor search for spatial network
  databases.
\newblock In {\em VLDB}, pages 840--851, 2004.

\bibitem{lee2012road}
K.~Lee, W.-C. Lee, B.~Zheng, and Y.~Tian.
\newblock Road: A new spatial object search framework for road networks.
\newblock {\em TKDE}, 24(3):547--560, 2012.

\bibitem{lee2009road}
K.~C.~K. Lee, W.-C. Lee, and B.~Zheng.
\newblock Fast object search on road networks.
\newblock In {\em EDBT}, pages 1018--1029, 2009.

\bibitem{mouratidis2006cknn}
K.~Mouratidis, M.~L. Yiu, D.~Papadias, and N.~Mamoulis.
\newblock Continuous nearest neighbor monitoring in road networks.
\newblock In {\em VLDB}, pages 43--54, 2006.

\bibitem{papadias2003ine}
D.~Papadias, J.~Zhang, N.~Mamoulis, and Y.~Tao.
\newblock Query processing in spatial network databases.
\newblock In {\em VLDB}, pages 802--813, 2003.

\bibitem{samet2005book}
H.~Samet.
\newblock {\em Foundations of Multidimensional and Metric Data Structures}.
\newblock Morgan Kaufmann, 2005.

\bibitem{samet2008distbrws}
H.~Samet, J.~Sankaranarayanan, and H.~Alborzi.
\newblock Scalable network distance browsing in spatial databases.
\newblock In {\em SIGMOD}, pages 43--54, 2008.

\bibitem{sankaranarayanan2005silc}
J.~Sankaranarayanan, H.~Alborzi, and H.~Samet.
\newblock Efficient query processing on spatial networks.
\newblock In {\em GIS}, pages 200--209, 2005.

\bibitem{shahabi2002rne}
C.~Shahabi, M.~Kolahdouzan, and M.~Sharifzadeh.
\newblock A road network embedding technique for k-nearest neighbor search in
  moving object databases.
\newblock {\em GeoInformatica}, 7(3):255--273, 2003.

\bibitem{sidlauskas2014impl}
D.~\v{S}idlauskas and C.~S. Jensen.
\newblock Spatial joins in main memory: Implementation matters!
\newblock {\em PVLDB}, 8(1):97--100, 2014.

\bibitem{wu2012shortest}
L.~Wu, X.~Xiao, D.~Deng, G.~Cong, A.~D. Zhu, and S.~Zhou.
\newblock Shortest path and distance queries on road networks: An experimental
  evaluation.
\newblock {\em PVLDB}, 5(5):406--417, Jan. 2012.

\bibitem{zhong2015gtree}
R.~Zhong, G.~Li, K.~Tan, L.~Zhou, and Z.~Gong.
\newblock G-tree: An efficient and scalable index for spatial search on road
  networks.
\newblock {\em TKDE}, 27(8):2175--2189, Aug 2015.

\bibitem{zhong2013gtree}
R.~Zhong, G.~Li, K.-L. Tan, and L.~Zhou.
\newblock G-tree: An efficient index for knn search on road networks.
\newblock In {\em CIKM}, pages 39--48, 2013.

\end{thebibliography}

\begin{appendix}

\section{Improved Algorithms}\label{sec:app:imprv}

To ensure each method performs as efficiently as possible, we carefully inspected and applied numerous optimisations to each of them. In contrast to implementation issues discussed in Section \ref{sec:impl}, these algorithmic improvements are applicable in any setting (in-memory or otherwise). In this Appendix we describe all changes, including pseudocode, and experiments for major improvements.

\subsection{Distance Browsing}

We have made several major changes to the DisBrw algorithm proposed in \cite{samet2008distbrws}. Here we discuss several minor optimisations, corrections of edge cases, and, in Sections \ref{sec:app:silc:junc} and \ref{sec:app:silc:enn}, two major improvements to DisBrw. The updated pseudocode from \cite{samet2008distbrws} is shown in Algorithm \ref{alg:optSILC}. Please refer to Section \ref{sec:methods:silc} and \cite{samet2008distbrws,sankaranarayanan2005silc} for a description of the algorithm and definition of subroutines.

One of the main improvements of DisBrw over the SILC $k$NN algorithm proposed in \cite{sankaranarayanan2005silc} was the pruning of inserts by computing an upper bound $D_k$ for the $k$th object. However, for any encountered object, DisBrw would still compute a distance interval (necessarily involving an $O(\log V)$ operation), before the insertion of that object could be pruned. Since DisBrw already uses Euclidean distance to compute intervals for Object Hierarchy nodes, we additionally compute the Euclidean distance as a cheaper $O(1)$ initial lower bound for newly encountered objects (as in line \ref{line:silc:prunerefine}). Thus we are able to prune insertion of many encountered objects, which cannot be better candidates, without computing distance intervals. Furthermore \cite{samet2008distbrws} omitted computation of distance interval upper bounds for Object Hierarchy nodes. However this is only a small additional expense when computing lower bounds. Instead, we compute upper bounds for nodes and use it to compute $D_k$ sooner. Our Object Hierarchies also store the number of objects contained in each node (a simple additional pre-processing step in object index construction), which allows us to update $D_k$ as shown in line \ref{line:silc:regiondk}. In this way, we also prune insertion of nodes, and avoid needless lower bound evaluations to prune object in regions that cannot contain objects.

The DisBrw algorithm of \cite{samet2008distbrws} does not handle several minor but possible edge-case scenarios, which we have corrected in Algorithm \ref{alg:optSILC}, as follows:
\begin{itemize}
	\item In the original DisBrw algorithm, the if condition at line \ref{line:silc:case1} was $\UB_e \geq Front(Q)$, i.e., test if the upper bound of the dequeued element ($\UB_e$) is greater than or equal to the next smallest lower bound in $Q$ ($Front(Q)$). If true, DisBrw attempts to refine the bounds for $e$ and re-insert it into $Q$. However if the interval for $e$ is fully refined (i.e., $\LB_e = \UB_e$) but we still have $\UB_e = Front(Q)$, then we re-insert $e$ into $Q$ only to dequeue it and again re-insert leading to an infinite loop. Therefore we change the condition at line \ref{line:silc:case1} to $\UB_e > Front(Q)$. The second part of the condition ensures that objects with the same upper bound are refined further, otherwise elements may be out of order in $L$.
	
	\item Let $x$ be the object associated with $D_k$. Consider the scenario where $x$ is at the front of $Q$ (i.e., has the smallest lower bound) with $\LB_x < \UB_x$. Now when $x$ is dequeued, if $\UB_x > Front(Q)$ it will be refined. If $Front(Q)$ is associated with another object $p$, then $p$ may be a $k$NN (as $\LB_p = Front(Q)$ and before refinement $\UB_x = D_k$, so $\LB_p < \UB_x$). But if $x$ is refined such that we then have $\LB_x = \UB_x = D_k$ (i.e., $\UB_x$ did not change), then by the original algorithm, $x$ will not be re-inserted into $Q$. This is because the if condition at line \ref{line:silc:case2} was $\LB_e < D_k$. But when $p$ is dequeued next, we may have $\UB_p < Front(Q)$ (as the next smallest lower bound is unknown). In this case $p$ is dropped implicitly, potentially losing a correct $k$NN. This means when $\LB_e = D_k$ it must still be inserted into $Q$, to ensure $x$ is the last element dequeued before termination, and we change line \ref{line:silc:case2} in the algorithm accordingly.
	
	\item At line \ref{line:silc:contains} we ensure $L$ contains the dequeued object before it is deleted from $L$. If there are objects with the same upper bound as $D_k$, then the dequeued object may not be in $L$.
\end{itemize}

\begin{algorithm}[!htbp]
	\SetInd{0.2em}{0.7em}
	\SetAlgoVlined 
	\SetFuncSty{textrm}
	\SetArgSty{textrm}
	\SetCommentSty{texttt}
	\small
	\caption{\bf~ kNN\_DisBrw($v_q,k,SILC,OH$) \cite{samet2008distbrws}}
	\label{alg:optSILC}
	\In{$v_q$: a query vertex, $k$: the number of NNs, $SILC$: the SILC quadtree for $v_q$, $OH$: an Object Hierarchy for the object set}
	\Out{$R$: the set of $k$NNs for $v_q$}
	\Local{$Q$: a min priority queue for vertices and object hierarchy nodes keyed by lower bound, $L$: a max priority queue for up to $k$ candidate objects keyed by upper bound, $Dk$: upper bound on the $k$th neighbor}
	\State{$L \gets \phi$}
	\State{$D_k \gets \infty$}
	\State{$Enqueue(Q,([OH.Root,0,\_,\_],0))$}
	\State{\tcp*[h]{Note: $Q$ elements also stores upper bound $\UB_e$ and (for objects) $v_n$ the next intermediary vertex in the shortest path from $v_q$ and $d$ the distance to $v_n$ from $v_q$}}
	\While{$Q \neq \phi$}
	{
		\State{$([e,\UB_e,v_n,d],\LB_e) \gets Dequeue(Q)$}
			
		\If{$\UB_e \geq D_k$}
		{
			\State{\textbf{break}}
		}
		\ElseIf{$IsObject(e)$}
		{
			\If{$\UB_e > Front(Q)$ \OR $(\UB_e = Front(Q)$ \AND $\UB_e \neq \LB_e$)\label{line:silc:case1}}
			{
				\If{$\UB_e \leq D_k$ \AND $Contains(L,e)$\label{line:silc:contains}}
				{
					\State{$Delete(L,e)$}
				}
				\State{$(v_n,d,\LB_e,\UB_e) \gets Refine(v_n,d,\LB_e,\UB_e)$}
				\State{\tcp*[h]{$Refine$ tightens bounds, and updates the next vertex and its distance (this only involves a single binary search on the Morton List for the current $v_n$)}}
				\If{$\UB_e \leq D_k$}
				{
					\State{$UpdateL(L,e,\UB_e,D_k)$}				
				}
				\If{$\LB_e \leq D_k$\label{line:silc:case2}}
				{
					\State{$Enqueue(Q,([e,\UB_e,v_n,d],\LB_e))$}
				}
			}
			\State{\tcp*[h]{Else $e$ is implicitly dropped as it need not to be refined further}}
		}
		\Else
		{
			\State{\tcp*[h]{$e$ must be an Object Hierarchy node}}
			\If{$IsLeaf(e)$}
			{
				\ForEach{object $v_o \in e$}
				{
					\State{$\LB_o \gets EuclideanDistance(v_q,v_o)k$\label{line:silc:prunerefine}}
					\If{$\LB_o < D_k$}
					{
						\State{$(v_n,d,\LB_o,\UB_o) \gets Refine(v_q,0,\LB_o,\inf)$}
						\State{\tcp*[h]{Initial $v_n$ is $v_q$ with $d{=}0$}}
						\If{$\LB_o < D_k$}
						{
							\State{$Enqueue(Q,([o,\UB_o,v_q,0],\LB_o))$}
							\If{$\UB_o < D_k$}
							{
								\State{$UpdateL(L,o,\UB_o,D_k)$}	
							}
						}
					}
				}	
			}
			\Else
			{
				\ForEach{child node $c \in e$}
				{
					\If{$NumObjects(c) > 0$}
					{
						\State{$(\LB_c,\UB_c) \gets ComputeInterval(\LB_e,\UB_e)$}
						\If{$\LB_c < D_k$}
						{
							\State{$Enqueue(Q,([c,\UB_c,\_,\_],\LB_c))$}
							\If{$NumObjects(c) >= k$ \AND $\UB_c < D_k$}
							{
								\State{$D_k \gets \UB_c$\label{line:silc:regiondk}}
							}
						}
					}					
				}				
			}
		}
	}
	\State{$Populate(R,L)$ \tcp*[h]{Deqeue from $L$ to populate $R$ so they are in order like other algorithms}}
	\State{\KwRet{$R$}}

	{\DontPrintSemicolon \;}
	\SetKwBlock{Begin}{\textbf{Function} $UpdateL(L,o,\UB_o,D_k)$}{end}
	\Begin{
		\State{$Enqueue(L,(o,\UB_o))$}
		\If{$|L| \geq k$}
		{
			\If{$|L| > k$}
			{
				\State{$Dequeue(L)$}
			}
			\State{$D_k \gets Front(L)$}
		}
	}
	
\end{algorithm}

\subsubsection{Distance Browsing via Euclidean NN}\label{sec:app:silc:enn}
The Object Hierarchy is a key component of DisBrw. It is most easily represented by a quadtree containing all objects from a particular object set. DisBrw visits the most promising branches of this quadtree first, by computing distance intervals to child blocks (i.e., Object Hierarchy nodes). As described in \cite{samet2008distbrws}, DisBrw retrieves all leaf blocks from the SILC quadtree of the query vertex intersecting with that Object Hierarchy node. It uses these blocks to compute lower and upper bounds on the distance from the query vertex to any object in that node. But computing intersections is not a trivial expense. In the worst-case, all SILC quadtree leaf blocks must be retrieved. Furthermore many of the same intersections must be recomputed whilst traversing down the hierarchy. This implies a trade-off between the ability to prune regions using the hierarchy and the height of the hierarchy. A larger height improves performance on very high densities, but penalises lower densities. We observed that very shallow Object Hierarchies (with leaf capacities of 500 objects) provided the best overall performance.

\begin{algorithm}[h]
	\SetInd{0.2em}{0.7em}
	\SetAlgoVlined 
	\SetFuncSty{textrm}
	\SetArgSty{textrm}
	\SetCommentSty{texttt}
	\small
	\caption{\bf~ kNN\_DB-ENN($v_q,k,SILC,Rt$)}
	\label{alg:dbENN}
	\In{$v_q$: a query vertex, $k$: the number of NNs, $SILC$: the SILC quadtree for $v_q$, $Rt$: an R-tree for the object set}
	\Out{$R$: the set of $k$NNs for $v_q$}
	\Local{$Q$: a min priority queue for vertices keyed by lower bound, $L$: a max priority queue for up to $k$ candidate objects keyed by upper bound, $Dk$: upper bound on the $k$th neighbor, $E$: a min priority queue for R-tree NN search}
	\State{$D_k \gets \infty$}
	\State{$K \gets GetEuclidean\mathit{KNN}s(E,Rt,k)$}
	\ForEach{object $v_o \in K$}
	{
		\State{$ProcessCandidate(Q,L,e,0,D_k))$}
	}
	\State{\tcp*[h]{$D_k$ will be set if $k \leq |O|$}}
	
	\While{$Q \neq \phi$ \OR $E \neq \phi$}
	{
		\If{$Front(E) < Front(Q)$}
		{
			\State{$(e,\LB_e) \gets GetNextEuclidean\mathit{NN}(E)$}
			\State{$ProcessCandidate(Q,L,e,\LB_e,D_k))$}
		}
		\Else{
			\State{$([e,\UB_e,v_n,d],\LB_e) \gets Dequeue(Q)$}
			\If{$\UB_e \geq D_k$}
			{
				\State{\textbf{break}}
			}
			\Else
			{
				\If{$\UB_e > Front(Q)$ \OR $(\UB_e = Front(Q)$ \AND $\UB_e \neq \LB_e$)}
				{
					\If{$\UB_e \leq D_k$ \AND $Contains(L,e)$}
					{
						\State{$Delete(L,e)$}
					}
					\State{$(v_n,d,\LB_e,\UB_e) \gets Refine(v_n,d,\LB_e,\UB_e)$}
					\If{$\UB_e \leq D_k$}
					{
						\State{$UpdateL(L,e,\UB_e,D_k)$}				
					}
					\If{$\LB_e \leq D_k$}
					{
						\State{$Enqueue(Q,([e,\UB_e,v_n,d],\LB_e))$}
					}
				}
				\State{\tcp*[h]{Else $e$ is implicitly dropped as it need not to be refined further}}
			}
		}
	}
	\State{$Populate(R,L)$ \tcp*[h]{Deqeue from $L$ to populate $R$ so they are in order like other algorithms}}
	\State{\KwRet{$R$}}
	
	{\DontPrintSemicolon \;}
	\SetKwBlock{Begin}{\textbf{Function} $ProcessCandidate(Q,L,o,\LB_o,D_k)$}{end}
	\Begin{
		\State{$(v_n,d,\LB_o,\UB_o) \gets Refine(v_q,0,\LB_o,\inf)$}
		\If{$\LB_o < D_k$}
		{
			\State{$Enqueue(Q,([o,\UB_o,v_n,d],\LB_o))$}
			\If{$\UB_o < D_k$}
			{
				\State{$UpdateL(L,o,\UB_o,D_k)$}	
			}
		}
	}
	
\end{algorithm}

{
	\setlength{\abovecaptionskip}{\abvfigskp}
	\setlength{\belowcaptionskip}{\belfigskp}
	\begin{figure}[!htbp]
		\subfigure[Varying $k$]{
			\includegraphics[width=0.5\linewidth]{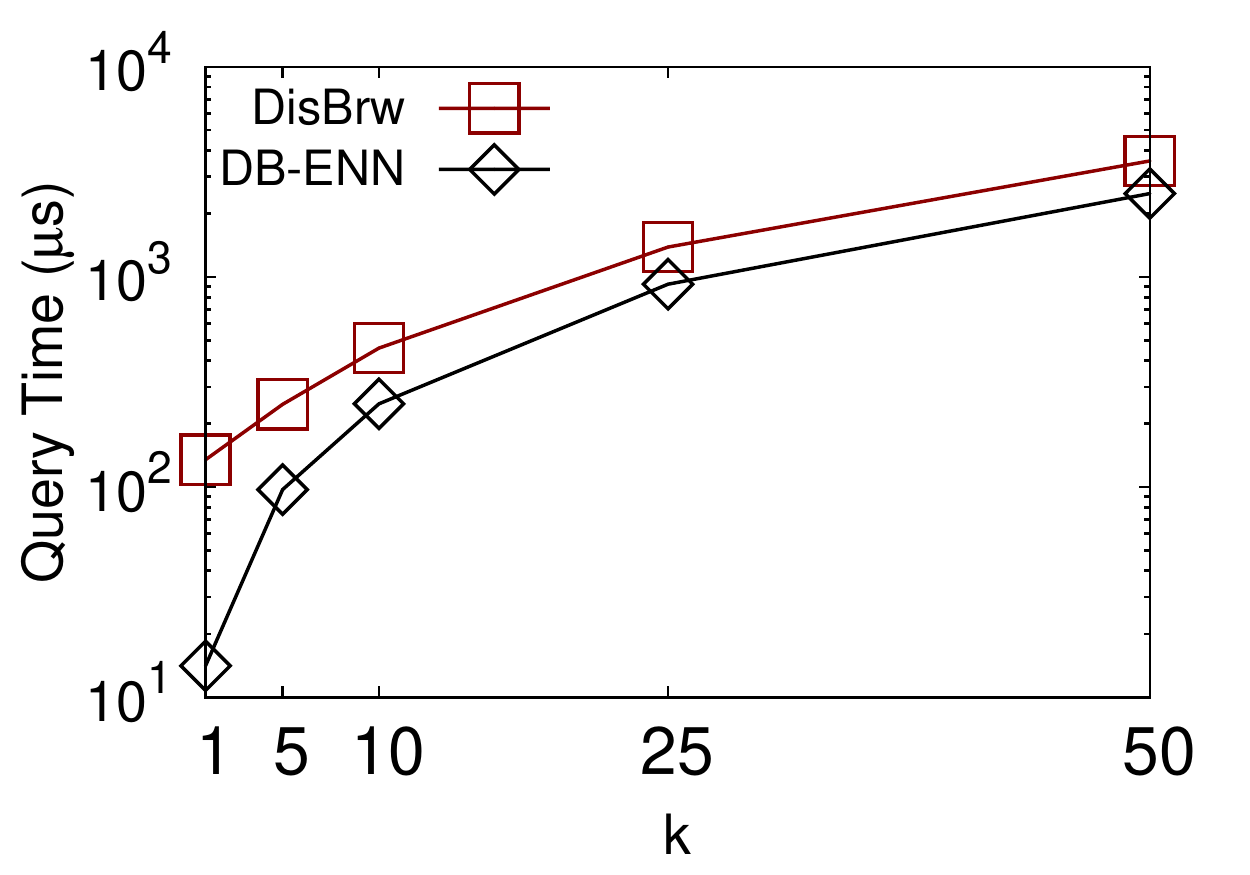}
			\label{exp:silc_dbenn:varyk_lin}
		}
		\hspace*{-5mm}
		\subfigure[Varying Density $d$]{
			\includegraphics[width=0.5\linewidth]{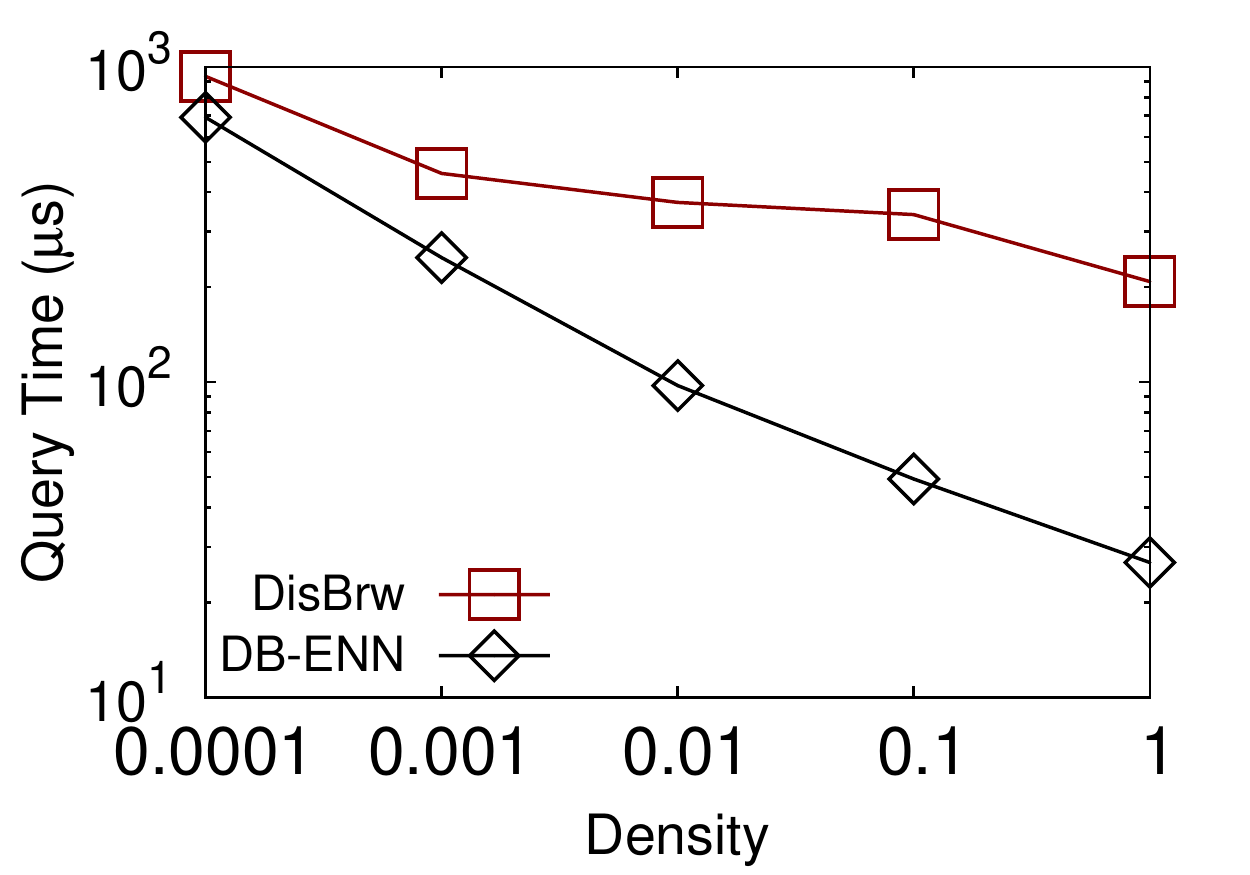}
			\label{exp:silc_dbenn:varyd}
		}
		\caption{DisBrw vs. DB-ENN \textmd{(NW, $d{=}0.001$, $k{=}10$)}}\label{exp:silc_dbenn}
		\vspace{-3mm}
	\end{figure}
}

To overcome this we propose a variant of DisBrw that eliminates computing intersections called DB-ENN presented in Algorithm \ref{alg:dbENN}. Essentially we replace the Object Hierarchy with Euclidean NNs to generate candidates (recall that Euclidean distances are already used to compute distance ratios \cite{samet2008distbrws}). We first retrieve Euclidean $k$NNs using an R-tree as the initial candidates and then suspend the search (i.e., we keep the priority queue $E$ used by the Euclidean $k$NN search). Now we compute distance intervals for each candidate and insert these candidates into queues $Q$ and $L$, setting $D_k$ as the largest upper bound. DisBrw proceeds as before, except before dequeuing an element from $Q$, we check if $Front(E) < Front(Q)$. If true, there may be a closer Euclidean NN, so we retrieve the next Euclidean NN from $E$. This object is handled in the same way as an object in a leaf node of the Object Hierarchy (i.e., potentially inserted into $Q$ and $L$). We compare DisBrw to DB-ENN in Figure \ref{exp:silc_dbenn}. DB-ENN's improvement increases with higher density and lower $k$ as this is when the overhead from the Object Hierarchy is highest. The improvement peaks at 1 order of magnitude. Since this suggests that Object Hierarchies do not use the SILC index to its full potential, we instead use DB-ENN in our experiments.

\subsubsection{Exploiting Vertices with Outdegree $\leq$ 2}\label{sec:app:silc:junc}
Real road network graphs consist of large numbers of degree-$2$ vertices. Generally 30\% of vertices have degree-2 for road networks in Table \ref{tab:datasets}, e.g., on the US dataset 30.3\% of vertices have degree-2 (another 19.9\% have degree-1). These may exist to capture details such as varying speed limits or curvature. This degree distribution can have a significant impact on computing shortest paths, and we demonstrate the potential improvement on DisBrw.

SILC uses the quadtrees and coloring scheme described in Section \ref{sec:methods:silc} to iteratively compute the vertices in a shortest path, at a cost of $O(\log |V|)$ for each vertex. We use \emph{chain} to refer to a path consisting only of vertices with degree-$2$ or less, e.g., a section of motorway with no exits. Let $v$ be the current vertex in the shortest path from $s$ to $t$ and $u$ be the previous vertex in the shortest path. If $v$ is on a chain, we do not need to consult the quadtree because the next vertex in the shortest path \textit{must} be the neighbor of $v$ that is not $u$. This saves $O(\log |V|)$ for each degree-$2$ vertex in the shortest path. In fact, if target $t$ is not on the chain, we can directly ``jump'' to the last vertex in the chain saving several $O(\log |V|)$ lookups. This observation can be easily exploited by storing the two ends of the chain for each vertex with degree less than $2$. 

{
	\setlength{\abovecaptionskip}{\abvfigskp}
	\setlength{\belowcaptionskip}{\belfigskp}
	\begin{figure}[htbp]%
		\vspace{-3mm}
		\subfigure[Varying $k$]{
			\includegraphics[width=0.5\linewidth]{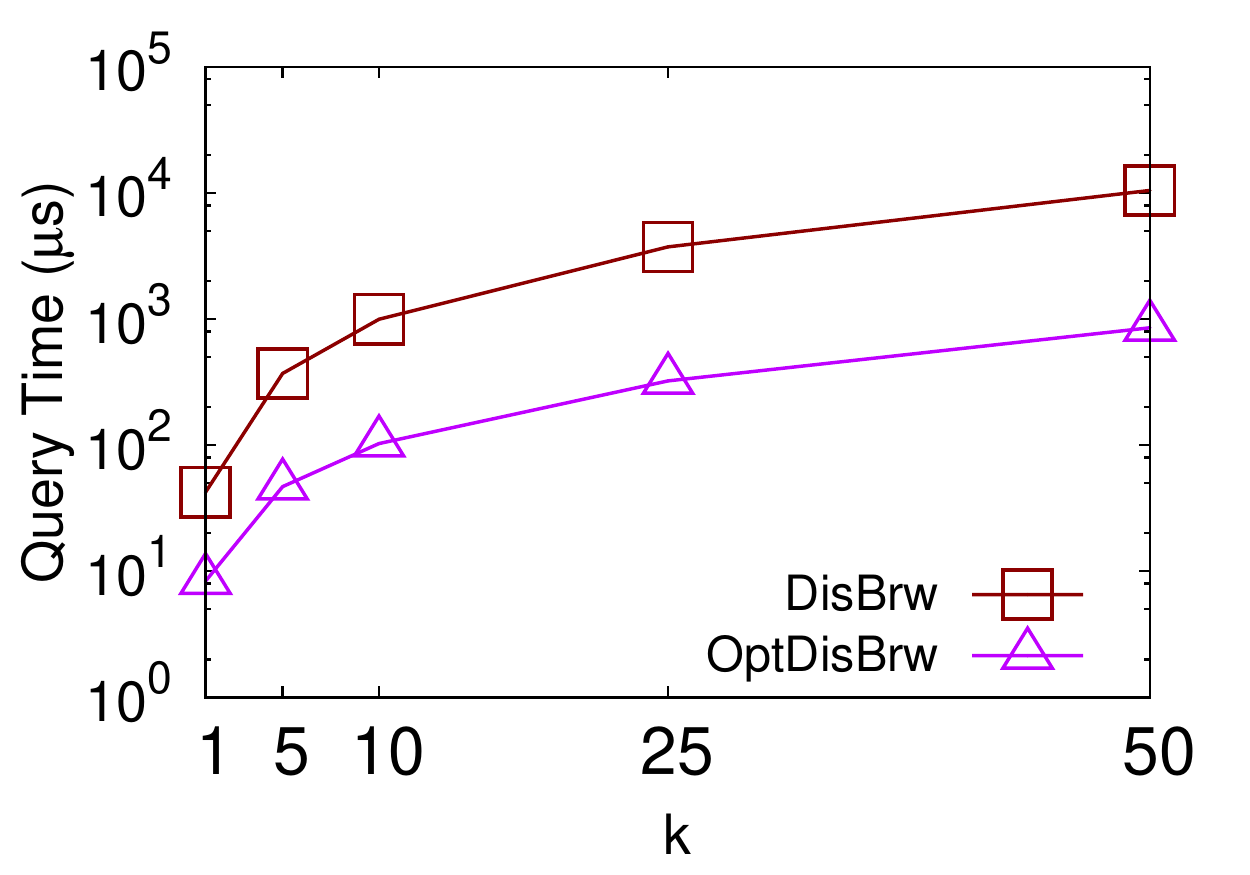}		
			\label{exp:silc_opt:k}
		}
		\hspace*{-5mm}
		\subfigure[Varying Density]{
			\includegraphics[width=0.5\linewidth]{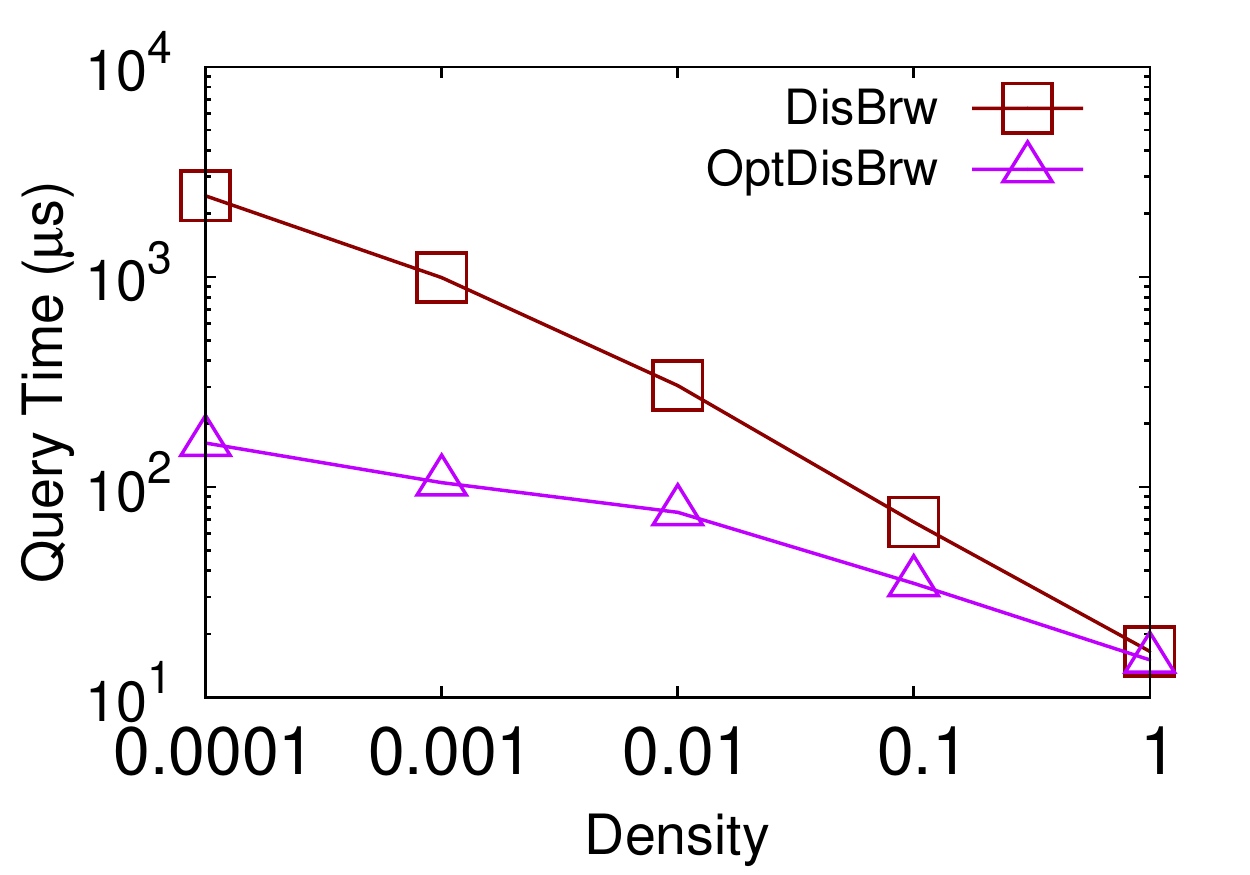}		
			\label{exp:silc_opt:d}
		}
		\caption{Deg-2 Optimisation \textmd{(NA-HWY, $d{=}0.001$, $k{=}10$)}}\label{exp:silc_opt_highway}
		\vspace{-1mm}
	\end{figure}
}

{
	\setlength{\abovecaptionskip}{\abvfigskp}
	\setlength{\belowcaptionskip}{\belfigskp}
	\begin{figure}[htbp]%
		\vspace{-5mm}
		\subfigure[Varying $k$]{
		\includegraphics[width=0.5\linewidth]{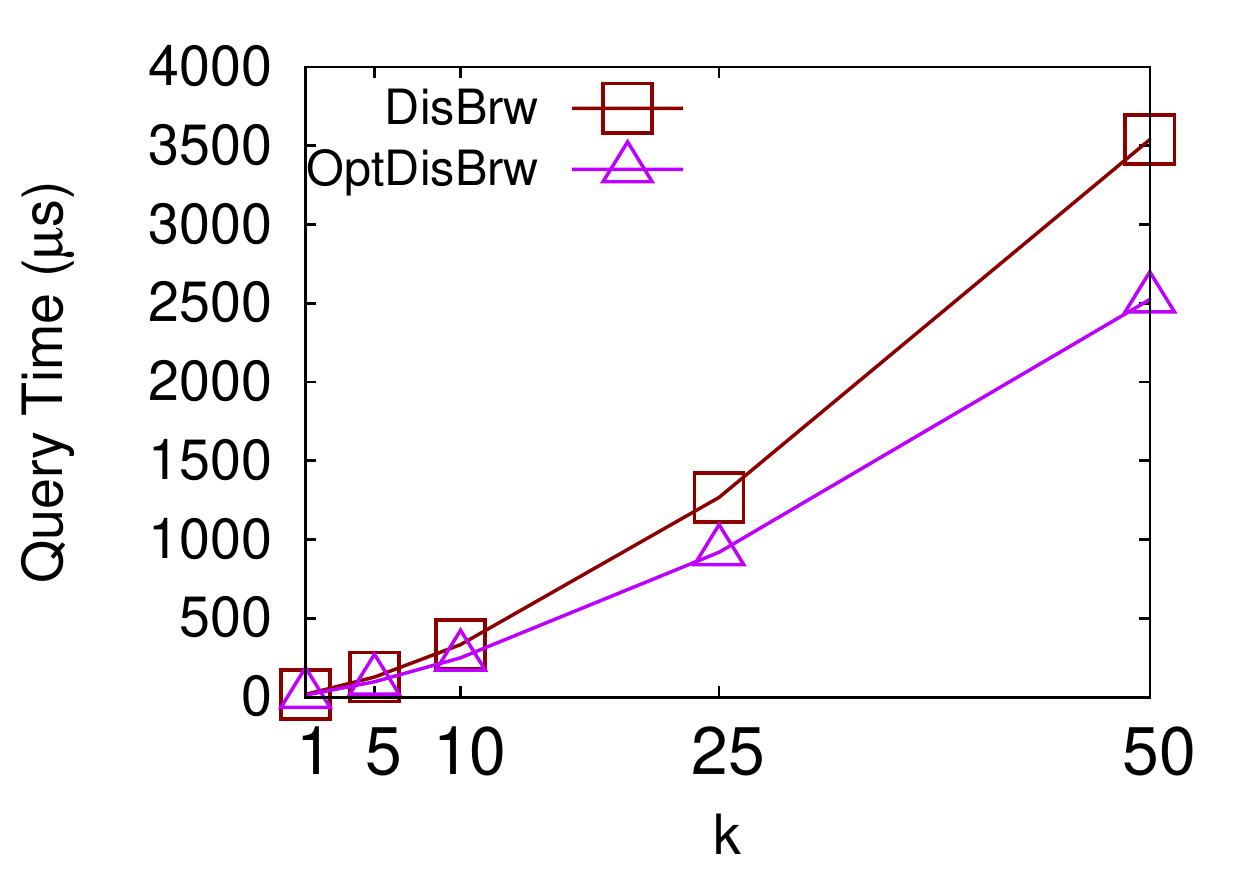}		
			\label{exp:silc_opt:hwy_k}
		}
		\hspace*{-5mm}
		\subfigure[Varying Density]{
		\includegraphics[width=0.5\linewidth]{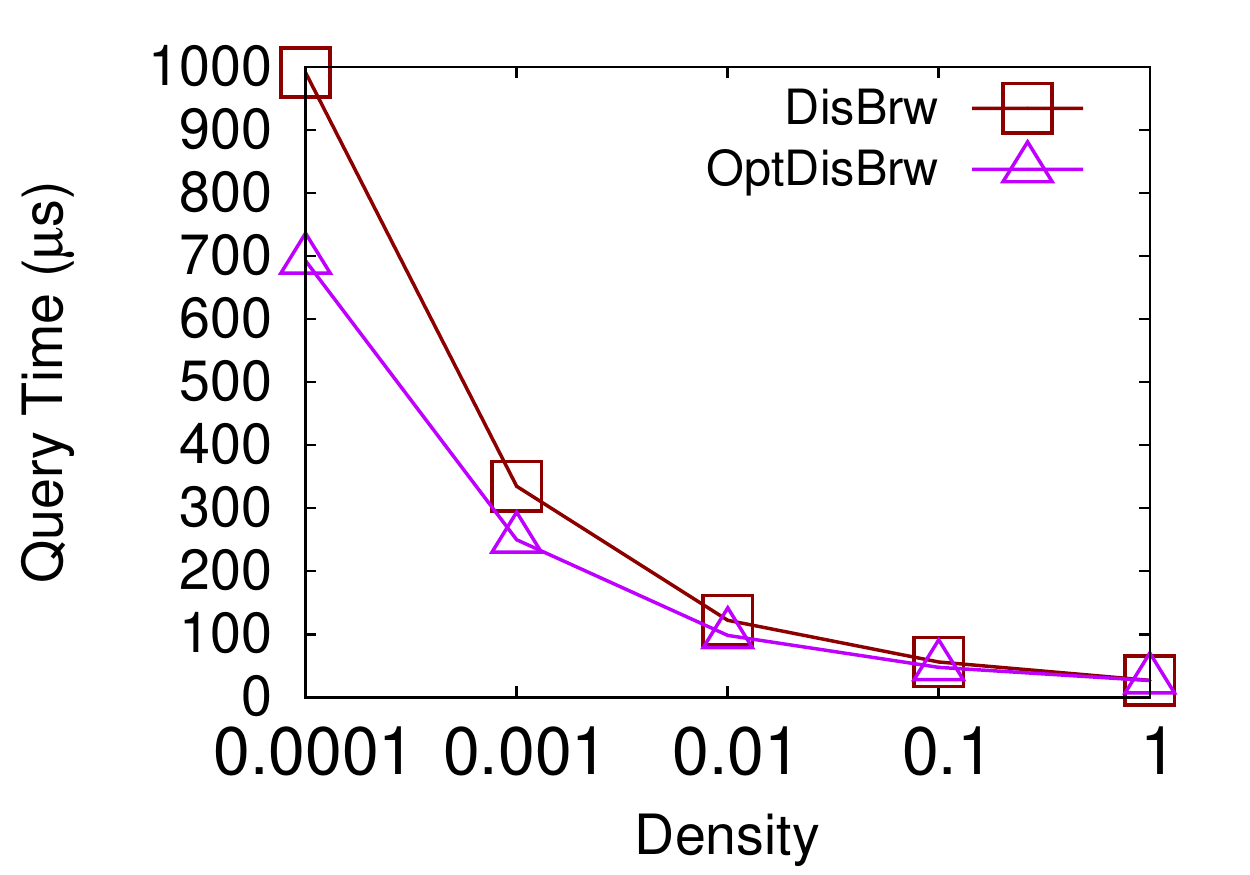}		
			\label{exp:silc_opt:hwy_d}
		}
		\caption{Deg-2 Optimisation \textmd{(NW, $d{=}0.001$, $k{=}10$)}}\label{exp:silc_opt}
		\vspace{-3mm}
	\end{figure}
}

This optimisation can significantly improve DisBrw query times. We refer to this version as OptDisBrw. For our default NW dataset this results in a 30\% improvement as in Figure~\ref{exp:silc_opt}, coinciding with the number of degree-2 vertices quoted above. However some road networks have an even larger proportion of degree-$2$ vertices, such as the highway road network for North America used in past studies \cite{lee2009road,lee2012road,samet2008distbrws} with $175,813$ vertices \cite{lifeifei}, 95\% of which are degree-$2$. In this case OptDisBrw is up to an order of magnitude faster than DisBrw as shown in Figure~\ref{exp:silc_opt_highway}, as the average chain length is significantly higher resulting in longer jumps. Accordingly future work must keep degree-$2$ vertices in mind for potential optimisations. Given these results we use chain optimised refinement for DisBrw in our experiments.

\subsection{G-tree}

The G-tree $k$NN algorithm is largely unchanged from the more recent G-tree study \cite{zhong2015gtree}, except for an improved leaf search algorithm we describe in Section \ref{sec:app:imprv:gtree:leaf_opt} below. Algorithm \ref{alg:optGtree} includes an additional guard condition at line \ref{line:gtree:guard} as the $UpdateT$ subroutine does not guarantee that $Q$ will not be empty (i.e., when the new $T_n$'s only child with an occurrence is the previous $T_n$, which we do not re-insert). Similar to how Association Directories are used by ROAD, we make a small modification to the algorithm to accept an Occurrence List $OL$ rather than the set of objects $O$. The original algorithm implies that $OL$ must be constructed for each query, which is not a trivial cost or how it would be used in practice (as seen in Section \ref{sec:exp:obj_indexes}). Note that again node refers to G-tree nodes and vertex refers to road network vertices. Please refer to Section \ref{sec:methods:gtree} and \cite{zhong2015gtree} for descriptions of the data structures and explanation of the main algorithm.

\begin{algorithm}[h]
	\SetInd{0.2em}{0.7em}
	\SetAlgoVlined 
	\SetFuncSty{textrm}
	\SetArgSty{textrm}
	\SetCommentSty{texttt}
	\small
	\caption{\bf~ kNN\_Gtree($v_q,k,Gt,OL$) \cite{zhong2015gtree}}
	\label{alg:optGtree}
	\In{$v_q$: a query vertex, $k$: the number of NNs, $Gt$: a G-tree, $OL$: an Occurrence List}
	\Out{$R$: the set of $k$NNs for $v_q$}
	\Local{$Q$: a minimum priority queue for vertices and nodes}
	\If{$|OL(Leaf(v_q))| > 0$}
	{
		\State{$GtreeLeafSearch(v_q,k,OL,Q,R)$}
	}
	\State{$T_n \gets Leaf(v_q)$ and set $T_{min}$ for $T_n$}
	\While{$|R| < k$ \AND $(Q \neq \phi$ \OR $T_n \neq Gt.Root)$}
	{
		\If{$Q = \phi$}
		{
			\State{$UpdateT(T_n,T_{min},OL,Q$}
		}
		\If{$Q \neq \phi$\label{line:gtree:guard}}
		{
			\State{$(e,d) \gets Dequeue(Q)$}
			\If{$d > T_{min}$}
			{
				\State{$UpdateT(T_n,T_{min},OL,Q)$}
				\State{$Enqueue(Q,(e,d))$}
			}
			\ElseIf{$e$ is a vertex}
			{
				\State{$R \gets R \cup e$}
			}
			\ElseIf{$e$ is a node}
			{
			  	\ForEach{node or vertex $c \in OL(e)$ }
			  	{
			  		\State{$Enqueue(Q,SPDist(v_q,c))$}
			  	}
			}
		}
	}
	\State{\KwRet{$R$}}
	
	{\DontPrintSemicolon \;}
	\SetKwBlock{Begin}{\textbf{Function} $UpdateT(T_n,T_{min},OL,Q)$}{end}
	\Begin{
	  	\State{$Tn \gets Tn.father$ and update $T_{min}$ for $T_n$}
	  	\ForEach{node $c \in OL(T_n)$}
	  	{
	  		\State{\tcp*[h]{Note: We exclude the previous $T_n$}}
	  		\State{$Enqueue(Q,SPDist(v_q,c))$}
	  	}
	}
	
\end{algorithm}

\subsubsection{G-tree Leaf Search Improvement}\label{sec:app:imprv:gtree:leaf_opt}

We note that G-tree's query performance plateaus and sometimes increases for very high densities. Given a query vertex $v_q$, let $G_q$ be the leaf subgraph containing $v_q$. Now the network distance to any object in $G_q$ is the minimum of two possible shortest paths (a) one consisting of only vertices within $G_q$ (b) one that leaves and re-enters $G_q$ through some its borders. To ensure correctness, the original G-tree algorithm performs Dijkstra's search limited to $G_q$ until all objects are found, capturing network distances of type (a). For each object found it then computes network distances for type (b) paths by using the distance matrix to compare distances through borders. Recall that a leaf node contains at most $\tau$ vertices (e.g. $\tau{=}256$ for NW and $\tau{=}512$ for US datasets as in Table \ref{tab:datasets}). For objects with density $d$ there are on average $d{\times}\tau$ objects in each leaf node. If $d\times\tau>k$, G-tree computes distances of each type for more objects than necessary. The penalty is worse with increasing $\tau$ and decreasing $k$. INE cannot be applied to $G_q$ because the $k$NNs within it may not be the global $k$NN and some objects within $G_q$ may be closer through paths that travel outside it.

\begin{algorithm}[h]
	\SetInd{0.2em}{0.7em}
	\SetAlgoVlined 
	\SetFuncSty{textrm}
	\SetArgSty{textrm}
	\SetCommentSty{texttt}
	\small
	\caption{\bf~ GtreeLeafSearch($v_q,k,OL,Q,R,G_q$)}
	\label{alg:optGtree:leafsearch}
	\In{$v_q$: a query vertex, $k$: the number of NNs, $OL$: an Occurrence List, $Q$: the queued used by Algorithm \ref{alg:optGtree}, $R$: the set of $k$NNs for $v_q$, $G_q$ leaf node containing $v_q$}
	\Local{$L$: a minimum priority queue for vertices}
	\State{$Enqueue(L,(v_q,0))$}
	\State{targetsFound $\gets 0$}
	\State{borderFound $\gets$ false}
	\While{$L\neq \phi$ \AND $|R| < k$ \AND targetsFound $< k$}
	{
		$(v_e,d) \gets Dequeue(L)$
		
		\If{$\neg IsVisited(v_e)$}
		{
			\If{$v_e \in OL(G_q)$}
			{
				\State{targetsFound++}
				\If{$\neg$ border\_found}
				{
					\State{$R \gets R \cup v_e$}
				}
				\Else
				{
					\State{$Enqueue(Q,(v_e,d))$}
				}			
			}
			\State{borderFound $\gets RelaxLeafVertex(v_e,d,L,G_q)$)}
			\State{$IsVisited(v_e) \gets$ true}
		}
	}
	\State{\KwRet{}}
	
	{\DontPrintSemicolon \;}
	\SetKwBlock{Begin}{\textbf{Function} $RelaxLeafVertex(v_e,d,L,G_q)$}{end}
	\Begin{
		\ForEach{vertex $v_a \in AdjacencyList(v_e)$}
		{
			\If{$\neg IsVisited(v_a)$ \AND $v_a \in G_q$}
			{
				\State{$Enqueue(L,(v_a,d+w(v_e,v_a)))$}
			}
		}
		\If{$v_e \in Borders(G_q)$}
		{
			\ForEach{border $v_b \in Borders(G_q)$}
			{
				\If{$\neg IsVisited(v_b)$}
				{
					\State{$Enqueue(L,(v_b,d+G_q.DistMatrix(v_e,v_b)))$}
				}
			}
			\State{\KwRet{true}}
		}
		\State{\KwRet{false}}
	}
	
\end{algorithm}

We modify Dijkstra's search within $G_q$ to capture paths of both types as shown in Algorithm \ref{alg:optGtree:leafsearch}. Let $L$ be the priority queue used by this search. Our search continues until the first $k$ leaf objects are settled (i.e. dequeued from $L$). Until the first border is settled, all settled objects are $k$NNs (like INE). This is correct as, without a closer border, no objects can have a shortest path that leaves $G_q$. But any subsequent object may not be a $k$NN, so we instead insert them into the  priority queue $Q$ used by the main G-tree algorithm. To ensure the distances take into account paths that leave $G_q$, whenever we settle a border $v_b$ we insert every other unsettled border $v_b'$ of $G_q$ into $L$. The distance to $v_b'$ from $v_q$ can be computed with the distance matrix. In Figure \ref{exp:imprv:gtree_leafsearch} we see a significant speed-up for $k=10$ and over an order of magnitude improvement for $k=1$ on both datasets for the highest density. The improvement is even noticeable at lower densities for $k=1$ on NW and both $k$ on the US dataset as the leaf still contains far more objects than $k$. Note that this improvement is also applicable to other object distributions with the same density as leaf nodes will contain the same number of objects, on average.

{
		\setlength{\abovecaptionskip}{\abvfigskp}
		\setlength{\belowcaptionskip}{\belfigskp}
	\begin{figure}[!htbp]
		\vspace{-4mm}
		\subfigure[Varying Density (NW)]{
			\includegraphics[width=0.49\linewidth]{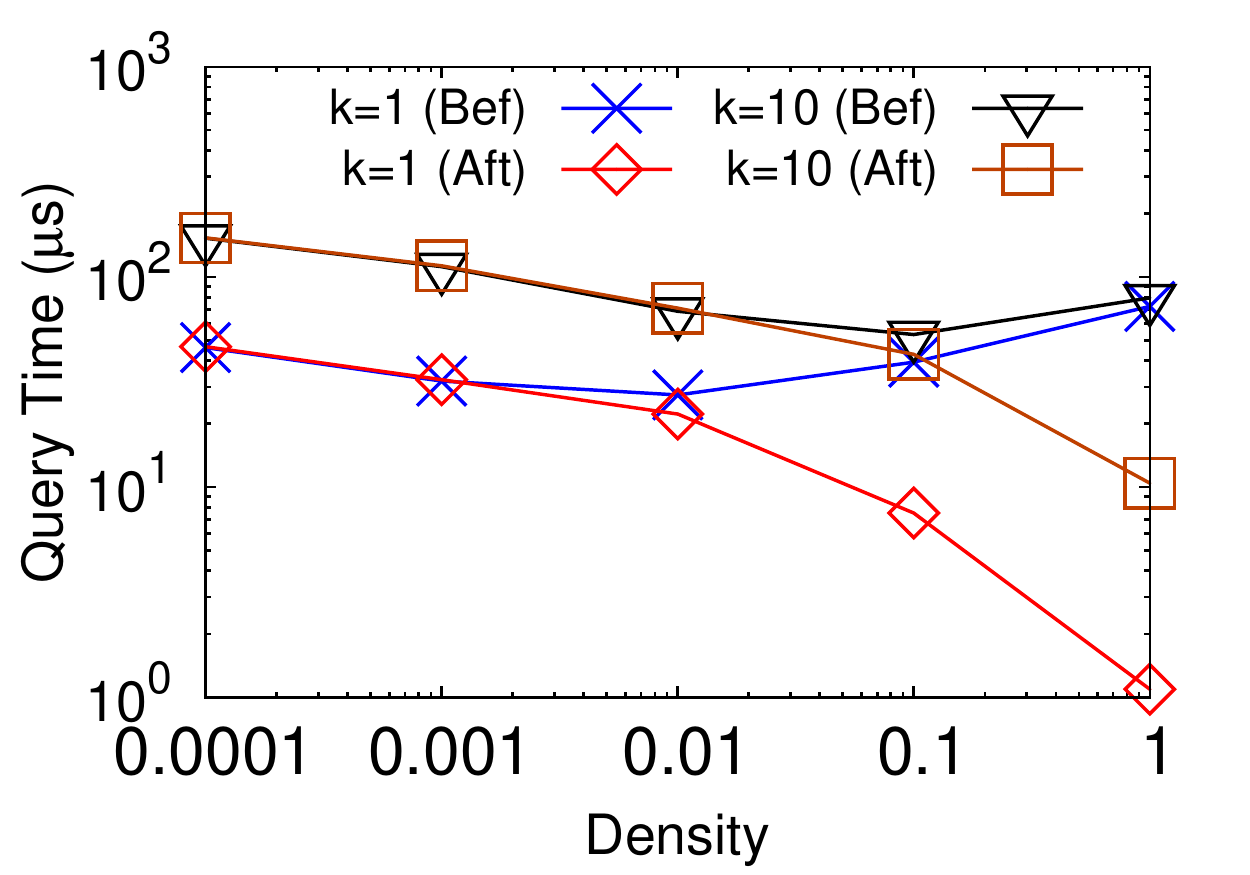}
			\label{exp:imprv:gtree_leafsearch:nw}
		}
		\hspace*{-5mm}
		\subfigure[Varying Density (US)]{
			\includegraphics[width=0.49\linewidth]{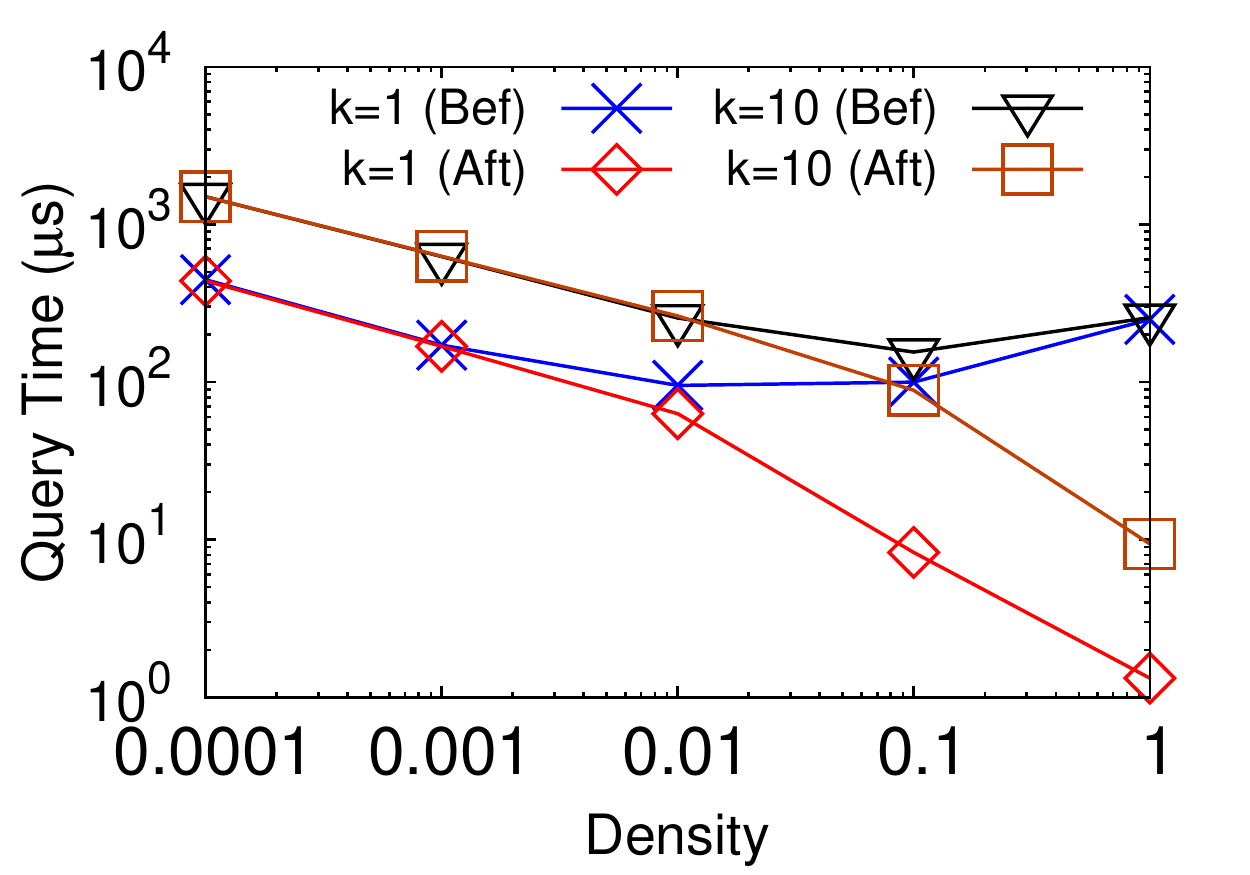}
			\label{exp:imprv:gtree_leafsearch:us}
		}
		\caption{Improved G-tree Leaf Search}\label{exp:imprv:gtree_leafsearch}
	\end{figure}
}
\subsection{ROAD}

The kNN search and supporting algorithms using the ROAD index are largely unchanged from \cite{lee2009road}. We simplify these for the scenario where objects occur on vertices. In addition we made a minor improvement by preventing unnecessary priority queue inserts for Rnet borders that have already been visited (see line \ref{line:road:visited_opt} of Algorithm \ref{alg:optROAD:relax}). The original algorithm re-inserts all borders into the priority queue $Q$, but then discards each of them immediately after dequeuing (as they are ``visited"). This can be particularly expensive for larger Rnets (as they tend to have more borders). Please refer to Section \ref{sec:methods:road} and \cite{lee2009road} for descriptions of the data structures and explanation of the algorithms.

\begin{algorithm}[h]
	\SetInd{0.2em}{0.7em}
	\SetAlgoVlined 
	\SetFuncSty{textrm}
	\SetArgSty{textrm}
	\SetCommentSty{texttt}
	\small
	\caption{\bf~ kNN\_ROAD($v_q,k,RO,AD$) \cite{lee2009road}}
	\label{alg:optROAD}
	\In{$v_q$: a query vertex, $k$: the number of NNs, $RO$: a Route Overlay index, $AD$: an Association Directory}
	\Out{$R$: the set of $k$NNs for $v_q$}
	\Local{$Q$: a minimum priority queue}
	\State{$Enqueue(Q,(v_q,0))$}
	\State{$R\gets\phi$}
	\While{$Q \neq \phi$ \AND $|R| < k$}
	{
		$(v_e,d) \gets Dequeue(Q)$
		
		\If{$\neg IsVisited(v_e)$}
		{
			\If{$IsObject(AD,v_e)$}
			{
				\State{$R \gets R \cup v_e$}
			}
			\State{$RelaxShortcuts(Q,RO,AD,v_e$)}
			\State{$IsVisited(v_e) \gets$ true}
		}
	}
	\State{\KwRet{$R$}}
\end{algorithm}

\begin{algorithm}[h]
	\SetInd{0.2em}{0.7em}
	\SetAlgoVlined 
	\SetFuncSty{textrm}
	\SetArgSty{textrm}
	\SetCommentSty{texttt}
	\small
	\caption{\bf~ RelaxShortcuts($v_e,d,Q,RO,AD$) \cite{lee2009road}}
	\label{alg:optROAD:relax}
	\In{$v_e$: current vertex, $d$: distance to $v_e$, $Q$: priority queue of unvisited vertices, $RO$: route overlay index, $AD$: association directory}
	\Local{$S$: stack}
	\State{$T \gets $ $RO.GetShortcutTree(v_e)$)}
	\State{$Push(S,T.Root)$}
	\While{$S \neq \phi$}
	{
		\State{$n \gets Pop(S)$}
		\If{$\neg IsLeaf(n)$}
		{
			\ForEach{$R \in Rnets(n)$}
			{
				\If{$\neg HasObject(AD,R)$}
				{
					\State{\tcp*[h]{Then this Rnet can be bypassed}}
					\ForEach{shortcut $S(v_e,v_b) \in R$}
					{
						\If{$\neg IsVisited(v_b)$\label{line:road:visited_opt}}
						{
							\State{$Enqueue(Q,(v_b,d+|S(v_e,v_b)|))$}
						}
					}
				}
				\Else
				{
					\ForEach{child tree node $c$ of $n$}
					{
						\State{$Push(S,c)$}
					}
				}
			}
		}
		\Else
		{
			\State{$R \gets Rnets(n)$ \tcp*[h]{Leaves have only one Rnet}}
			\ForEach{edge $e(v_e,v_a) \in R$}
			{
				\If{$\neg IsVisited(v_a)$}
				{
					\State{$Enqueue(Q,(b,d+w(v_e,v_a)))$}
				}
			}			
		}
	}
	\State{\KwRet{}}
\end{algorithm}

\section{Repeated Experiments}\label{sec:app:tt}

A $k$NN query returns the $k$ closest objects to a query vertex by their network distance. However network distance may not necessarily be a physical distance. A common scenario in practice is travel times. We presented the experimental results for queries on the travel time graph for the US in Section \ref{sec:exp:tt}. In this appendix we present results for other experiments on travel time graphs. Note that we do not test DisBrw on travel times, as the additional information (i.e., distance ratios) stored in the SILC index relies heavily on Euclidean distance, making it more complex to adapt than IER and likely to perform significantly slower than on travel distances.

{
	\setlength{\abovecaptionskip}{\abvfigskp}
	\setlength{\belowcaptionskip}{-13pt}
	\begin{figure*}[ht]
		\subfigure[Varying $k$]{
			\includegraphics[width=0.33\linewidth]{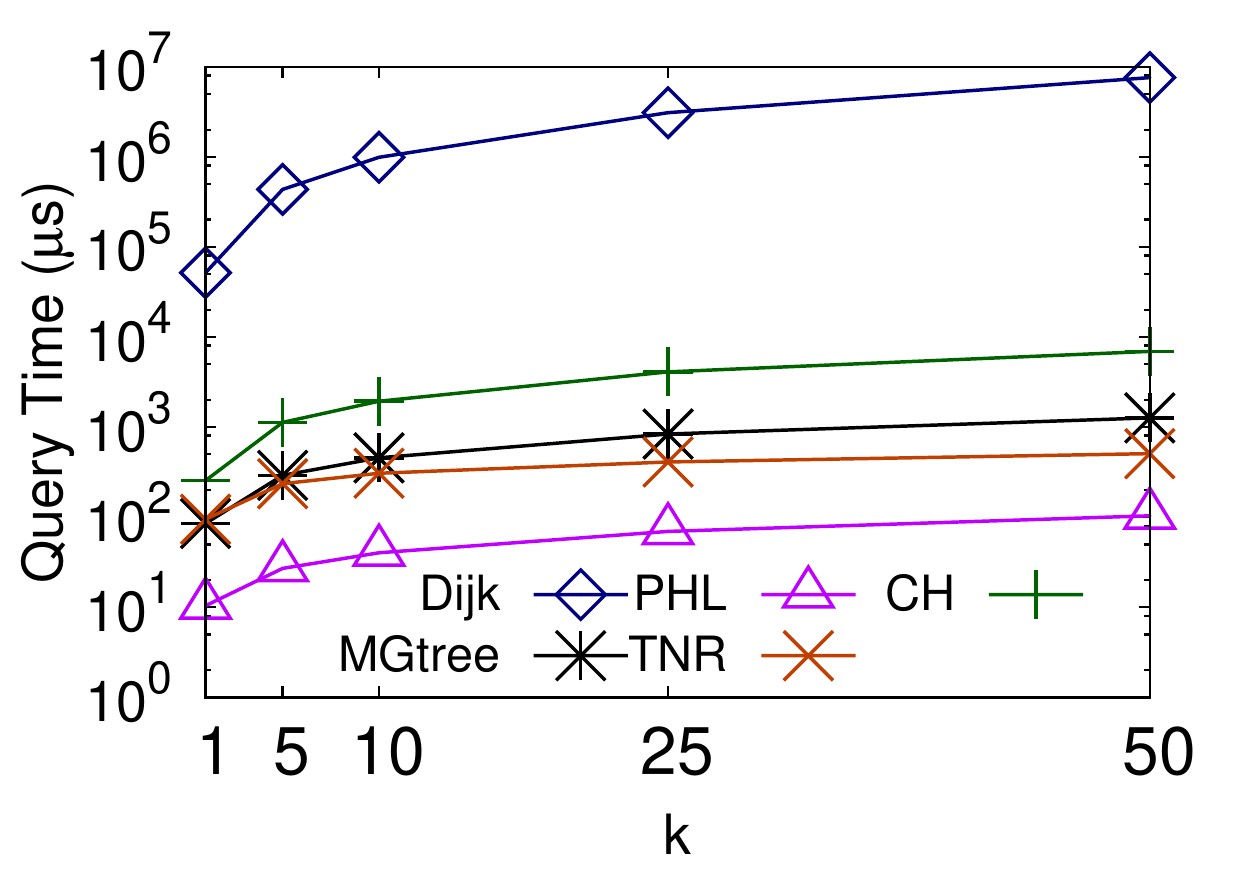}
			\label{app:exp:tt:ier_comparison::varyk}
		}
		\hspace*{-5mm}
		\subfigure[Varying Density]{
			\includegraphics[width=0.33\linewidth]{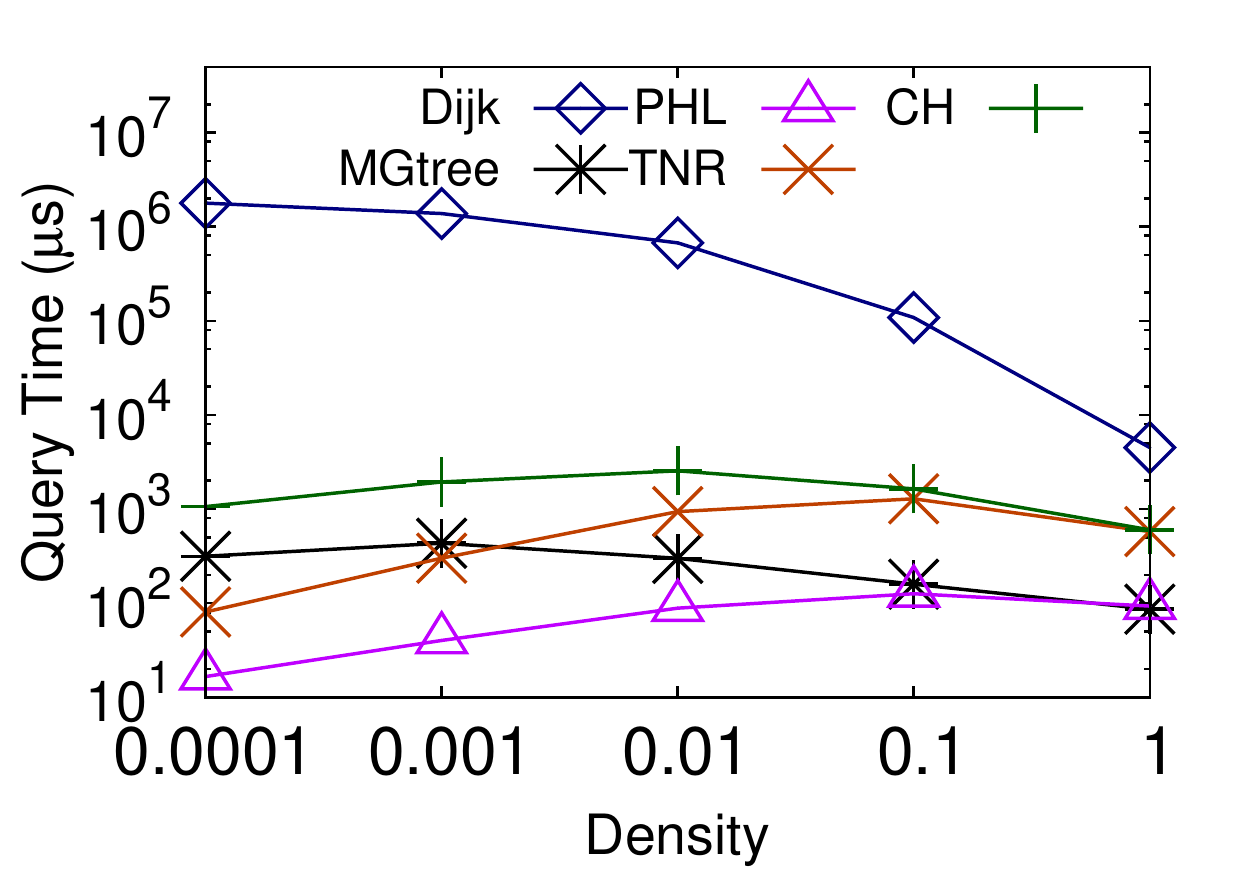}
			\label{app:exp:tt:ier_comparison:varyd}
		}
		\hspace*{-5mm}
		\subfigure[Varying Density]{
			\includegraphics[width=0.33\linewidth]{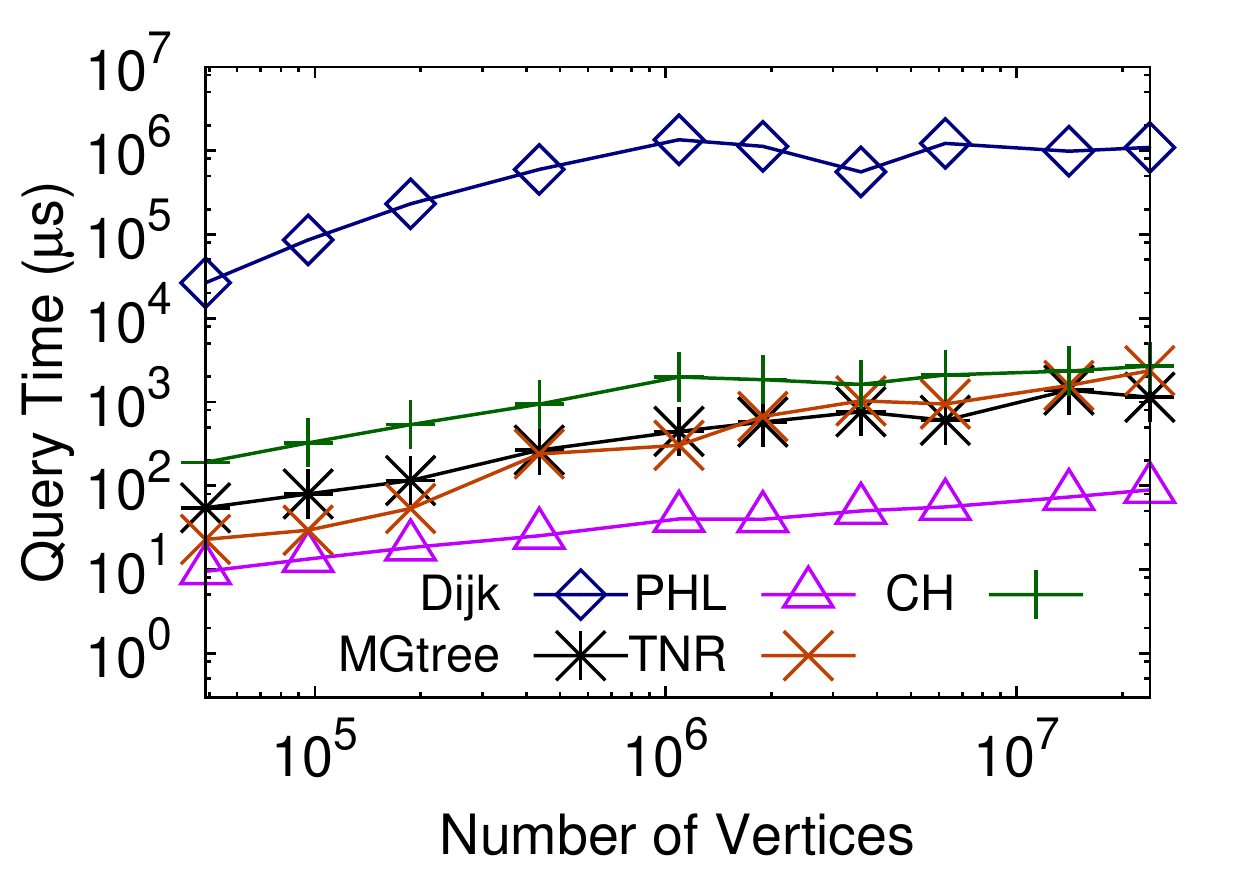}
			\label{app:exp:tt:ier_comparison:varyn}
		}
		\caption{IER Variants on Travel Time Graphs \textmd{(NW, $d{=}0.001$, $k{=}10$, uniform objects)}}\label{app:exp:tt:ier_comparison}
	\end{figure*}
}

{
	\setlength{\abovecaptionskip}{\abvfigskp}
	\setlength{\belowcaptionskip}{-13pt}
	\begin{figure*}[ht]%
		\subfigure[Varying $k$]{
			\includegraphics[width=0.25\linewidth]{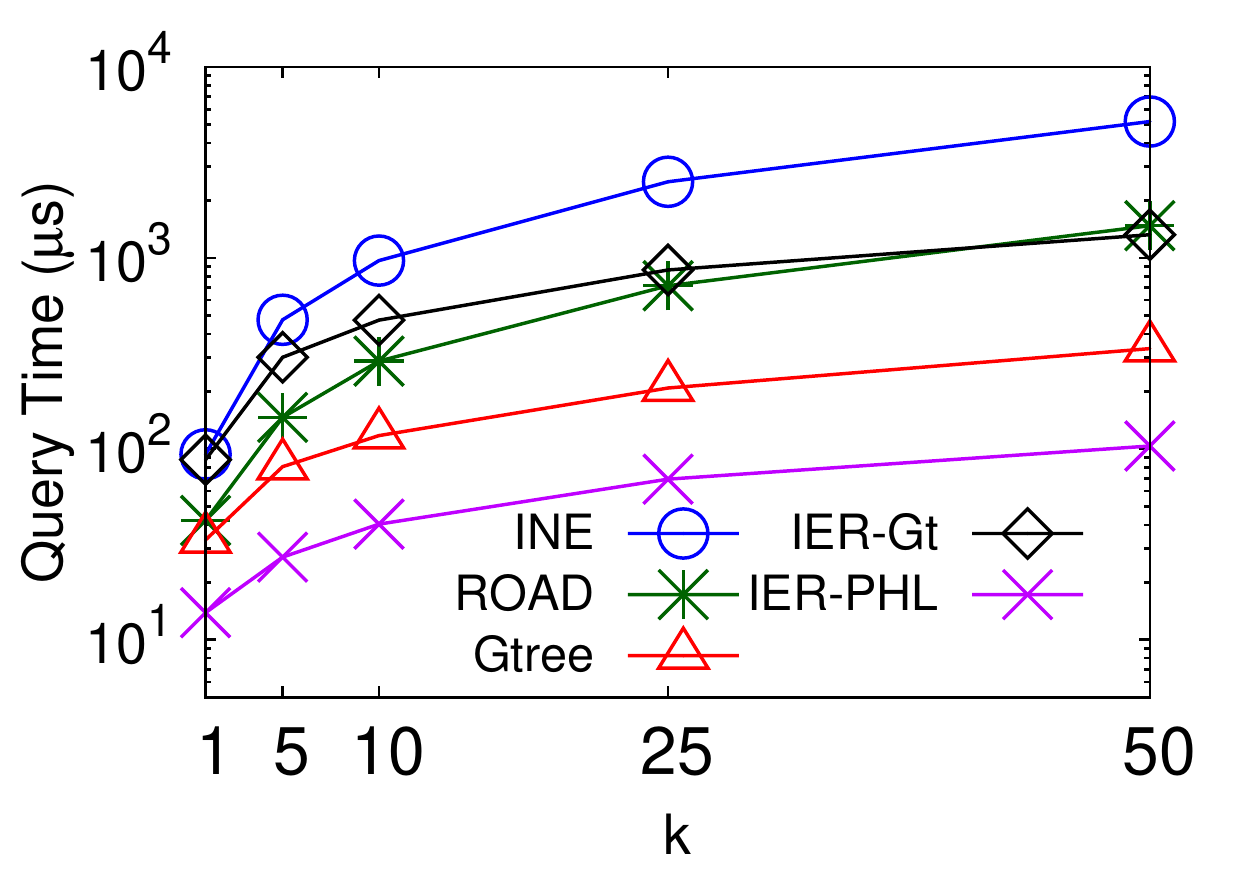}
			\label{app:exp:tt:tt_nw_k_density001}
		}
		\hspace*{-5mm}
		\subfigure[Varying Density $d$]{
			\includegraphics[width=0.25\linewidth]{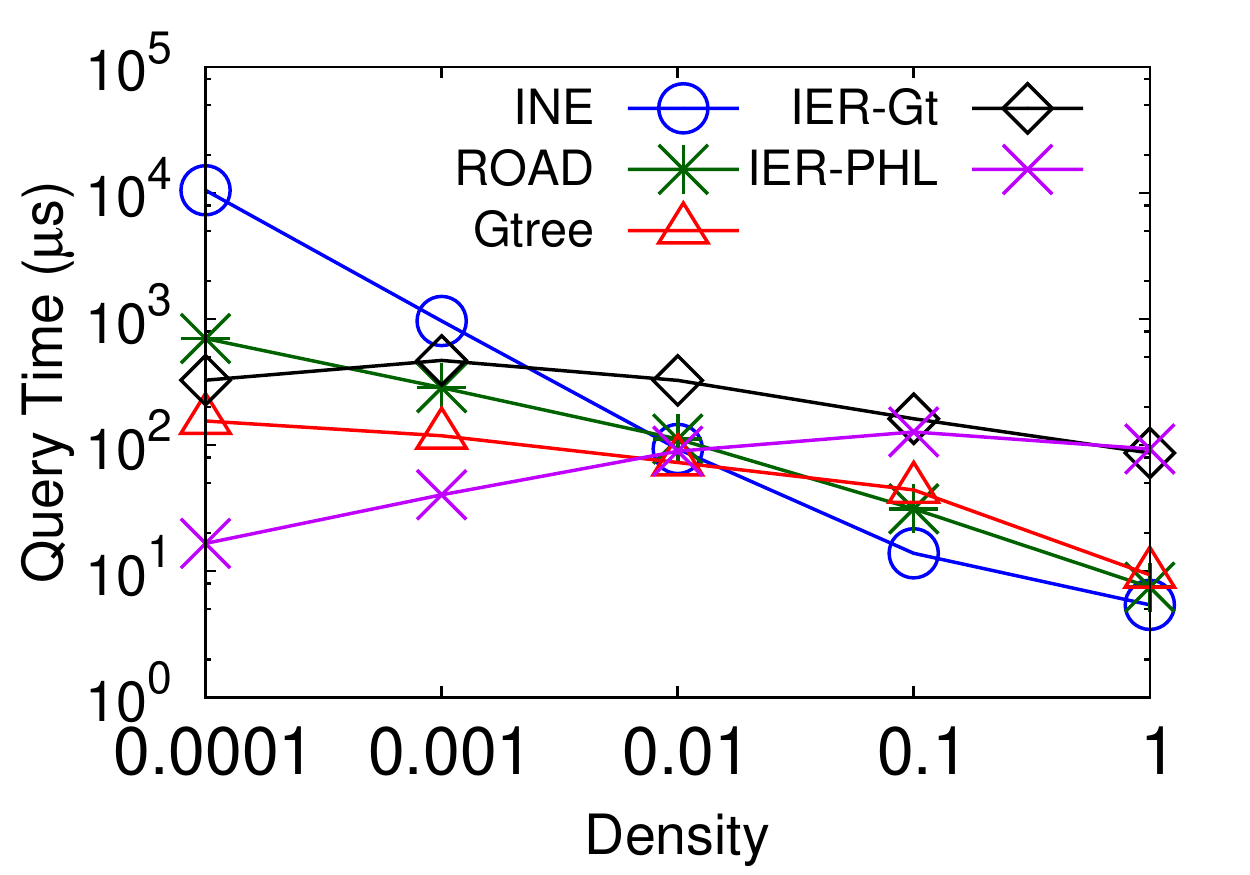}
			\label{app:exp:tt:tt_nw_density_k10}
		}		
		\hspace*{-5mm}
		\subfigure[Varying Min. Obj. Distance]{
			\includegraphics[width=0.25\linewidth]{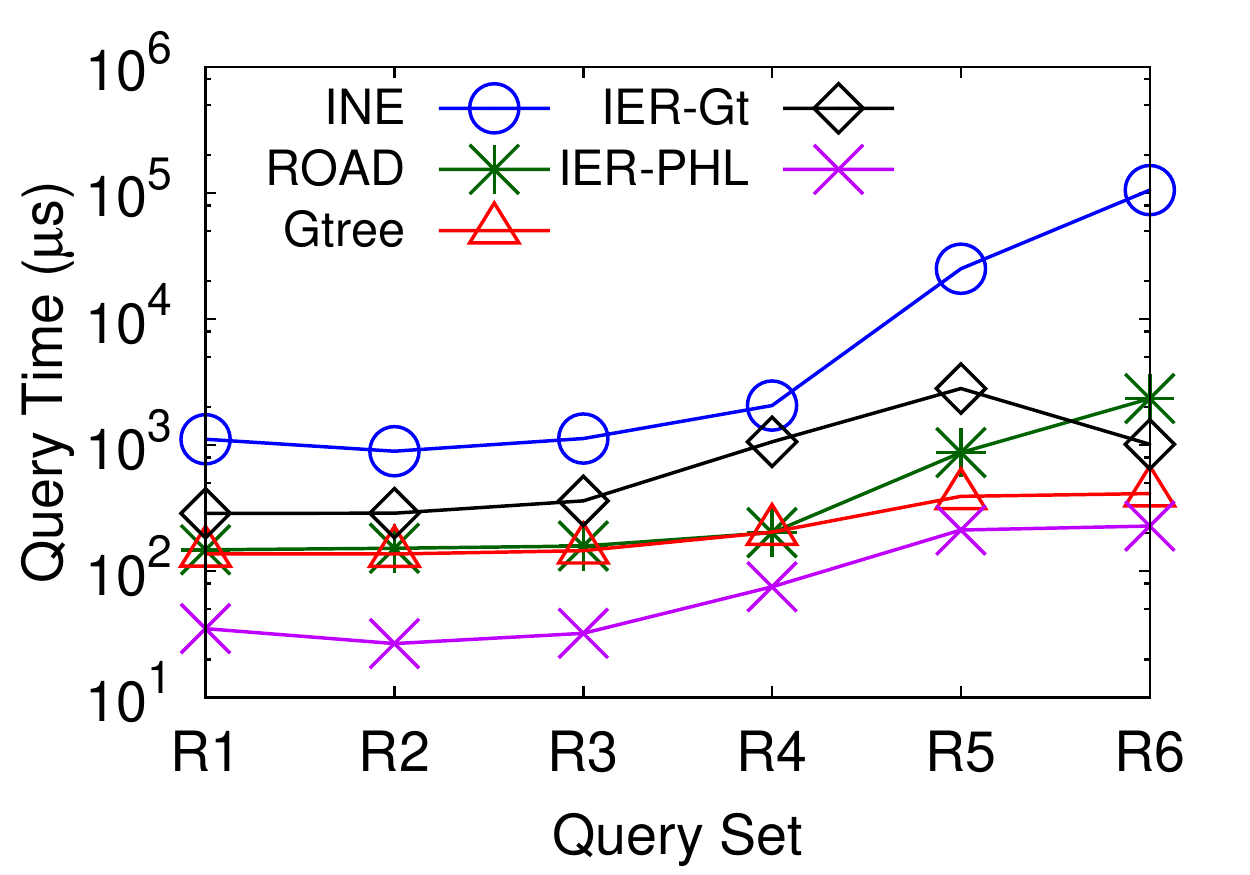}
			\label{app:exp:tt:tt_nw_obj_mindist}
		}		
		\hspace*{-5mm}
		\subfigure[Varying No. of Clusters]{
			\includegraphics[width=0.25\linewidth]{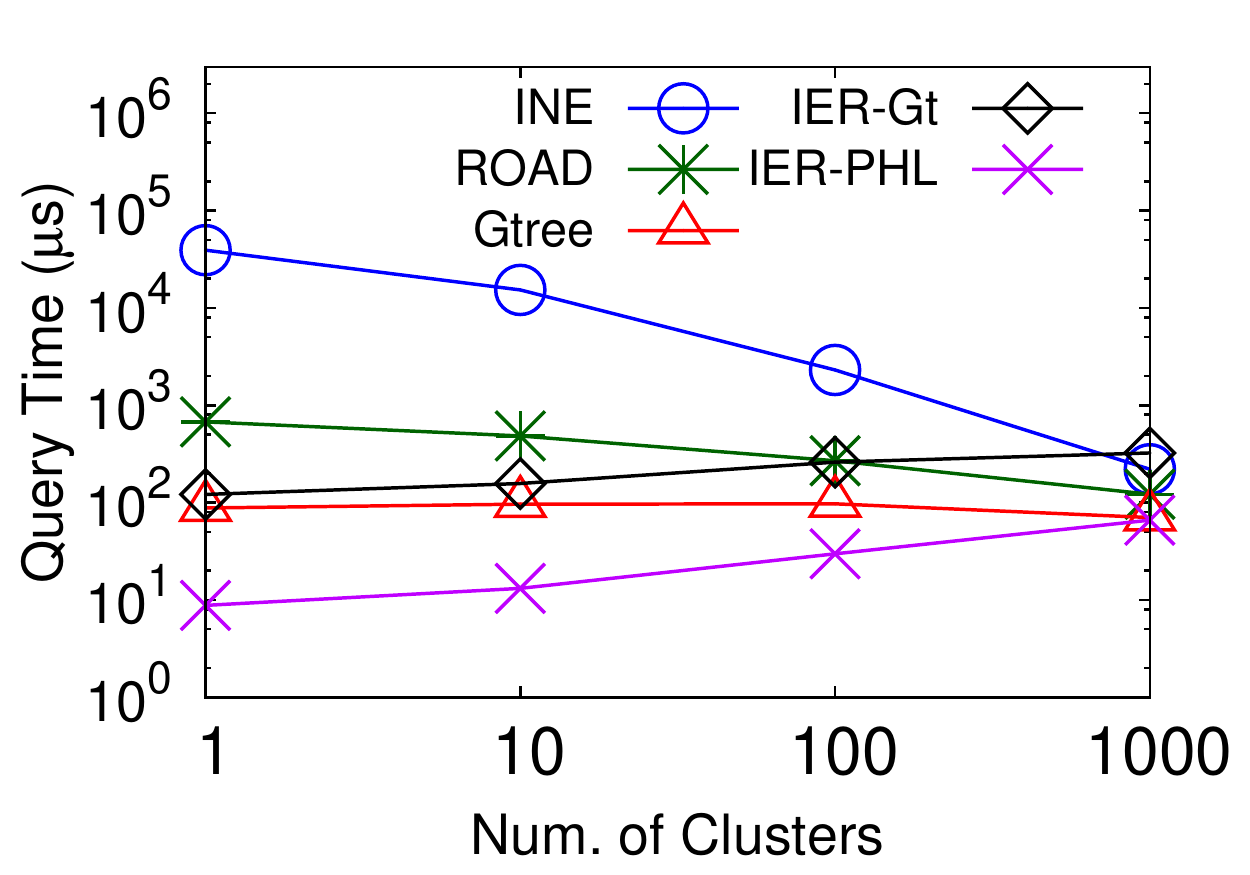}
			\label{app:exp:tt:tt_nw_clusters}
		}
		\caption{Query Performance on Travel Time Graphs \textmd{(NW, $d{=}0.001$, $k{=}10$, uniform objects)}}\label{app:exp:tt:nw}
	\end{figure*}
}

{
	\setlength{\abovecaptionskip}{\abvfigskp}
	\setlength{\belowcaptionskip}{-13pt}
	\begin{figure*}[ht]
		\subfigure[NW Road Network]{
			\includegraphics[width=0.5\linewidth]{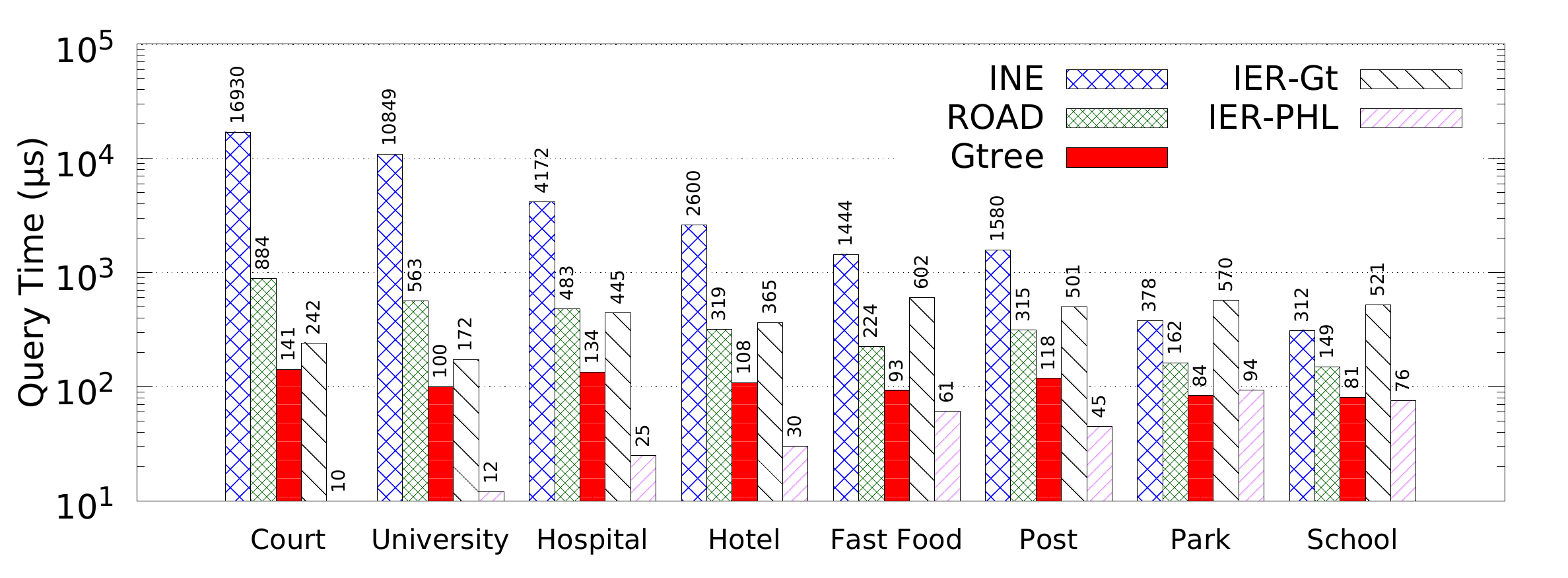}
			\label{app:exp:tt:rw:NW}
		}
		\hspace*{-5mm}
		\subfigure[US Road Network]{
			\includegraphics[width=0.5\linewidth]{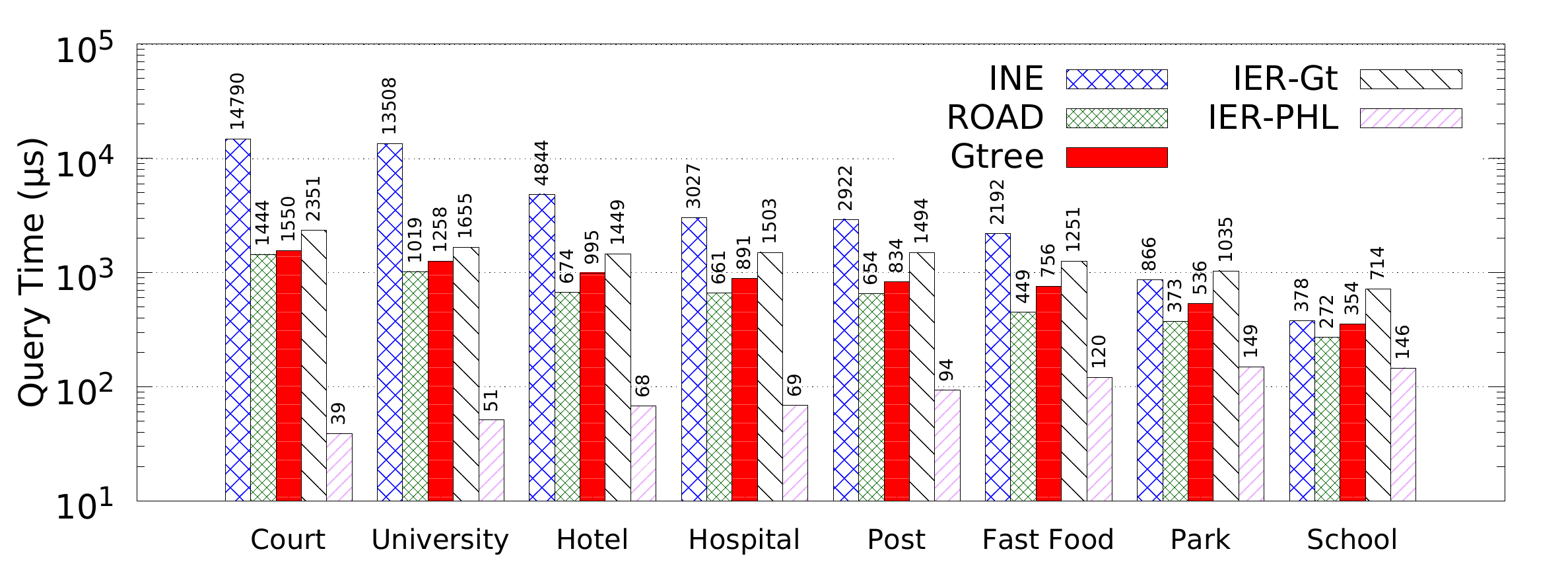}
			\label{app:exp:tt:rw:USA}
		}		
		\caption{Varying Real-World Object Sets for Travel Time Graphs \textmd{(Defaults: $k{=}10$)}}\label{app:exp:tt:rw}
	\end{figure*}
}

\subsection{IER Variants on Travel Times }\label{sec:app:tt:ier}

As discussed in Section \ref{sec:exp:tt}, IER can also be adapted for use on travel time road networks. The performance of shortest path techniques is known to vary between travel distance and travel time graphs. For example CH and TNR has been seen to perform $5{-}20\times$ worse on travel distances \cite{bauer2010chg}. This is because travel distances do not display hierarchies as prominently as travel times. For example, highways may not always provide the shortest travel distance, but generally provide faster travel time. Methods that rely on such properties, such as CH and TNR, are more effective when they are present (as in travel time graphs).

In Figure \ref{app:exp:tt:ier_comparison} we again compare IER with different network distance techniques on NW but with travel time edge weights. All methods perform worse at high densities as IER encounters more false hits as the Euclidean lower bound becomes looser. Interestingly, CH and TNR query times actually do not change significantly from the travel distance case for lower densities while MGtree's does. As mentioned before, despite the greater number of false hits, both these methods are faster on travel time graphs. As a result TNR actually performs better than MGtree on travel times for low densities. In fact all methods perform better on density $0.0001$ than density $0.001$ because there are fewer objects and therefore a smaller chance of having similar distances, leading to fewer false hits. MGtree's performance degrades by the smallest amount on high densities, as its optimised repeated computations make it more robust to the increase in false hits. If TNR were to be combined with MGtree to answer local queries (rather than CH), it may be a better option on NW than just MGtree. Regardless of this, PHL performs significantly better than TNR across the board. Additionally we also compare IER methods for increasing network size $|V|$ in Figure \ref{app:exp:tt:ier_comparison:varyn}. The most notable observation is that TNR deteriorates more rapidly than other methods because, for the same grid size, TNR is able to answer fewer queries using transit nodes with increasing $|V|$. With increasing $|V|$, grid cells contain more vertices, and as a result distances to more $k$NNs must be computed using the slower local method.

\subsection{Road Network Pre-Processing and Space }\label{sec:app:tt:indexes}

Figure \ref{app:exp:tt:indexes} shows the index construction time and index size for the travel time edge weight versions of the road networks in Table~\ref{tab:datasets}. The key difference to travel distances is that PHL is constructed faster (in fact faster than the other methods) and uses significantly less memory allowing it to be constructed for all datasets up to and including the US dataset with 24 million vertices. Travel time graphs display better hierarchies allowing for more effective pruning, leading to smaller label sizes on average. Note we do not need to repeat object index comparisons for travel times as they will be the same as for travel distances. E.g., the same partitioning of the road network is used to construct a G-tree (resp. ROAD) index in either case, which means Occurrence Lists (resp. Association Directories) will be identical.

{
	\setlength{\abovecaptionskip}{\abvfigskp}
	\setlength{\belowcaptionskip}{\belfigskp}
	\begin{figure}[!htbp]
		\vspace{-5mm}
		\subfigure[Construction Time]{
			\includegraphics[width=0.48\linewidth]{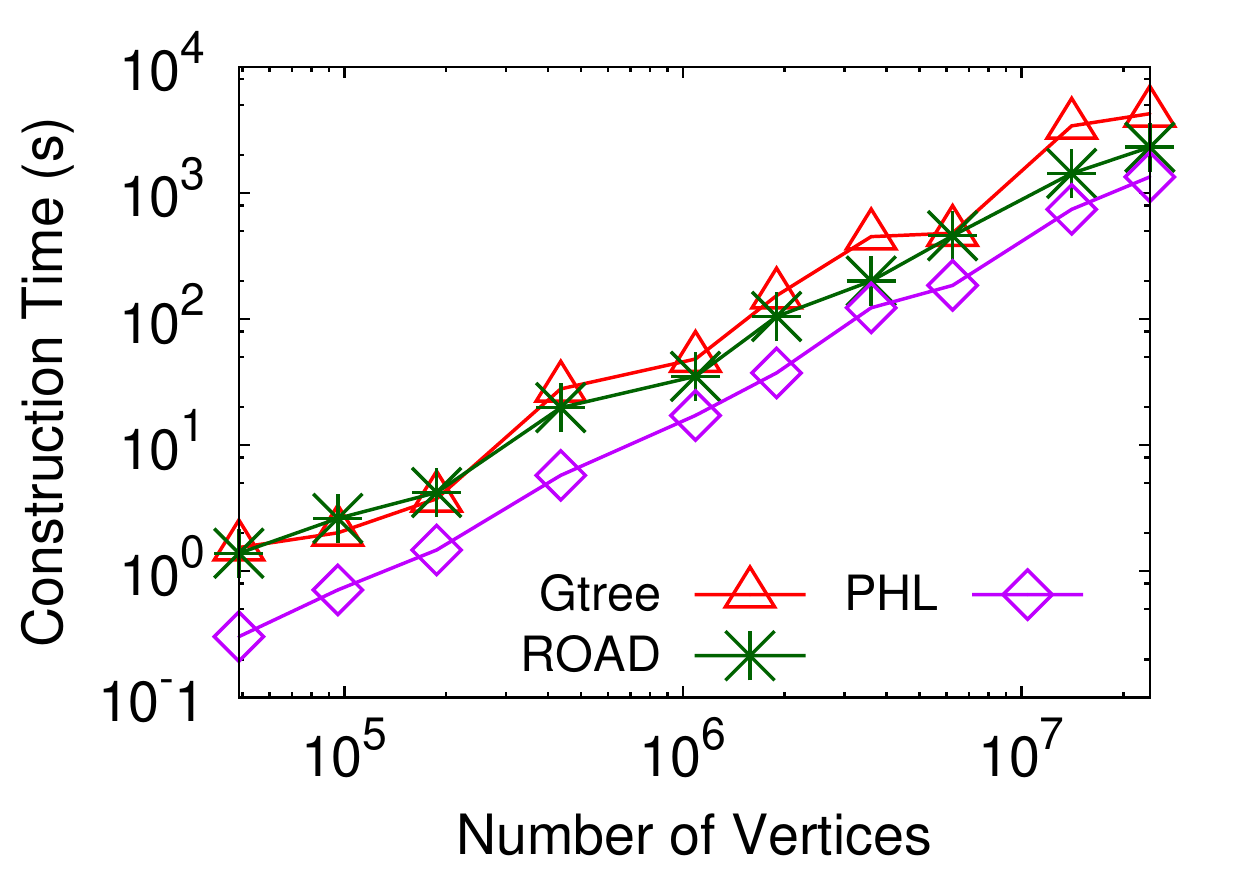}
			\label{app:exp:tt:indexes:time}
		}
		\hspace*{-5mm}
		\subfigure[Index Size]{
			\includegraphics[width=0.48\linewidth]{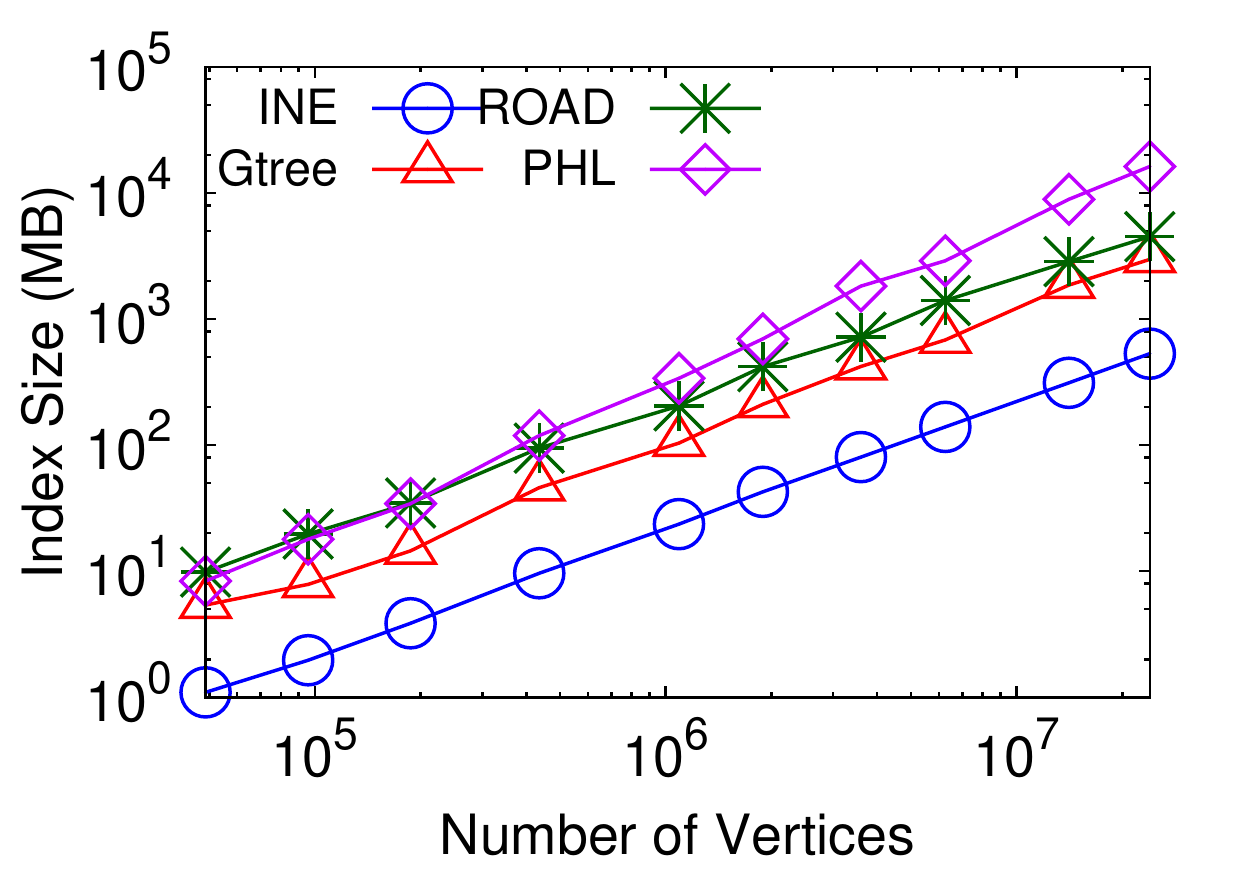}
			\label{app:exp:tt:indexes:size}
		}
		\caption{Effect of Road Network Size $|V|$ (Travel-Time)}\label{app:exp:tt:indexes}
		\vspace{-3mm}
	\end{figure}
}

\subsection{Query Performance }\label{sec:app:tt:query}

In addition to comparison of several varying query settings on the US dataset in Section \ref{sec:exp:tt}, in this section we present results for other experiments on query performance.

\newpage
\subsubsection{Varying Parameters}

Figure \ref{app:exp:tt:nw} shows the equivalent experimental results for the NW dataset for varying parameters on travel time graphs, namely $k$, uniform object density, minimum object distances and number of clusters. We verify that the observations made in Section \ref{sec:exp:tt} are also observed here. For example, IER-PHL is again generally the best performing method except in the case of densities greater than $0.01$ in Figure \ref{app:exp:tt:tt_nw_density_k10}. As before, high densities imply there are more objects at a closer distance. This generates greater numbers of false hits, which is only made worse by the looser lower bound provided by Euclidean distance on travel time graphs.

\subsubsection{Real-World Object Sets}

{
	\setlength{\abovecaptionskip}{\abvfigskp}
	\setlength{\belowcaptionskip}{\belfigskp}
	\begin{figure}[!htbp]
		\vspace{-5mm}
		\subfigure[Hospitals]{
			\includegraphics[width=0.49\linewidth]{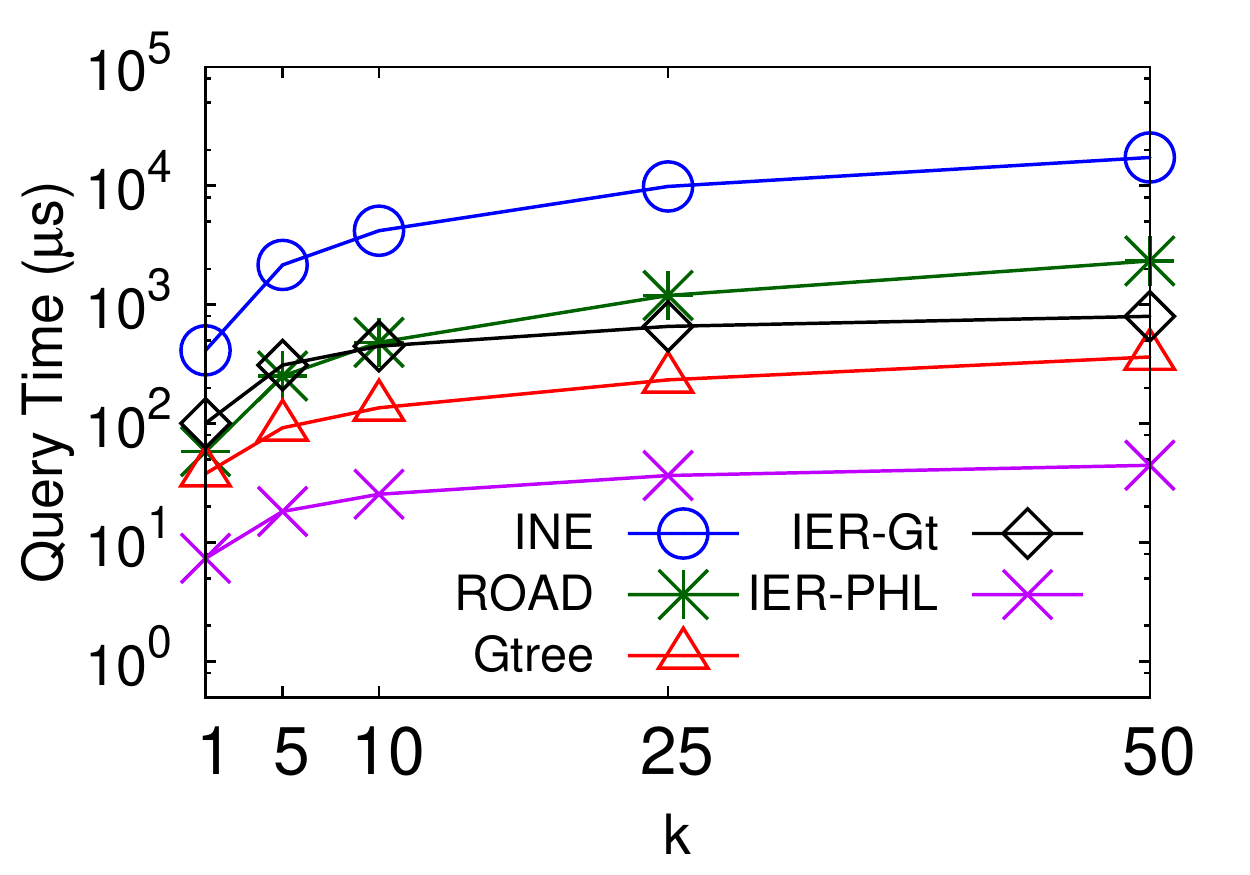}
			\label{app:exp:tt:rw_k:NW_rw_k_hospital}
		}
		\hspace*{-5mm}
		\subfigure[Fast Food]{
			\includegraphics[width=0.49\linewidth]{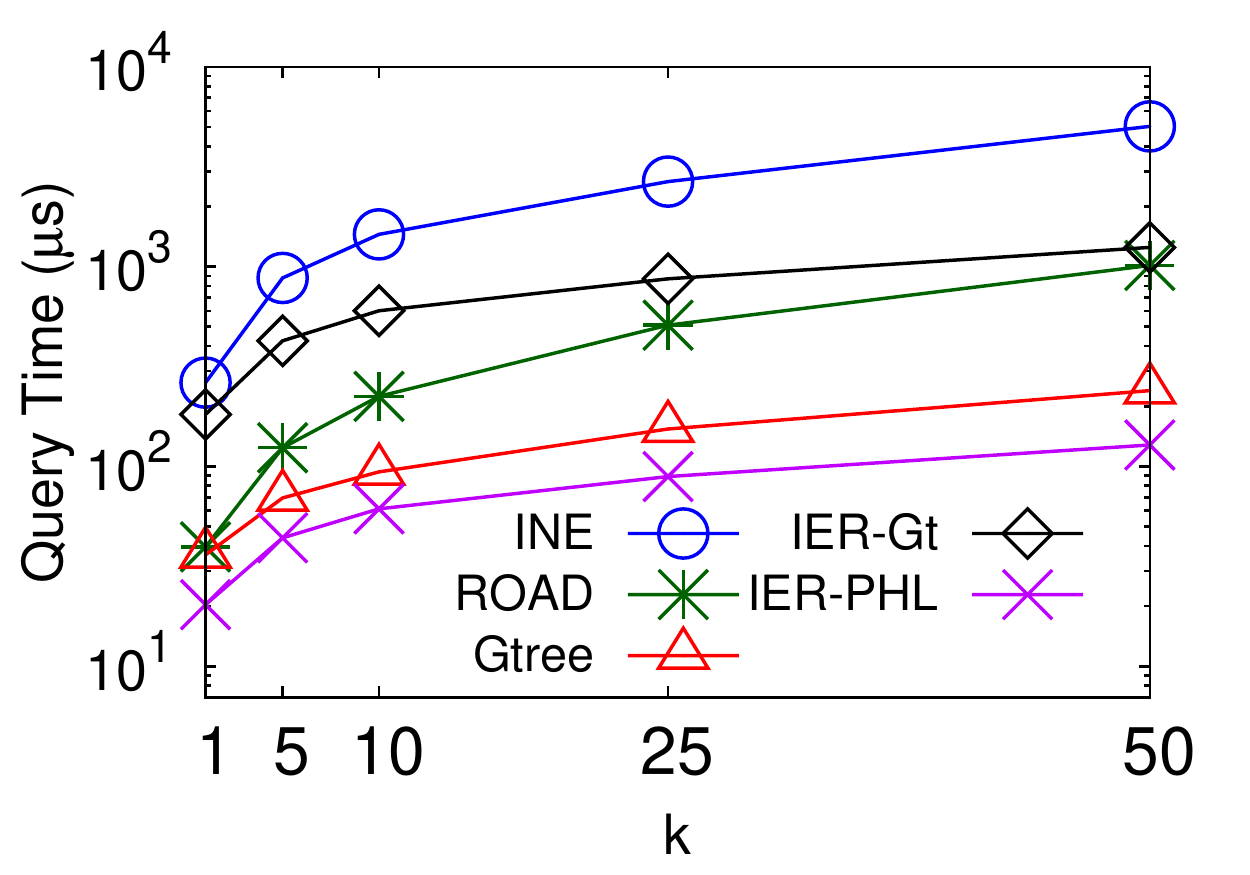}
			\label{app:exp:tt:rw_k:NW_rw_k_fastfood}
		}
		\caption{Varying $k$ for Real POIs (Travel-Time) (\textmd{NW, $k{=}10$)}}\label{app:exp:tt:rw_k}
	\end{figure}
}

We repeat the experiments for real-world object sets in Figures \ref{app:exp:tt:rw} and \ref{app:exp:tt:rw_k}. All observations we have made so far are similarly observed here. For example we observe the same trends for increasing real-world object set size in Figure \ref{app:exp:tt:rw} as for increasing object density. We also observe that G-tree again performs worse on the US dataset than on NW. One big difference is that IER-PHL is included in comparisons for the US dataset in Figure \ref{app:exp:tt:rw:USA} as its index can be constructed for all datasets. This offsets the degraded performance of IER-Gt. The sparse object set (Hospitals) and the clustered object set (Fast Food) again show similar trends, with IER having degraded performance on the clustered object set as it is less able to distinguish candidates. Again this effect is aggravated for travel times.

\newpage
\section{Further Related Work}\label{sec:app:rw}

Road Network Embedding (RNE) \cite{shahabi2002rne} was the first work to propose a solution to the $k$NN problem specifically for road networks. It proposes a better approximation of NNs in a road network than the Euclidean NNs by pre-computing certain shortest path distances and then computing $k$NNs in higher dimensional space.

The Voronoi-based Network Nearest Neighbor (VN\textsuperscript{3}) \cite{kolahdouzan2004vknn} computes the network equivalent of a Voronoi diagram for a given object set to partition all road network vertices based on its nearest neighbor. This partitioning produces borders, and VN\textsuperscript{3} computes and stores network distances between borders of the same region. It then answers $k$NN queries by observing that the next NN must be in some Voronoi region adjacent to the currently visited regions. But lower object set densities result in larger numbers of borders and significant pre-computation overhead.

Other methods such as the Nearest Descendent \cite{hu2006nd}, UNICONS \cite{cho2005unicons} and Islands \cite{huang2005islands} involve pre-computation of the nearest neighbors for some or all vertices. UNICONS and Islands use a parameter to limit how many NNs are computed for each vertex, and UNICONS also limits pre-computation to only some vertices. The performance of both methods degrades when $k$ exceeds the number of pre-computed NNs. While Nearest Descendent does not, it must still store the nearest object (using a different graph representation) for each road network vertex.

A ``full index" essentially stores, for each road network vertex $v$, all objects in order of the network distance from $v$. $k$NN can be answered by simply performing $k$ look-ups. This involves computing the network distances from each $v$ to each object resulting in significant space overhead. The Distance Index \cite{hu2006distidx} provided a means to compress it. This is achieved by categorising network distance into a set of pre-defined ranges and storing only a bit-wise representation of each range. This results in a 1-order of magnitude space reduction over the full index. 

A common theme among these methods, in stark contrast to those we have evaluated, is that $k$NN queries are answered by first creating a single index combining the road network and an object set. This is highly disadvantageous as it will not scale with greater numbers of object sets. We must also reprocess the entire road network for changes to an object set, which may be quite frequent. Thus the current state-of-art has moved towards decoupling the object set from the road network. As a result these methods are unlikely to be useful in practice. These issues were discussed in detail in Section \ref{sec:bg:scope}.

The methods we have mentioned so far have been concerned with static data. While beyond the scope of this study, we briefly identify several continuous $k$NN (C$k$NN) problems for the interested reader. One variation is when the query vertex traverses some path $P$ \cite{cho2005unicons}, and $k$NNs are computed for each vertex in $P$. A related variation is finding the \textit{path nearest neighbors} \cite{chen2009pathknn}, i.e. the single set of $k$ objects that are nearest to the entire path $P$. However these methods assume a static object set, we may also want to find the C$k$NNs given moving objects \cite{mouratidis2006cknn}.

\end{appendix}

\end{document}